\documentclass[aps,prA,amsmath,amsfonts,superscriptaddress,twocolumn]{revtex4}
\usepackage{epsfig,graphicx}
\usepackage{epsfig,psfig,graphicx}
\usepackage[english]{babel}
\usepackage{amsfonts}
\usepackage{amsmath}
\usepackage{latexsym}
\usepackage{epsfig}
\usepackage{graphics}
\usepackage{graphics,bm}
\usepackage{subfigure}
\usepackage{dcolumn}
\usepackage{bm}
\usepackage{rotating}

\def\bk{{\bf k}}

\newcommand{\uk}{u_{\mathbf{k}}}
\newcommand{\vk}{v_{\mathbf{k}}}

\begin{document}
\flushbottom
\raggedbottom

\title{Dressed-molecules in resonantly-interacting ultracold atomic Fermi gases}

\author{G.M. Falco}

\affiliation{Institute for Theoretical Physics, Utrecht
University, Leuvenlaan 4, 3584 CE Utrecht, The Netherlands}

\author{H.T.C. Stoof}

\affiliation{Institute for Theoretical Physics, Utrecht
University, Leuvenlaan 4, 3584 CE Utrecht, The Netherlands}

\begin{abstract}
We present a detailed analysis of the two-channel atom-molecule effective
Hamiltonian for an ultracold two-component homogeneous Fermi gas interacting near a Feshbach resonance. 
We particularly
focus on the two-body and many-body properties of the dressed molecules in such a gas.
An exact result for the many-body ${\rm T}$-matrix of the two-channel theory is 
derived by both considering coupled vertex equations 
and the functional integral methods. The field theory
incorporates exactly the two-body physics of the Feshbach scattering
by means of simple analytical formulas without any fitting parameters. 
New interesting many-body effects are discussed in the case 
of narrow resonances. We give also a description of the BEC-BCS crossover
above and below $T_C$. The effects of different approximations for the selfenergy of the dressed
molecules are discussed.
The single-channel results are derived as a special limit for broad resonances.
Moreover,
through an analytic analysis of the BEC limit,
the relation between the composite boson of the single-channel model and the dressed-molecule
of the two-channel model is established.  
\end{abstract}


\maketitle

\section{Introduction.} 
Since the achievement of the first Bose-Einstein condensate 
in a trapped dilute gas of alkali atoms in 1995 \cite{Anderson95,Davis95,Hulet95}, the 
area of ultracold atomic gases is one of the most active
fields of research in physics. 
The experimental realization of this new state of matter has been made possible 
by trapping an atomic gas cloud in a  magnetic trap and lowering its temperature
down to about 10-100 nK by means of a combination of laser-cooling and 
evaporative-cooling techniques.

The extraordinary progress in the technology of 
trapping and cooling alkali atoms
has been applied in the last two years also to study superfluidity in fermionic gases. 
However, 
in a weakly-attracting mixture of two species of fermionic atoms,
the standard evaporative cooling stops to work in the
degenerate regime and one cannot reach the superfluid transition directly.
For this reason, it is more difficult to reach the fermionic analog of the Bose-Einstein condensation
transition temperature,
i.e., the Bardeen-Cooper-Schrieffer (BCS) temperature for 
Bose-Einstein condensation of 
Cooper pairs.
In a BCS superconductor,
the critical temperature of the superfluid transition
is exponentially small with respect to the Fermi temperature, 
which is the temperature scale associated
with the onset of degeneracy in the gas. 
Moreover, the BCS phase does not affect the density distribution
of the gas in the same manner as in a trapped Bose gas. 
As a result it is not easy to detect the pair
correlations of  the Cooper pairs characterizing the superfluid properties of the gas.
Fortunately, the use of Feshbach resonances 
in atomic Fermi gases {\cite{Feshbach62,Demarco99,Trusco01,Schreck01,Dieck02,Bourdel03}},  
has solved {\cite{Jin04,Zwierlein04a}} both of 
these problems simultaneously.

Near a Feshbach resonance two atoms can virtually form a long-lived bound molecular state
during an $s-$wave collision. The scattering process consists of two incoming atoms 
in an energetically open channel, which has a different hyperfine state than the
bound state in the closed channel. The coupling between the open and the closed
channels is physically provided by the exchange interaction,
i.e., the difference between the singlet $^1\Sigma^+_g$ and triplet
$^3\Sigma^+_u$ potentials of the two alkali atoms in the electronic ground state. 
Moreover, the two channels have a different Zeeman shift in a magnetic
field because of the difference in hyperfine state. Therefore, the energy
difference between the closed-channel bound molecular state and the two-atom continuum 
threshold is experimentally adjustable by tuning the magnetic field.
As a result, the $s-$wave scattering length, and hence the magnitude and sign of the
interatomic interactions, 
can be precisely controlled over a wide range.
 
The possibility of tuning the interactions in a system
is rather unusual in condensed-matter physics and has triggered
many new experimental developments.
In particular, Feshbach resonances provide a remarkable opportunity to study
strongly-interacting fermions. By varying the interaction strength between
the fermionic atoms, the crossover from a Bose-Einstein condensate of molecules
to a BCS superfluid of loosely bound pairs 
\cite{Eagles69,Leggett80,Nozieres85,SadeMelo93,Haussmann94}
has been explored 
{\cite{Jin04,Zwierlein04a,Kinast04a,Bartenstein04,Bourdel04a,Partridge05,Zwierlein05,Zwierlein05b,Partridge05b}}.
These experiments have revealed evidence of fermionic pair 
condensation {\cite{Jin04,Zwierlein04a}} and superfluidity \cite{Zwierlein05}.
Moreover, in a magnetized Fermi gas 
the phase separation between a superfluid paired component and the normal unpaired fermions
has also been observed \cite{Partridge05b,Zwierlein05b}.

Feshbach resonance in atomic and nuclear physics is a multichannel problem. 
Already in 1962, Feshbach \cite{Feshbach62}
developed a general method to 
solve the multichannel elastic scattering equations in the presence of a bound state,
using a formalism based on projection operators techniques. His approach reduces the solution of the
multichannel problem
of the colliding particles to a single-channel problem, namely the scattering of a particle from an optical potential.
In general, the price of this simplification is a complicated non-local 
and energy-dependent optical potential. 
Neglecting the possible occupation of the bound state energy level
and the energy dependence of the optical potential ultimately leads to 
a many-body theory with a local interaction described by a single adjustable parameter,
namely the $s-$wave scattering length $a$. 
This is usually called the {\it{single-channel}} approximation 
{\cite{Eagles69,Leggett80,Nozieres85,SadeMelo93,Haussmann94,RanderiaBook95,
StoofHoubiers96,Heiselberg00,Heiselberg01,Combescot03a,Combescot03b,Combescot05,PieriStrinati03,
Pieri04a,Perali04,Pieri04b,Viverit04,Adenkov04,Ho03,Kagan05,Levinsen05}}
in the literature related to the study of
cold atoms near Feshbach resonances.
A different many-body approach, which 
allows for the population of the bound molecular state and thus
preserves the multichannel nature of the 
Feshbach problem, has also been considered by many authors.
This is based on a {\it{two-channels}} low-energy effective Hamiltonian 
{\cite{Drummond98,Timmermans99,Timmermans01,HollandKok01,Milstein02,Duine04,Ohashi0203,Falco04a,
Bruun03,Stajic03,Falco04b,Drummond04,Romans03,Levin04,Falco05,Romans05,Mackey05,Xia-Ji05}}, 
where the 
effects of the resonant scattering are described by a bosonic molecular field
coupled to the atoms in the Fermi sea. This bosonic field describes
the bound state in the closed channel
near the continuum of the open-channel scattering states.  
As an effect of the coupling with the atoms, 
the physical bosonic degree of freedom 
in the system is not 
this bare molecule but the so-called dressed molecule \cite{Duine04}, namely the bare molecule "dressed"
by the coupling to the continuum of scattering states.
The wavefunction of the latter is always a superposition sharing a component in the closed 
channel and one in the open channel. The probability of the bare closed-channel component
is denoted by $Z$. Note that in this language the single-channel model puts $Z=0$ from the outset. 

In this paper we begin a systematic 
analysis of a two-channel many-body theory 
for a mixture of fermionic atoms near a Feshbach resonance 
based on the dressed-molecule picture.
The paper is organized as follows.
In Sec.~\ref{atomic} we introduce the physics of Feshbach resonant
scattering in ultracold atomic gases.
This preliminary analysis is based on a microscopic atomic physics Hamiltonian
with spin-dependent interactions in the presence of an external magnetic field.

In Sec.~\ref{efffith} we introduce 
the low-energy effective quantum field theory 
above the superfluid critical temperature based on the atom-molecule Hamiltonian.
The field theory is then solved in the so-called many-body ${\rm T}$-matrix approximation. 
We show that this
approach incorporates exactly the two-body scattering properties in the presence of
a Feshbach resonance. In the two-body limit, 
we find new and general analytic expressions for the binding energy and the closed-channel 
component $Z$ of the dressed-molecules, which reproduce experimental data
without any fitting parameters.
Next we rederive the many-body ${\rm T}$-matrix approximation by means of functional integral
techniques. In the limit of a very broad resonance the  many-body ${\rm T}$-matrix  
of the single-channel approximation is recovered. 
Moreover, new interesting many-body effects are also discussed 
in particular in connection with the possibility of having narrow resonances.

In Sec.~\ref{tc} and Sec.~\ref{crossover} the physics of the BEC-BCS crossover  
is investigated.
Section~\ref{tc} considers the problem of determining the superfluid critical temperature $T_c$
approaching the transition from above, while in Sec.~\ref{crossover} 
we turn our attention to the description of the crossover in the superfluid phase below $T_c$.
In both regimes, we give  
a systematic analysis of the crossover equations both in the mean-field approximation \cite{Leggett80} and beyond mean-field,
at the level of the Gaussian fluctuations \cite{Nozieres85,SadeMelo93,Pieri04a} and 
according to a self-consistent approach \cite{Haussmann94}.
We focus expecially on the BEC limit of the crossover, 
where analytical calculations are possible.
In the case of broad resonances, we show that the dressed-molecule approach \cite{Falco04b,Romans05} is 
consistent with the single-channel model results. 
The relation between the {\it{composite boson}}
of the single-channel model and the dressed-molecule 
of the two-channel approach is exstablished also at the mathematical level.

\section{Feshbach resonances in atomic Fermi gases.}
\label{atomic}
At low temperatures only $s$-wave scattering is important in the gas, since
the centrifugal barrier prevents higher-order partial waves
to enter the interaction region. 
Due to the Pauli principle occurring
in a gas of fermionic particles, interaction effects can therefore only be observed 
when we consider a mixture of two spin states.
For this reason we only consider such mixtures in the following.

Feshbach resonances occur in atomic physics when studying 
collisions of ultracold atoms under
variation of an external magnetic field \cite{Stwalley76,Tiesinga93}.
In the presence of an external magnetic field $B$ the 
spin-dependent part of the single-atom Hamiltonian is
\begin{equation}
\label{eq:SingleAtomSpinHamiltonian}
{H}
_{\rm{int}}=\left(\frac{a_{\rm{hf}}}{\hbar^2}\right)
{{\mathbf{s}}\cdot{\mathbf{i}}}+\mathbf{B}\cdot\frac
{2 \mu_e {\mathbf{s}}-\mu_N {\mathbf{i}}}{\hbar}, 
\end{equation}
where the energy $a_{\rm{hf}}$ depends on the atomic species and 
characterizes the hyperfine interaction between the electronic spin
${\mathbf{s}}$ and the nuclear spin ${\mathbf{i}}$.
The two constants $\mu_e$ and $\mu_N$ are respectively the 
electronic and nuclear magnetic moments
of the alkali atom of interest
and we always have that $\mu_e\gg\mu_N$.

The diagonalization of this Hamiltonian
yields the hyperfine states,
which, at zero magnetic field, have 
the magnitude $f$ of the total atomic spin $\mathbf{f}=\mathbf{s}+\mathbf{i}$
and its projection on the $z-$axis $m_f$ as good quantum numbers.
At high magnetic fields $B\gg a_{\rm{hf}}/
\mu_e\hbar$, the hyperfine interaction can be treated as a perturbation,
and, to lowest order, the electronic and nuclear spin projections
$m_s$ and $m_i$ are the good quantum numbers.
At intermediate values of the magnetic field, 
where the Feshbach resonances are often observed,
the atomic eigenstates are a linear combination 
of the latter states, that schematically can be written as
\begin{align}
\label{eq:HyperStates}
c_{-\frac{1}{2}}|m_s=-\frac{1}{2};m_i=m+\frac{1}{2}>+\nonumber\\
+c_{\frac{1}{2}}|m_s=\frac{1}{2};m_i=m-\frac{1}{2}>\nonumber. 
\end{align}
This implies that only the total spin projection $m_f$ is a good quantum number.
Without much loss of generality, we consider a gas of $^6$Li atoms from now on.
This species is most widely used in
current experiments with trapped ultracold Fermi gases.
The $^6$Li isotope has nuclear spin $i=1$, which gives 
in increasing order of energy, the six hyperfine states 
\begin{eqnarray}
\label{eq:hyperfinestates}
|1\rangle & = & \sin \theta_+ |1/2\:0\rangle - 
                \cos \theta_+|-\!1/2\:1\rangle \\
|2\rangle & = & \sin \theta_- |1/2\:-\!1\rangle -
                \cos \theta_-|-\!1/2\:0\rangle \nonumber\\
|3\rangle & = & |-\!1/2\:-\!1\rangle \nonumber\\
|4\rangle & = & \cos \theta_- |1/2\:-\!1\rangle +
                \sin \theta_- |-\!1/2\:0\rangle \nonumber\\
|5\rangle & = & \cos \theta_+ |1/2\:0\rangle + 
                \sin \theta_+|-\!1/2\:1\rangle \nonumber\\
|6\rangle & = & |1/2\:1\rangle,\nonumber
\end{eqnarray}
with $\sin \theta_{\pm} = 1/\sqrt{1+(Z^{\pm} + R^{\pm})^2/2}$,
$Z^{\pm} = (\mu_n + 2\mu_e)B/a_{hf} \pm 1/2$, and  $R^{\pm} = \sqrt{(Z^{\pm})^2 + 2}$. 
Here the states
$|m_s\: m_i\rangle$ have a well-defined
projection of the electron and nuclear spin, respectively. 

So far, we have considered only the spin degrees of freedom
of a single atom. The spin-dependent
Hamiltonian of two atoms is \cite{Stoof86}
\begin{equation}
\label{eq:TwoAtomSpinHamiltonian}
{H}_{\rm{int}}=\left(\frac{a_{\rm{hf}}}{\hbar^2}\right)
\left(
{{\mathbf{s}}_1\cdot{\mathbf{i}}_1}+{{\mathbf{s}}_2\cdot{\mathbf{i}}_2}
\right)
+\mathbf{B}\cdot\frac{2 \mu_e \cdot{\mathbf{S}}-\mu_N\cdot{\mathbf{I}}}{\hbar}, 
\end{equation}
where $\mathbf{S}=\mathbf{s}_1+\mathbf{s}_2$ and $\mathbf{I}
=\mathbf{i}_1+\mathbf{i}_2$ are the total electronic and nuclear spin, respectively.
\begin{figure}
\begin{center}
\vskip 1 cm
\resizebox{8cm}{4.6cm}{
 \includegraphics{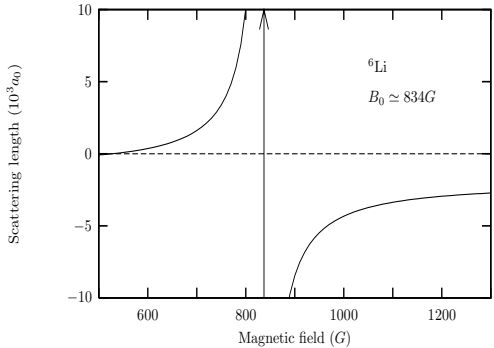}
 \hspace{2 cm}} 
\end{center}
  \caption{Scattering length as a function of the magnetic field
  for two $^6$Li 
  atoms in the hyperfine states $|1\rangle$ and  $|2\rangle$
  near the broad resonance at $834~G$ \cite{Houbiers98}.}
  \label{resa}
\end{figure}
In view of the ongoing experiments, 
we consider in the following a hyperfine mixture
of the states $|1\rangle$ and $|2\rangle$.
Hence, the incoming channel for 
two-atom scattering is the antisymmetric state
\begin{equation}
|\{1 2\}\rangle\equiv \frac{1}{\sqrt{2}} \left[ |1\rangle_1|2\rangle_2
-|2\rangle_1|1\rangle_2 \right].
\end{equation}
This combination can only be trapped in 
a far-off resonance optical trap \cite{KetterleVarenna99}
and not in a magnetic trap. It is, however,
the most favourable mixture experimentally, because it cannot decay
through inelastic two-body collisions.
For future reference it is useful to rewrite this state as
the linear superposition
\begin{align}
|\{12\}\rangle  = & \sin \theta_+\sin \theta_-|1\:1; 1\:-1\rangle+
\\
&
\sin \theta_+\cos \theta_-\left(
\frac{1}{\sqrt{3}}|0\:0; 0\:0\rangle-\sqrt{\frac{2}{3}}|0\:0; 2\:0\rangle
\right)+\nonumber\\
&\cos \theta_+\sin \theta_-\nonumber\\
&\times
\left(
\frac{1}{\sqrt{3}}|0\:0; 0\:0\rangle+
\frac{1}{\sqrt{6}}|0\:0; 2\:0\rangle-
\frac{1}{\sqrt{2}}|1\:0; 1\:0\rangle
\right)
\nonumber\\
&
+\cos \theta_+\cos \theta_-|1\:-1; 1\:1\rangle\nonumber
\end{align}
of the basis states $|SM_S;IM_I\rangle$, which diagonalize the second term in
the right-hand side of 
Eq.~(\ref{eq:TwoAtomSpinHamiltonian}). 
The quantum numbers $M_S$ and $M_I$ represent the total electronic and nuclear spin projection
along the direction of the magnetic field, respectively.

The central interaction for two colliding atoms depends on the magnitude
of the total electronic spin. 
It can be written as \cite{Stoof86}
\begin{equation}
{\rm{V}}_{\rm C}={\rm{V}}_{\rm S}\cdot{\rm{P}}_{\rm S}+{\rm{V}}_{\rm T}\cdot{\rm{P}}_{\rm T},
\end{equation}
by using the projection operators
onto states of definite total electronic spin $S$. 
The singlet and triplet potentials ${\rm{V}}_{\rm{S},\rm{T}}$ are 
by now rather accurately known from
atomic-physics measurements.
The central interaction ${\rm{V}}_{\rm C}$ is diagonal in
the basis $|SM_S;IM_I\rangle$ but induces transitions between different hyperfine levels.
More precisely, the central interaction can be rewritten as
\begin{equation}
{\rm{V}}_{\rm C}={\rm{V}}_{\rm{dir}}+{\rm{V}}_{\rm{exch}},
\end{equation}
with the direct interaction given by
\begin{equation}
{\rm{V}}_{\rm{dir}}=\frac{({\rm{V}}_{\rm S}+{\rm{V}}_{\rm T})}{2},
\end{equation}
and the exchange interaction defined as 
\begin{equation}
{\rm{V}}_{\rm{exch}}=\frac{({\rm{P}}_{\rm S}-{\rm{P}}_{\rm T})
({\rm{V}}_{\rm S}-{\rm{V}}_{\rm T})}{2}.
\end{equation}
If we describe a collision in the hyperfine basis,
it is this latter interaction that is responsible for the
coupling of different hyperfine states in the quantum
collision of cold alkali atoms.
Only transitions where the quantum number $M_S+M_I$ is conserved are allowed,
since the central interaction cannot change the total electron or nuclear spin.
Therefore, the state $|\{1 2\}\rangle$ couples 
only to the states
\begin{align}
\label{eq:canali}
 |\{14\}\rangle &=\sin \theta_+\cos \theta_-
 \frac{1}{\sqrt{2}}|1\:1; 1\:-1\rangle\\
  +&\left(
  \sqrt{\frac{2}{3}}\sin \theta_+\sin \theta_- +\frac{1}{\sqrt{6}}
  \cos \theta_+\cos \theta_-\right)|0\:0; 2\:0\rangle
 \nonumber\\
 +&\left(
 \frac{1}{\sqrt{3}}\cos \theta_+\cos \theta_- -\frac{1}{\sqrt{3}} 
 \sin \theta_+\sin \theta_-
  \right)|0\:0; 0\:0\rangle\nonumber\\
  -&\cos \theta_+\cos \theta_-
  \frac{1}{\sqrt{2}}|1\:0; 1\:0\rangle \nonumber\\-&
  \cos \theta_+\sin \theta_-|1\:-1; 1\:1\rangle,  
\nonumber\\ 
 |\{2 5\}\rangle&=\cos \theta_+\sin \theta_-|1\:1; 1\:-1\rangle\nonumber\\
 -&\left(
 \frac{1}{\sqrt{3}}\sin \theta_+\sin \theta_- -\frac{1}{\sqrt{3}} 
 \cos \theta_+\cos \theta_-
  \right)|0\:0; 0\:0\rangle\nonumber\\
  -&\left(
  \sqrt{\frac{2}{3}}\cos \theta_+\cos \theta_- +\frac{1}{\sqrt{6}}
  \sin \theta_+\sin \theta_-\right)|0\:0; 2\:0\rangle
 \nonumber\\
 +&\frac{1}{\sqrt{2}}\sin \theta_+\sin \theta_-|1\:0; 1\:0\rangle\nonumber\\-&
 \sin \theta_+\cos \theta_-|1\:-1; 1\:1\rangle,
\nonumber\\ 
 |\{4 5\}\rangle&=\cos \theta_+\cos \theta_-|1\:1; 1\:-1\rangle\nonumber\\
 -&\left(
 \frac{1}{\sqrt{3}}\cos \theta_+\sin \theta_- +\frac{1}{\sqrt{3}}
 \sin \theta_+\cos \theta_-
 \right)|0\:0; 0\:0\rangle\nonumber\\
 +&\left(
 \sqrt{\frac{2}{3}}\cos \theta_+\sin \theta_- -\frac{1}{\sqrt{6}}
 \sin \theta_+ \cos \theta_-
 \right)|0\:0; 2\:0\rangle\nonumber\\
 +&\frac{1}{\sqrt{2}}\sin \theta_+ \cos \theta_-|1\:0; 1\:0\rangle\nonumber\\+&
 \sin \theta_+\sin \theta_-|1\:-1; 1\:1\rangle, 
 \nonumber\\
 |\{3 6\}\rangle&=\frac{1}{\sqrt{3}}|0\:0; 0\:0\rangle+
 \frac{1}{\sqrt{6}}|0\:0; 2\:0\rangle+\frac{1}{\sqrt{2}}|1\:0; 1\:0\rangle.
 \nonumber
\end{align}
The magnetic field dependence of the $s-$wave scattering length $a(B)$
in the $|\{1 2\}\rangle$ channel is 
obtained by solving
the system of $5$ coupled-channels equations for these channels.
The scattering length exhibits several resonances 
as an effect of the interplay between
the different channels in the scattering process.
A very broad resonance occurring at $834~ G$ is shown in Fig.~\ref{resa}.

Physically, this Feshbach resonance arises when the energy of 
the atoms with the incoming spin
wavefunction $|\{1 2 \}\rangle$, which is almost purely triplet near resonance, coincides with
the energy of a bound-state energy in the singlet potential ${\rm{V}}_{S}$.
It is convenient for this purpose to rewrite the 
interaction piece of the 
total Hamiltonian as
\begin{align}
\label{eq:bisTwoAtomSpinHamiltonian}
{H}_{\rm{spin}}=&
\mathbf{B}\cdot\frac{2 \mu_e \cdot{\mathbf{S}}-\mu_N\cdot{\mathbf{I}}}{\hbar}+
\frac{a_{\rm{hf}}}{\hbar^2}
\left({\mathbf{s}}_1\cdot
{\mathbf{i}}_1+{\mathbf{s}}_2\cdot
{\mathbf{i}}_2\right)
+{\rm{V}}_{\rm C}
\nonumber\\
\equiv&{H}_{\rm{Z}}+
\frac{a_{\rm{hf}}}{\hbar^2}
\left({\mathbf{s}}_1\cdot
{\mathbf{i}}_1+{\mathbf{s}}_2\cdot
{\mathbf{i}}_2\right)
+{\rm{V}}_{\rm C}
\nonumber\\
\equiv&{H}_{\rm{Z}}+{H}_{\rm{hf}}+{\rm{V}}_{\rm C}.
\end{align}
because $|SM_S;IM_I\rangle$,
diagonalizes the ${H}_{\rm{Z}}+{\rm{V}}_{\rm C}$ operator.
The hyperfine interaction ${H}_{\rm{hf}}$ is not diagonal on the $|SM_S;IM_I\rangle$ states.
The latter couples states with $I$ and $S$ quantum numbers differing by one. 
Classically, it gives rise to the independent single-atom precessions
of $\mathbf{s}_i$ and $\mathbf{i}_i$ about their sum vector $\mathbf{f}_i$.
In describing the collision, we project the spin of the 
asymptotic incident state $|\{1 2\}\rangle$ onto the 
new collision basis $|SM_S;IM_I\rangle$
and approximate the spin state with the one having almost probability one.
During the collision,
the hyperfine coupling ${H}_{\rm{hf}}$
can induce transitions to others, spin-flipped collision channels.
A Feshbach resonance occurs when one of these supports
a bound molecular state $\phi_{\rm{m}}\left(\mathbf{r}\right)|S'M_S';I'M_I'\rangle$ with energy
$E_{\rm{m}}$, which lies near the continuum level of the incident $|\{1 2\}\rangle$ channel.
\begin{figure}
\begin{center}
 \resizebox{8cm}{!}{
 \includegraphics{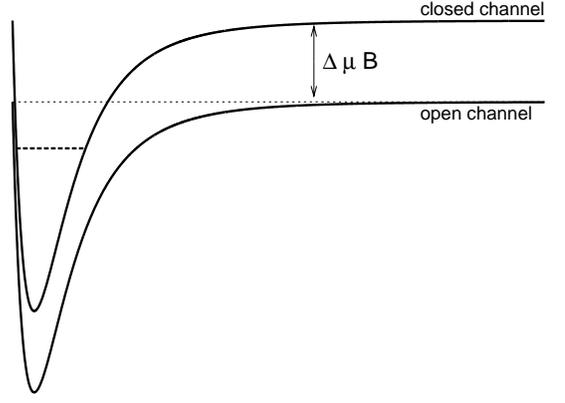}
 } 
  \caption{Illustration of a Feshbach resonance. The upper
  potential curve corresponds to the closed channel interaction potential
  that contains the bound state responsible for the Feshbach resonance.
  The lower potential curve corresponds to the open-channel interaction potential.} 
  \label{feshplot}
   \end{center}
\end{figure}
The channel of the metastable bound state is said to be closed because,
due to energy conservation,
atoms can be observed asymptotically only in the incident open channel.
The situation is illustrated in Fig.~\ref{feshplot}.

At the high magnetic fields where
the resonance occurs, the hyperfine interaction represents a small 
perturbation, and the total spin projections $M_S$ and $M_I$ are rather
good quantum numbers. Therefore, we can approximate the basis
$|\{ij\}\rangle$ with the basis $|SM_S;IM_I\rangle$.
More precisely, in the limit of high magnetic field
we have $\sin \theta_{\pm}\simeq0$
and the six states in Eq.~(\ref{eq:hyperfinestates}) can be approximated
by
\begin{eqnarray}
\label{eq:spinstates}
|1\rangle & \simeq &  
                |-\!1/2\:1\rangle \\
|2\rangle & \simeq & 
                |-\!1/2\:0\rangle \nonumber\\
|3\rangle & \simeq & |-\!1/2\:-\!1\rangle \nonumber\\
|4\rangle & \simeq & |1/2\:-\!1\rangle  \nonumber\\
|5\rangle & \simeq & |1/2\:0\rangle \nonumber\\
|6\rangle & \simeq & |1/2\:1\rangle,\nonumber
\end{eqnarray}
which implies that the scattering channel $|\{1 2\}\rangle$ is almost the pure
triplet state
\begin{align}
|\{12\}\rangle  \simeq-|1\:-1; 1\:1\rangle.
\end{align}
When the magnetic field is tuned  between $800~ G$ and $900~G$,
the energy of the incoming channel $|\{1 2\}\rangle$ 
lies near the $v=38$ molecular bound state energy of the singlet potential 
$\rm{V}_{\rm{S}}$.
As a result a very broad Feshbach resonance occurs.
The closed channel $|00;00\rangle$, supporting the bare molecular state,
can be written by using Eq.~(\ref{eq:canali}) and taking the limit
of high magnetic field as 
\begin{eqnarray}
\label{eq:baremolstate}
|00;00\rangle\simeq \frac{1}{\sqrt{3}}\left(
|\{36\}\rangle+|\{25\}\rangle
+|\{14\}\rangle
\right).
\end{eqnarray}
The width of the resonance is thus determined by 
the matrix element 
\begin{eqnarray}
\label{eq:resstrength}
\langle00;00|{H}_{\rm{hf}}
|1\:-1; 1\:1\rangle\nonumber,
\end{eqnarray}
which describes the overlap 
induced by the hyperfine coupling between the 
spin wavefunctions of the closed and 
the open channel. The total atom-molecule coupling involves of course
also the overlap between the relevant spatial wavefunctions of
the atomic potentials. 

Another resonance, a very narrow one, occurs at about $ 549 ~G$ when the 
energy of the open channel approaches the
energy of the same $v=38$
bound state of the singlet potential, but now in the channel
\begin{eqnarray}
\label{eq:narrowres}
|00;20\rangle\!\simeq \!\frac{1}{\sqrt{3}}\left(
\frac{1}{\sqrt{2}}|\{36\}\rangle-\sqrt{2}|\{25\}\rangle
+\frac{1}{\sqrt{2}}|\{14\}\rangle
\right)\!\!,
\end{eqnarray}
calculated also from Eq.~(\ref{eq:canali}) in the limit
of high magnetic field, i.e.,
approximating $\sin \theta_{\pm}\simeq0$ and $\cos \theta_{\pm}\simeq1$
at lowest order in $a_{\rm{hf}}/\mu_{\rm{e}} B$. 
The matrix element related to the width of the resonance is in this approximation
\begin{eqnarray}
\label{eq:resstrength1}
\langle00;20|{H}_{\rm{hf}}
|1-1;11\rangle\nonumber,
\end{eqnarray}
which is of the same order as the matrix element in Eq.~(\ref{eq:resstrength}).
Hence, the different width of these two resonances 
does not originate from the overlap in the spin parts of the wavefunctions but 
in the overlap of the wavefunctions of
the spatial degrees of freedom.

In the absence of the hyperfine coupling the energy difference between the closed 
and the open channels is linear in the magnetic field, i.e., 
\begin{align}
\label{eq:zerocouplibindenergy}
\epsilon_{\rm{m}}(B)=E_{\rm{m}}+2\mu_{B} B,
\end{align}
where we have used that the difference in magnetic moment 
$\Delta\mu$ between the closed singlet channel and the open triplet channel is 
twice the Bohr's magneton or $\Delta\mu=2\mu_{B}$.
The hyperfine coupling 
produces a shift in the 
position of the magnetic field at which the resonance occurs \cite{Moerdelijk95}.
It is this shifted value $B_0$ that is observed experimentally.
The exact location of the resonance can be used to redefine
the energy difference between the bound state in the closed channel
and the threshold continuum. This is 
\begin{align}
\label{eq:detuning}
\delta(B)=\Delta\mu \left(B-B_0\right),
\end{align}
and is called the detuning because it represents the variable that can be tuned
experimentally by varying the magnetic field. 
At large negative detuning, the binding energy 
$\epsilon_{\rm{m}}(B)$ of the diatomic
molecules,
i.e., the energy you need to break up a molecule,
goes linearly in the magnetic field
$B$ as $\epsilon_{\rm{m}}(B)\simeq \delta(B)$.
Near resonance, the hyperfine coupling leads to a quadratic magnetic-field dependence 
of the binding energy
according to 
Wigner's formula 
\cite{Sakurai82}
\begin{eqnarray}
\label{eq:WignLaw}
\epsilon_{\rm{m}}(B)=-\frac{\hbar^2}{m a^2(B)},
\end{eqnarray}  
where the total scattering length $a(B)$ diverges near resonance as
\begin{eqnarray}
\label{eq:ascat}
a(B)=a_{\rm{bg}}\left(1-\frac{\Delta B}{B-B_0}\right).
\end{eqnarray} 
The width of the resonance $\Delta B$ is a phenomenological parameter
and the scattering length $a_{\rm{bg}}$ describes the nonresonant scattering
in the open channel. 
At positive detuning, there no longer exists a bound state because the molecule decays into
two free atoms due to the coupling
 with the atomic continuum.

In conclusion, Feshbach resonances in atom-atom scattering involve intermediate states
that are molecular bound states. 
For these molecules, the electronic and nuclear spins have been
flipped  from the spin of the colliding atoms by virtue of the 
exchange interaction.  
Next we give a brief account of some recent experiments in ultracold Fermi gases
that make use of Feshbach resonances.
This account is, however, by no means complete.
We discuss only the experimental results which are relevant in connection with the theory we want to discuss
in the next.

By sweeping an external magnetic field 
across a Feshbach resonance, it is possible to
create bosonic diatomic molecules in an ultracold gas of 
fermionic atoms trapped in an optical trap
\cite{Greiner03,Strecker03,Cubizolles03,Jochim03a,Jochim03b,Zwierlein03}.
The optical potential has the advantage to trap atoms in any spin state,
as well as the molecules created from these atoms.
The Fermi temperature $T_F$ of these systems 
is usually of the order of $0.1-1\mu K$. 
Cooling fermionic atoms down to temperatures $T\ll T_F$
below quantum degeneracy is problematic under normal conditions,
because the standard evaporative cooling stops to work in the
degenerate regime.
However, in a mixture with two different spin states,
the large cross section for elastic scattering near the Feshbach resonance can be used
for efficient evaporative cooling, expecially at negative
scattering lengths above the resonance where inelastic loss is negligible
\cite{O'Hara02}.
This permits experiments to reach temperatures down to $0.05 T_F$.

Trapped molecules can be created adiabatically, by sweeping slowly
through a Fesh\-bach resonance.
If the temperature of the gas is comparable or lower than the binding energy of the
molecular state, an almost pure ultracold molecular gas forms. At even lower
temperatures, the molecules form a Bose-Einstein condensate.
The process is reversible because the gas can be converted back to atoms by reversing
the sweep.
These molecules are vibrationally highly-excited states and would usually undergo
fast relaxational decay into deeply-bound states by molecule-molecule 
and atom-molecule collisions.
This is true for 
Feshbach molecules made with bosonic atoms \cite{Herbig03,Durr04}. In that case, the lifetime
of the molecules 
can be increased only by trapping the sample in an optical lattice \cite{Thalhammer05,Dennis05b}.
However, 
$^6$Li$_2$ and $^{40}$K$_2$ molecules,
consisting of fermionic atoms, 
show very long lifetimes \cite{Greiner03,Strecker03,Cubizolles03,Jochim03a}.

The origin of such a long lifetime has been explained by Petrov 
{\it{et al.}}
in \cite{Petrov04} as follows.
The size of the weakly-bound molecules 
with binding energy $\hbar^2/ma(B)^2$
is of order $a(B)$. This length is much 
larger than the characteristic radius of interaction $r_0$. The relaxation requires
the presence of at least three fermions at distances 
of about $r_0$ from each other.
As there are only two different spin states, two particles are 
necessarily identical and the Pauli exclusion principle 
prevents this from happening and thus suppresses
the relaxation probability.

The bound molecule has a magnetic moment which differs from that of the unbound atoms pairs.
The difference in magnetic moment facilities Stern-Gerlach selection of molecules
and atoms. Jochim {\it{et al.}}~\cite{Jochim03a} describe a purification scheme based on a Stern-Gerlach
technique that efficiently removes all the atoms,
while leaving all molecules trapped.
The basic idea of this method consists in applying a magnetic field gradient $B'$ perpendicularly
to the axis of the optical dipole trap. The magnetic field pulls 
out of the trap particles for which the magnetic force is larger than the trapping force.
The value $B'_{\rm{at}}$ at which all the atoms are lost turns out to be smaller than the 
value $B'_{\rm{mol}}$ at which molecules 
start to spill out of the
the trap. This means that 
in principle one can remove all the atoms while keeping the 
total number of molecules constant.
The magnetic moment of the molecules can be estimated from the relation
$\mu_{\rm{mol}}=2 \mu_{\rm{at}}
B'_{\rm{mol}}/
B'_{\rm{at}}$,
where $2\mu_{\rm{at}}$ is the magnetic moment of two unbound atoms in the open channel.
In a gas 
of $^6$Li atoms at low density and  
at high magnetic fields, the latter is almost completely in the triplet state which 
implies $2\mu_{\rm{at}}=2 \mu_{B}$.
We have mentioned this experiment because the magnetic moment of the Feshbach molecule
at negative detuning near the resonance it will be one of the crucial quantities in order to verify
that our theory incorporates the two-body physics exactly.

The long lifetime of the molecules 
makes it possible to observe a Bose-Einstein condensate of 
K$_2$ \cite{Greiner03} and Li$_2$ \cite{Jochim03b,Zwierlein03}
Feshbach molecules.
Zwierlein {\it{et al.}} create in Ref.~\cite{Zwierlein03}, for example, 
the Bose-Einstein condensate of Li$_2$ Feshbach molecules 
by operating additional evaporative-cooling cycles on
the molecular side of the resonance.
It is important to point out that in all these experiments,
the condensed molecules are not detected directly by standard
imaging techniques. In  Ref. \cite{Zwierlein03} the molecular density is inferred from absorption
images at zero magnetic field taken with and without dissociating
the molecules on the positive-detuning side of the resonance. 
The onset of Bose-Einstein condensation is characterized by an abrupt change
from a smooth distribution to a bimodal distribution \cite{Anderson95,Davis95}.
Differently, in Ref.~\cite{Greiner03}  Greiner {\it{et al.}} deduce
the existence of a molecular component 
from radio-frequency spectroscopy by 
coupling the closed-channel component to a
third hyperfine state initially unoccupied.
In all cases,
quantitative information about the temperature, the total number of atoms 
and the condensate fraction, is obtained by fitting the density profiles
using a Bose-Einstein distribution for the broad normal component and
a Thomas-Fermi \cite{Goldman81,BaymPethick96}
distribution for the narrow condensate component.

The success in creating a molecular condensate depends also on a favourable
ratio of collisional rates for formation and decay of molecules.
Atomic $^6$Li is optimal in this respect. Hence it
is particularly suitable to explore the thermodynamics of a gas
under variable interaction conditions. This is crucial in view
of the new possibility of exploring the BEC-BCS crossover in ultracold atomic Fermi gases
by tuning the magnetic field near a Feshbach resonance.
In a dilute gas 
at very low temperatures the scattering length describes the interactions of the atoms.
Hence, changing the scattering length by sweeping the magnetic field across the resonance,
makes it possible to change the sign and the strength of interactions in the gas. 

When the background scattering length $a_{\rm{bg}}$ is small and negative, 
the total scattering length is small and negative
for large positive detuning. 
In this case, the gas undergoes
a Bardeen-Cooper-Schrieffer phase transition
when lowering the temperature
far below the degeneracy temperature $T_F$. 
This state is characterized by a Bose-Einstein condensate of
Cooper pairs that are largely delocalized in coordinate space.
Their extension is much larger than the interparticle distance
between two atoms.

When approaching the resonance, the scattering length
diverges and the gas enters a strong-coupling
regime, defined by the condition 
$k_F |a|>1$, where $k_F$ is the Fermi momentum of the gas. 
At resonance,
the scattering length jumps from $-\infty$
to $+\infty$, which is the hallmark of a formation of a zero-energy 
bound state.
At large negative detuning the superfluid gas turns into a Bose-Einstein
condensate of tightly-bound molecules. 
The evolution between
the two limits is called a {\it{crossover}} because the same $U(1)$ symmetry
is always broken and there is therefore no difference between the 
symmetries of the two phases.
The molecular condensate obtained in \cite{Greiner03,Jochim03b,Zwierlein03,Partridge05} 
represents one extreme of the BEC-BCS crossover. The investigation
of the BCS side is more difficult because the fermionic
superfluid transition hardly affects 
the density profile of the gas in the trap \cite{StoofHoubiers96}. Therefore, it cannot be detected by simple
imaging techniques.


 

Several ideas have been put forward in order
to point out a clear signature of the superfluid transition
\cite{StoofHoubiers96,Zhang99,Torma01,Pedri02,Orso04,Zwerger02}.
However, it turns out that the Feshbach resonance itself makes it possible to study
pairing effects in momentum space directly. 
The method, introduced by Regal {\it{et al.}} at JILA in 
Ref.~\cite{Jin04}, uses the Feshbach resonance
to transfer,
slowly compared to the two-body physics but fast compared to the many-body physics,
the fermionic $^{40}$K pairs onto the bare molecular state.
This enables them to detect the condensate component of the gas
also on the positive side of the resonance.
The results of the experiment of Regal {\it{et al.}}
give the molecular condensate fraction after the sweep 
as a function
of temperature and magnetic field. 
The experiments reveals a few-percent condensate fraction on the positive side
of the resonance. The fraction depends on the initial temperature $T$
of the gas and seems to have a threshold at about $0.5~ G$.
This is the magnetic field at which $k_F |a(B)|$ is about $1$.

The condensate component has been interpreted by Regal {\it{et al.}}
as a fermionic condensate. They make use the absence of a two-body bound 
state on the positive side of the resonance. Hence the   
fermionic condensation must be understood as a
"{\it{macroscopic occupation of a single quantum state, in which 
the underlying Fermi statistic of the paired particles plays an essential role}}".
A similar experiment has been realized a few months later by 
Zwierlein {\it{et al.}}~\cite{Zwierlein04a} at MIT
using $^6$Li atoms. 
Surprisingly, they observe a much larger molecular condensate fraction
than in the experiment at JILA.
The origin of this difference still lacks an explanation.
Both experiments are performed in axially elongated traps, which should
exclude trapping effects as a main reason.

\section{Atom-molecule theory  
for Feshbach-resonant interactions.}
\label{efffith}


For an incoherent equal mixture of 
fermionic atoms in two different hyperfine states,
and in the vicinity of a Feshbach resonance at low temperatures and densities, 
it is possible to derive \cite{Drummond98,Timmermans99,HollandKok01,Duine04} a low-energy effective quantum field theory
in terms of the atom-molecule Hamiltonian
\begin{align}
\label{eq:Ham}
{H}=&\!\!\!\sum_{\mathbf{k},\sigma 
\in \{\uparrow,\downarrow\}}\!\!\! 
\left(\epsilon_{\mathbf{k}}-\mu\right)a^{\dagger}_{\mathbf{k},\sigma} a_{\mathbf{k},\sigma}
\nonumber\\
&
+\frac{1
}{V}\sum_{\mathbf{k},\mathbf{k'},\mathbf{q}}
V_{\rm{bg}}
\left({\mathbf{q}}\right)
~a^{\dagger}_{\mathbf{k+q},\uparrow}
a^{\dagger}_{\mathbf{k'-q},\downarrow}a_{\mathbf{k'},\downarrow}a_{\mathbf{k},\uparrow}
\nonumber\\
&+\sum_{\mathbf{k}}\left(\frac{\epsilon_{\mathbf{k}}}{2}+\delta_{\rm{bare}}-2\mu
\right) b^{\dagger}_{\mathbf{k}}
b_{\mathbf{k}}
\nonumber\\
&+
\frac{1}{\sqrt{V}}
\sum_{\mathbf{k},\mathbf{q}}
\left(g^*\left({\mathbf{q}}\right)~ b^{\dagger}_{\mathbf{k}}
a_{\frac{\mathbf k}{2}+\mathbf{q},\downarrow}a_{\frac{\mathbf k}{2}-\mathbf{q},\uparrow}+{\rm h.c.}\right).
\end{align}
Here, $a^{\dagger}_{\mathbf{k},\sigma}$ is the creation operator
of an atom with momentum $\hbar\mathbf{k} $ in the hyperfine
state $|\sigma\rangle$ with chemical
potential $\mu$, and $b^{\dagger}_{\mathbf{k}}$ is the
creation operator of a bare Feshbach molecule
with momentum $\hbar\mathbf{k} $. In addition,
$V_{\rm{bg}}$ is the nonresonant or
background interaction between the atoms in the open channel.
The last two terms of the Hamiltonian in Eq.~({\ref{eq:Ham}}) describe the 
formation of a molecule from two atoms and the time-reverse process.
The strength of this coupling is given by the
bare atom-molecule coupling $g({\mathbf{k}})$.
Both $g$ and $V_{\rm{bg}}$ depend in principle also
on the external magnetic field $B$.
The energy $\delta_{\rm{bare}}=\epsilon_{\rm{bare}}
+\Delta\mu B$ is the bare detuning, where
$\epsilon_{\rm{bare}}$ is the energy of a bare molecule with zero total
momentum when the external magnetic
field is absent, and $\Delta \mu$ is the difference in magnetic moment
between the bare molecule and two atoms in the states $|\uparrow\rangle$ and $|\downarrow\rangle$.
Bare quantities 
must be eliminated with a renormalization procedure
in favour of experimentally known parameters in order 
to apply this Hamiltonian to a realistic gas. 
In the next section such a procedure is introduced
for the normal state of the atomic gas in the framework
of a many-body formalism based on Matsubara Green's function techniques
\cite{Fetter64}. 

\subsection{Many-body ${\rm{T}}-$matrix.}

The dynamical and thermodynamical 
properties of an atomic gas interacting near a Feshbach resonance
are determined by the exact (four-point) 
interaction vertex function $\Gamma_{\uparrow,\downarrow}$.
The latter describes
the collisions of two atoms with spin $|\uparrow\rangle$ and  $|\downarrow\rangle$,
in the presence of many-body correlations introduced by the medium.
Unfortunately, in the language of Matsubara Green's function techniques, 
the exact $\Gamma_{\uparrow,\downarrow}$ is given by 
an infinite series of Feynman diagrams 
and cannot be obtained in general.
Therefore, we have to introduce some approximations and to 
restrict ourselves to the relevant subset of 
Feynman graphs.
We assume that the average distance between the fermions is much larger
than the range of the interactions $r_0$, i.e., $k_F r_0\ll 1$.
Hence only 
two-particle collisions are expected to be important
and we can neglect three-particle and more particle collisions.
For a dilute system of fermions with short-range interactions
we thus have to consider
the ladder approximation \cite{Beliaev57,Galitskii57,Stoof96},
which sums exactly over all two-body processes.
In this approximation 
only the particle-particle channel is relevant and the
vertex function $\Gamma$
is reduced to the so-called many-body $\rm{T}-$matrix.
This satisfies a Bethe-Salpeter equation of the type
\begin{widetext}
\begin{align}
\label{eq:totMB}
{\rm T}^{\rm{MB}}&\left(\mathbf{k}_f,\mathbf{k}_i,\mathbf{K},\Omega_n\right)
={\rm V}_{\rm{eff}}\left(\mathbf{k}_f,\mathbf{k}_i,\mathbf{K},\Omega_n\right)+
\\
-&\frac{1}{\hbar}\frac{k_{\rm B} T}{\hbar V}\sum_{\mathbf{k},\omega_n}
{\rm V}_{\rm{eff}}\left(\mathbf{k}_f,\mathbf{k},\mathbf{K},\Omega_n\right)
G_{\uparrow}
\left(\frac{\mathbf{K}}{2}+\mathbf{k},\frac{\Omega_n}{2}+\omega_n\right)
G_{\downarrow}
\left(\frac{\mathbf{K}}{2}-\mathbf{k},\frac{\Omega_n}{2}-\omega_n\right)
{\rm T}^{\rm{MB}}\left(\mathbf{k},\mathbf{k}_i,\mathbf{K},\Omega_n\right)\nonumber,
\end{align}
\end{widetext}
where $G_{\sigma}$ is the exact atomic Green's function.
The effective atom-atom interaction is given by
\begin{align}
\label{eq:effint}
\rm{V}_{\rm{eff}}
\left(\mathbf{k}_f,\mathbf{k}_i,\mathbf{K},\Omega_n\right)&=\\
\rm{V}_{\rm{bg}}&\left(\mathbf{k}_f-\mathbf{k}_i\right)
+g^*({\mathbf{k}_f})
\frac{G_{0}\left(\mathbf{K},\Omega_n\right)}{\hbar} 
g({\mathbf{k}_i})\nonumber,
\end{align}
where 
\begin{eqnarray}
\label{eq:baremolprop}
\frac{G_{0}\left(\mathbf{K},\Omega_n\right)}{\hbar}=
\left(i\hbar\Omega_n-\frac{\epsilon_{\mathbf{K}}}{2}+2\mu-\delta_{\rm{bare}}
\right)^{-1}
\end{eqnarray}
is the bare molecular propagator.
Here $\hbar \mathbf{k}_f$ and $\hbar \mathbf{k}_i$ are the relative momenta
of the incoming and outcoming particles of the scattering processes,
$\hbar \mathbf{K}$ is the total momentum of the two fermions,
and $\Omega_n=2\pi n/\hbar\beta$ is the total bosonic Matsubara frequency of the 
fermion pair propagator and of the molecule.

In general, it is very difficult to solve 
this equation directly because the interaction
$V_{\rm{bg}}\left(\mathbf{k}\right) $ 
and the atom molecule coupling $g({\mathbf{k}})$ are included with
all their details and are known only numerically from very complicated
atomic physics measurements.  
Nevertheless, as we will show, 
in a dilute gas ($k_F\ll r_0^{-1}$) 
at very low temperatures,
the vertex function ${\rm T}^{\rm{MB}}$ can be calculated without
explicitly solving the Bethe-Salpeter equation in all its complexity.
This is achieved in a number of steps. First, we replace the complicated physical
atomic potentials with a pseudopotential that
leads to a more simple structure of the scattering equations.
Then we try to eliminate the
parameters of this pseudopotential from the theory
in favour of an equal number of parameters that
can be measured experimentally.
The theory eventually works because all the relevant information
to describe the low-energy physics of the system has been condensed
into the fenomenological parameters. 

When atoms are 
colliding at ultra-low temperatures,
the problem simplifies considerably.
At these temperatures they move
so slowly  that their kinetic energy is much smaller than
the energy  
$\hbar^2/m r_0^2$, related to the range of the interaction $r_0$.
This condition can be expressed as
\begin{eqnarray}
\lambda_{\rm{th}}(T)\gg r_0, 
\end{eqnarray}
where $\lambda_{\rm{th}}(T)=\left(2\pi \hbar^2/m k_{\rm B} T\right)^{1/2}$
is the thermal de Broglie wavelength.
As a result, 
the wave numbers of
the relative momenta $\hbar \mathbf{k}_f$ and $\hbar \mathbf{k}_i$ are 
much smaller than the characteristic wave number $r_0^{-1}$, i.e.,
\begin{eqnarray}
|\mathbf{k}_f|, |\mathbf{k}_i|\ll r_0^{-1},
\end{eqnarray}
and from scattering theory it is known that for low energy
the ${\rm T}-$matrix of a two-body collision 
becomes independent of the relative
momenta $ \mathbf{k}_f, \mathbf{k}_i$.
Therefore, we can try to neglect their explicit dependence
in Eq.~(\ref{eq:totMB}) by
replacing the physical potentials with pointlike interactions
$V_{\rm{bg}}\delta(\mathbf{x}-\mathbf{x'})$ and 
$g_{\rm{bare}}~\delta(\mathbf{x}-\mathbf{x'})$. 
This implies that the effective interaction in Eq.~(\ref{eq:effint})  can
be rewritten as
\begin{eqnarray}
\label{eq:effpotbare}
\rm{V}_{\rm{eff}}
\left(\mathbf{K},\Omega_n\right)
=\rm{V}_{\rm{bg}}
+g_{\rm{bare}} 
G_{0}\left(\mathbf{K},\Omega_n\right) 
g_{\rm{bare}}.
\end{eqnarray}
It turns out that under
these conditions the many-body ${\rm T}-$matrix is also
independent of $\mathbf{k_f}$ and $\mathbf{k_i}$ and may be
taken out of the integral. This gives
\begin{align}
\label{eq:totMBbis}
{\rm T}^{\rm{MB}}\left(\mathbf{K},\Omega_n\right)=&
{\rm V}_{\rm{eff}}\left(\mathbf{K},\Omega_n\right)+\\
{\rm V}_{\rm{eff}}&\left(\mathbf{K},\Omega_n\right)
\Pi\left(\mathbf{K},\Omega_n\right)
{\rm T}^{\rm{MB}}\left(\mathbf{K},\Omega_n\right)\nonumber,
\end{align}
where the kernel
\begin{align}
\label{eq:kernel}
\hbar\Pi\left(\mathbf{K},\Omega_n\right)=&
-\frac{k_{\rm B} T}{\hbar V}\sum_{\mathbf{k},\omega_n}
G_{\uparrow}
\left(\frac{\mathbf{K}}{2}+\mathbf{k},\frac{\Omega_n}{2}+\omega_n\right)\nonumber\\
\times&G_{\downarrow}
\left(\frac{\mathbf{K}}{2}-\mathbf{k},\frac{\Omega_n}{2}-\omega_n\right)
\end{align}
contains an unphysical ultraviolet divergence because
the internal integration over $\mathbf{k}$ in the 
Bethe-Salpeter equation in Eq.~(\ref{eq:totMBbis}) 
gives a contribute for every momenta
while in the original expression in Eq.~(\ref{eq:totMB})
the natural continuum cut-off 
around $\hbar/ r_0^{-1}$ of the 
physical potentials cuts off the high-momentum tail.
The $\rm{T}-$matrix renormalization procedure
consists of eliminating this divergency by means of the
the bare unknown constants $\delta_{\rm{bare}}$, 
$V_{\rm{bg}}$ and $g_{\rm{bare}}$
from Eq.~(\ref{eq:totMBbis}).
To this purpose we introduce the full molecular propagator
\begin{eqnarray}
\label{eq:DysMol}
G\left(\mathbf{q},\omega_n\right)^{-1}=G_0^{-1}
\left(\mathbf{q},\omega_n\right)-\Sigma_{\rm{m}}
\left(\mathbf{q},\omega_n\right)
\end{eqnarray}
by using the molecular selfenergy 
defined as
\begin{eqnarray}
\label{eq:molselfenbis}
\hbar\Sigma_{\rm{m}}
\left(\mathbf{q},\omega_n\right)=
g_{\rm{bare}}
\Pi\left(\mathbf{q},\omega_n\right)g^{\rm{MB}}
\left(\mathbf{q},\omega_n\right).
\end{eqnarray}
Here we have used for the momentum dependence of the
exact atom-molecule coupling
$g^{\rm{MB}}$ the same assumptions 
as for ${\rm T}^{\rm{MB}}$.
For the same reason as above, there is also an
unphysical ultraviolet divergency in the 
selfenergy of the molecular 
propagator that can be eliminated by
a simple subtraction in the selfenergy. 
This leads to the renormalization of the bare detuning 
$\delta_{\rm{bare}}$ to the physical detuning $\delta(B)=\Delta\mu(B-B_0)$
defined in the previous chapter.

\begin{figure}
\begin{center}
 \resizebox{85mm}{!}{ \epsfig{figure=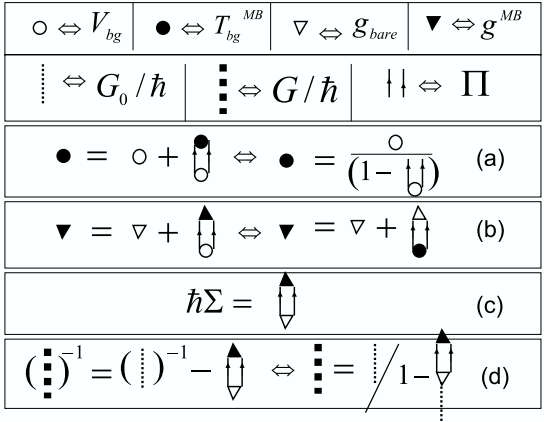,angle=-90}}
  \caption{{The first two line define the symbol we use in our diagrammatic
  notation. The last four lines represent respectively (a) Eq.~(\ref{eq:TMBGg}),
  (b) Eq.~(\ref{eq:gMBtris}),
  (c) Eq.~(\ref{eq:molselfenbis}),
  (d) Eq.~(\ref{eq:DysMol}).
  }} 
   \label{StoofA}
  \end{center}
\end{figure}

The desired renormalized atom-molecule coupling $g^{\rm{MB}}$ is connected to
the many-body ${\rm{T}}-$matrix by the relation \cite{Duine04}
\begin{align}
\label{eq:gMBtris}
g^{\rm{MB}}\left(\mathbf{K},\Omega_n\right)= g_{\rm{bare}}+
g_{\rm{bare}}
\Pi\left(\mathbf{K},\Omega_n\right){\rm T}
_{\rm{bg}}
^{\rm{MB}}
\left(\mathbf{K},\Omega_n\right)\!,
\end{align} 
where the many-body $\rm{T}$-matrix for the background processes satisfies
\begin{align}
\label{eq:TMBGg}
{\rm T}_{\rm{bg}}^{\rm{MB}}\left(\mathbf{K},\Omega_n\right)
={\rm V}_{\rm{bg}}+
\sum_{\mathbf{k}}
{\rm V}_{\rm{bg}}
\Pi\left(\mathbf{K},\Omega_n\right){\rm T}_{\rm{bg}}
^{\rm{MB}}\left(\mathbf{K},\Omega_n\right).
\end{align}
This means that we have dressed the bare
atom-molecule coupling  with all the two-body background interactions.
The set of equations~(\ref{eq:DysMol})-(\ref{eq:TMBGg}) is represented
in Fig.~\ref{StoofA} by means of a diagrammatic notation.

\begin{figure}
\begin{center}
 \resizebox{85mm}{!}{ \epsfig{figure=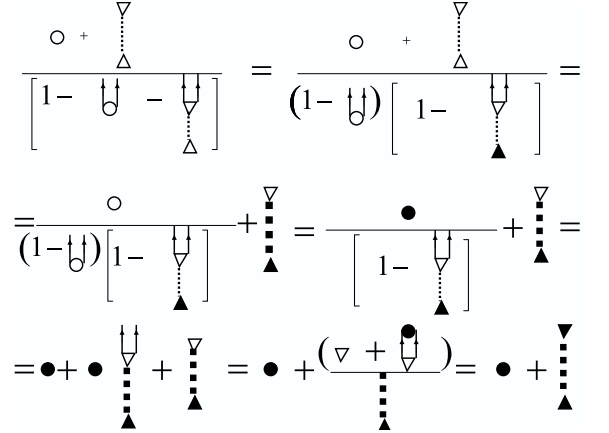,angle=-90}}
  \caption{{Diagrammatic calculation of the many-body $\rm{T}$-matrix.
  The derivation starts from the definition of the many-body $\rm{T}$-matrix
  in Eq.~(\ref{eq:totMBbis}) in terms of the bare quantities of the theory.
  The many-body $\rm{T}$-matrix in terms of the renormalized couplings and propagators
  follows exactly from the algebra defined in Fig.~\ref{StoofA} based
  on the exact relations in Eqs.~(\ref{eq:DysMol})-(\ref{eq:TMBGg}).
  }} 
   \label{StoofB}
  \end{center}
\end{figure}

By formally solving Eq.~(\ref{eq:totMBbis})
and using Eqs.~(\ref{eq:DysMol},~\ref{eq:gMBtris}) and~(\ref{eq:TMBGg})
to eliminate $G_0,\rm{V}_{\rm{bg}}$ and $g_{\rm{bare}}$, it 
is possible, by some straightforward algebra,
to obtain the desired result
\begin{widetext}
\begin{align}
\label{eq:theo}
{\rm T}^{\rm{MB}}\left(\mathbf{K},\Omega_n\right)&={\rm T}_{\rm{bg}}^{\rm{MB}}
\left(\mathbf{K},\Omega_n\right)+
+g^{\rm{MB}*}\left(\mathbf{K},\Omega_n\right)
\frac{G\left(\mathbf{K},\Omega_n\right)}{\hbar}
g^{\rm{MB}}\left(\mathbf{K},\Omega_n\right),
\end{align}
\end{widetext}
where all the bare quantities have been eliminated
from the theory. A nice derivation of this formula 
by means of an algebra of diagrams
is shown in Fig.~\ref{StoofB}.

This result represents a generalization of
the solution of the standard many-body ${\rm T}-$matrix
for a dilute ultracold gas of atoms and molecules
at the Feshbach resonance 
\cite{Beliaev57,Galitskii57,Stoof96}.
This solution describes the scattering of a pair of atoms 
at (complex) energy $i\hbar\Omega_n$, while the dependence
on the center-of-mass momentum $\mathbf{K}$ incorporates 
the Pauli-blocking effects of the medium.
As we have anticipated,
at this level the solution is still somewhat formal because we have only
eliminated the bare coupling constants in terms of the renormalized ones.
The next step consists in relating the renormalized quantities, 
such as ${\rm T}_{\rm{bg}}^{\rm{MB}}$ and
$g^{\rm{MB}}$, to the two-body parameters, which can be 
measured by experiments.

\subsection{Two-atom properties of the many-body theory.}
\label{sec:twobodyphys}
In the two-body limit \cite{Kokkelmans02,Mackie02,Kohler03} there are only two fermions present
so that in an infinite volume, the particle density
tends to zero, i.e., $k_F=0$, and there are no many-body effects anymore.
This limit can be achieved by letting $T\rightarrow 0$ and
 $i\hbar\Omega_n\rightarrow E$
while the chemical potential is fixed to 
$\mu=0$.
The dressed 
fermion propagator
$G_{\sigma}$
can be replaced by the free atomic propagator
\begin{eqnarray}
\label{eq:freeprop}
G_{\sigma,0}\left(\mathbf{k},E\right)=
\frac{\hbar}{E-\epsilon_{\mathbf{k}}},
\end{eqnarray}
and the dependence on the total momentum $\mathbf{K}$ drops out everywhere.
The many-body ${\rm T}-$matrix in Eq.~(\ref{eq:theo}) becomes 
the two-body $\rm{T}$-matrix
and Eq.~(\ref{eq:theo}) can be rewritten as
\begin{eqnarray}
\label{eq:twobody}
{\rm T}^{\rm{2B}} \left(E\right)={\rm T}_{\rm{bg}}^{\rm{2B}}\left(E\right)
+g^{\rm{2B}*}\left(E\right)
\frac{G^{\rm{2B}}\left(E\right)}{\hbar}
g^{\rm{2B}}\left(E\right),
\end{eqnarray}
where the two-particle dressed molecular propagator at zero momentum is given by
\begin{align}
\label{eq:2Bprop}
G^{\rm{2B}}\left(E\right)
^{-1}\equiv
G^{\rm{2B}}\left(\mathbf{0},E\right)
^{-1}=G_0\left(\mathbf{0},E\right)^{-1}-\Sigma^{\rm{2B}}_{\rm{m}}\left(\mathbf{0},E\right).
\end{align}
The on-shell energy-dependent $\rm{T}_{\rm{bg}}$-matrix 
\begin{eqnarray}
\label{eq:2BTbg}
{\rm T}_{\rm{bg}}^{\rm{2B}}\left(E\right)=
\frac{4\pi\hbar^{2}a_{\rm{bg}}(B)}{m}\frac{1}{1-a_{\rm{bg}}\sqrt{\frac{-E m}{\hbar^{2}}}},
\end{eqnarray}
is obtained by solving 
Eq.~(\ref{eq:TMBGg})
in the vacuum, which reduces to the well-known Lippmann-Schwinger equation \cite{Sakurai82}
\begin{eqnarray}
\label{eq:Lipp}
{\rm T}_{\rm{bg}}^{\rm{2B}}\left(E\right)
={\rm V}_{\rm{bg}}+  
\sum_{\mathbf{k}}
{\rm V}_{\rm{bg}}
\frac{1}{E-2 \epsilon_{\mathbf{k}}}
{\rm T}_{\rm{bg}}^{\rm{2B}}
\left(E\right).
\end{eqnarray}
The background scattering length  $a_{\rm{bg}}(B)$ is 
the quantity which is measurable experimentally
and is defined by ${\rm T}_{\rm{bg}}^{\rm{2B}}(0)=4\pi\hbar^2 a_{\rm{bg}}/m$.
Noting that Eqs.~(\ref{eq:gMBtris}) and~(\ref{eq:TMBGg}) lead to
$g^{\rm{2B}}\left(E\right)=g_{\rm{bare}} {\rm T}_
{\rm{bg}}^{\rm{2B}}\left(E\right)\rm /V_{\rm{bg}}$,
we conclude that 
the same kind of energy dependence 
can also be used
for the atom-molecule coupling constant, i.e.,
\begin{eqnarray}
\label{eq:gexp} 
g^{\rm{2B}}\left(E\right)=g \frac{1}{1-a_{\rm{bg}}\sqrt{\frac{-E m}{\hbar^{2}}}},
\end{eqnarray} 
where $g=g^{\rm{2B}}\left(0\right)$ 
is also inferred from experiments. 
The evaluation of the molecular selfenergy in the vacuum will complete the definition of 
the $\rm{T}^{\rm{2B}}$ in terms of measurable quantities.

As we have seen in the previous section, 
the molecular selfenergy in the dressed molecular propagator contains an unphysical
ultraviolet divergency.
This can be eliminated by rewriting the 
two-body propagator in Eq.~(\ref{eq:2Bprop}) as
\begin{align}
\label{eq:2Bpropren}
G^{\rm{2B}}\left(\mathbf{k},E\right)
^{-1}=&G_0\left(\mathbf{k},E\right)^{-1}
+\Sigma^{\rm{2B}}_{\rm{m}}\left(\mathbf{0},0\right)\nonumber\\
&-\Sigma^{\rm{2B}}_{\rm{m}}\left(\mathbf{k},E\right)
-\Sigma^{\rm{2B}}_{\rm{m}}\left(\mathbf{0},0\right)\nonumber\\
=&\frac{E-\epsilon_{\mathbf{k}}/2-\delta(B)}{\hbar}\nonumber\\
&-\Sigma^{\rm{2B}}_{\rm{m}}\left(\mathbf{k},E\right)
+\Sigma^{\rm{2B}}_{\rm{m}}\left(\mathbf{0},0\right),
\end{align}
where the energy-independent but infinite shift $\Sigma^{\rm{2B}}\left(\mathbf{0},0\right)$
has renormalized the bare detuning $\delta_{\rm{bare}}$ to the physical
detuning  $\delta(B)$ according to
\begin{eqnarray}
\label{eq:rendet}
\delta(B)=\delta_{\rm{bare}}+\hbar\Sigma^{\rm{2B}}_{\rm{m}}\left(\mathbf{0},0\right)
\equiv \Delta\mu(B-B_0),
\end{eqnarray}
in such a manner that the position of the resonance in the magnetic field is now precisely at
the observed magnetic field value $B_0$.
The renormalized molecular selfenergy can now be calculated by performing a simple
integration \cite{Duine04}
\begin{align} 
\label{eq:self}
\hbar\Sigma^{\rm{2B}}_{\rm{m}}\left(\mathbf{k},E\right)
-\hbar\Sigma^{\rm{2B}}_{\rm{m}}\left(\mathbf{0},0\right)=
\frac{\eta
(B)}{1+|a_{\rm{bg}}|\sqrt{\frac{-E m}{\hbar^{2}}}}\sqrt{-E},
\end{align}
where 
the quantity $\eta^2
=\left(g^2 m^{\frac{3}{2}}/4\pi\hbar^{3}\right)^2
$ 
defines an important energy scale in the problem which is related to the width
of the Feshbach resonance. 
Note that
as a result of the internal integration on $\mathbf{k}$
the background scattering length appears as an absolute value in
the selfenergy \cite{Duine04}. This turns out to be very important in the case of negative
background scattering length when we calculate the binding energy of the molecule. 
The two-body molecular propagator takes the final form
\begin{align}
\label{eq:2Bproprenbis}
\hbar G^{\rm{2B}}\left(E^+\right)^{-1}=E^+-\delta(B)-
\frac{\eta(B)}{1+|a_{\rm{bg}}|\sqrt{\frac{-E m}{\hbar^{2}}}} 
\sqrt{-E}.
\end{align}
Using this expression in Eq.~(\ref{eq:twobody}) we recover
the effective atom-atom 
resonant interaction in the limit of zero-energy scattering.
This is given by
\begin{align}
\label{eq:sclen}
{\rm T}^{\rm{2B}} \left(0\right)=&\frac{4\pi\hbar^{2}a(B)}{m}=
\frac{4\pi\hbar^{2}a_{\rm{bg}}(B)}{m}-\frac{2 g^2(B)}{\delta(B)}\nonumber\\
&\equiv\frac{4\pi\hbar^{2}}{m}\left(a_{\rm{bg}}(B)+a_{\rm{res}}(B)\right),
\end{align}  
where $a(B)$ is the total effective scattering length that diverges at $B=B_0$.

When the background scattering length can be considered as a constant
for the magnetic field range across the Feshbach resonance, such as
is certainly the case for a narrow resonance, the total effective scattering
length can be rewritten as 
\begin{eqnarray}
\label{eq:sclen1}
a(B)=a_{\rm{bg}}\left(1-\frac{\Delta B}{B-B_0}\right)
\end{eqnarray} 
with the width of the resonance $\Delta B$ defined 
through the relation $\Delta B=m g^2/4\pi \hbar^2 |a_{\rm{bg}}|\Delta\mu$.
For the broad resonance at $834 G$ in an atomic $^6$Li gas, however, the 
background scattering length exhibits a strong dependence on the external
magnetic field $B$ \cite{Marcelis04}
and it is not sufficiently accurate to use
Eq.~(\ref{eq:sclen1}) \cite{Falco05}. 
The energy of the molecule is given by $E_{\mathbf{k}}=\epsilon_{\rm{m}}+
\epsilon_{\mathbf{k}}/2$, 
where the energy of the zero-momentum molecular level $\epsilon_{\rm{m}}$ 
is determined by the pole of the retarded molecular propagator
given in Eq.~(\ref{eq:2Bproprenbis}) at zero kinetic energy. 
This is equivalent to 
finding the zero of the equation 
\begin{eqnarray}
\label{eq:root}
\epsilon_{\rm{m}}-\delta(B)-
\frac{\eta(B)}{1+|a_{\rm{bg}}(B)|\sqrt{\frac{-\epsilon_{\rm{m}}^+ m}{\hbar^{2}}}} 
\sqrt{-\epsilon_{\rm{m}}^+}=0.
\end{eqnarray}
For negative detuning $\delta(B)<0$ the molecular 
propagator has a real and negative pole corresponding to the bound-state energy.
The real zero of Eq.~(\ref{eq:root}) can be calculated analytically 
by solving the equivalent algebraic equation. We find
\begin{widetext}
\begin{eqnarray}
\label{eq:analroot}
\epsilon_{\rm{m}}(B)=-\frac{1}{9}\left(
-\sqrt{\epsilon_{\rm{bg}}}
-\frac{2^{1/3}\alpha}{(\gamma+\sqrt{4 \alpha^3+\gamma^2})^{1/3}}
+\frac{(\gamma+\sqrt{4 \alpha^3+\gamma^2})^{1/3}}{2^{1/3}}
\right)^2,
\end{eqnarray}
\end{widetext}

\begin{figure*}
  \begin{center}
    \begin{tabular}{cc}
    \\
      \resizebox{70mm}{!}{\includegraphics{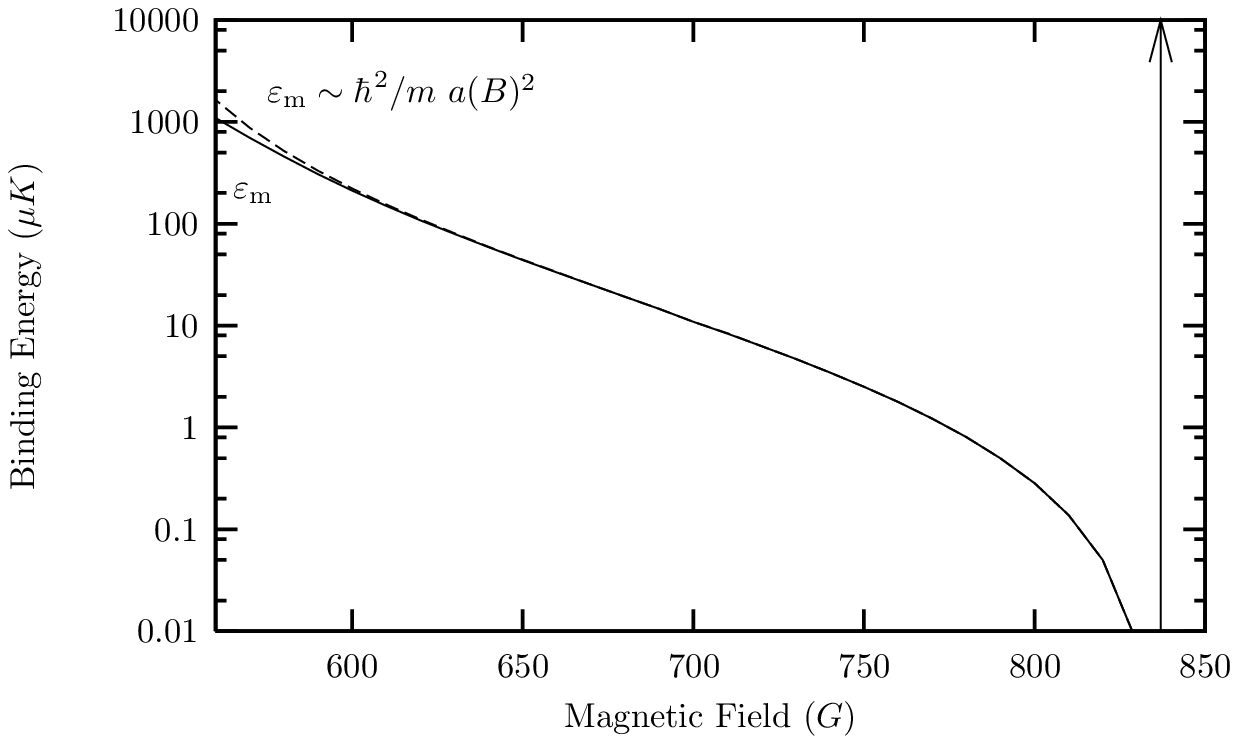}} & \hspace{1 cm}
       \resizebox{62mm}{!}{\includegraphics{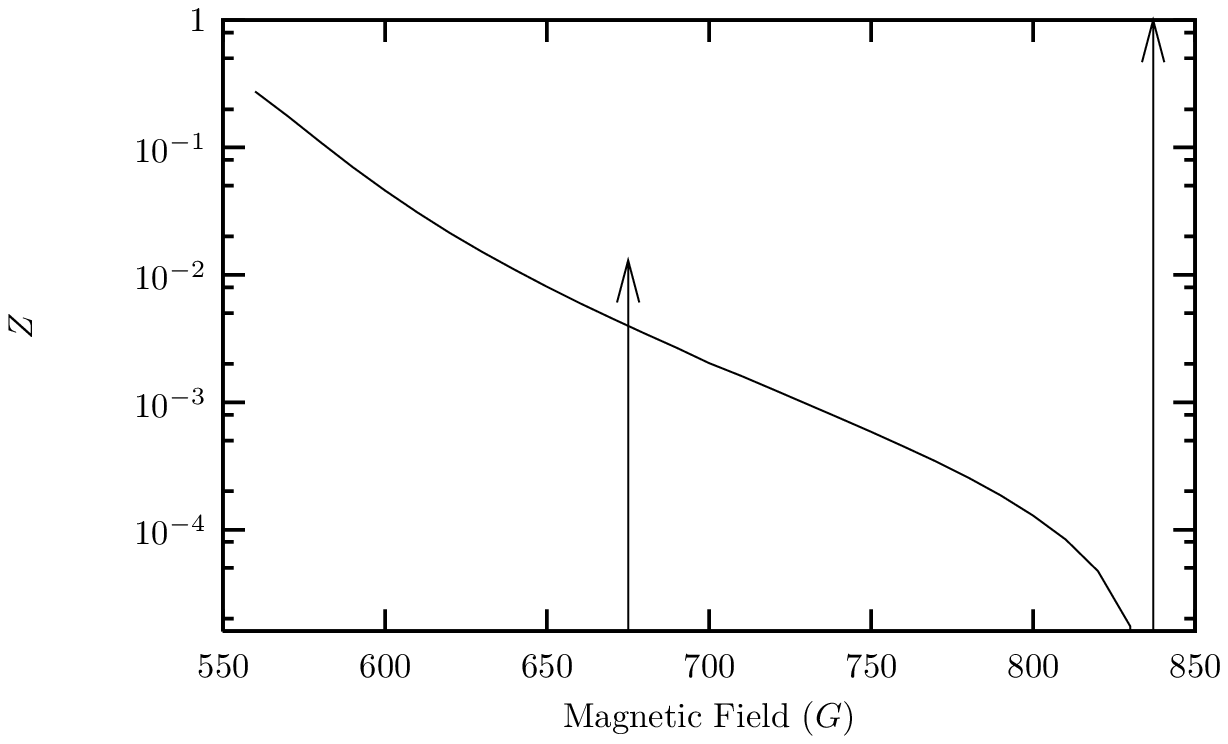}} \\
    \end{tabular}
    \caption{
    Binding energy and the probability
    Z as a function of the magnetic
    field near the broad resonance at $834~G$ for a mixture of $^6$Li atoms
    in the hyperfine states $|1\rangle$ and $|2\rangle$.
    The dashed line on the left picture represents the binding energy 
    approximated according to the Wigner formula in Eq.~(\ref{eq:bind}). The 
    arrow at about $675~G$
    on the right plot indicates the magnetic-field scale associated with
    the energy $\epsilon_{\rm{bg}}=\hbar^2/m a_{\rm{bg}}^2$.
    The curves for the binding energy and Z agree perfectly
    with the coupled-channel calculations \cite{Jochim03a,Partridge05} and 
    the experimental data 
    \cite{Jochim03a}.
    }
    \label{Bind1}
  \end{center}
\end{figure*}

\noindent where the coefficients are $\alpha=3\delta-\epsilon_{\rm{bg}}+3\eta\sqrt{\epsilon_{\rm{bg}}}$
and $\gamma=-18\delta\sqrt{\epsilon_{\rm{bg}}}-2\epsilon_{\rm{bg}}^{3/2}
+9\eta\epsilon_{\rm{bg}}$. 
The energy $\epsilon_{\rm{bg}}$, 
associated to the background scattering length $a_{\rm{bg}}$,
is defined as
\begin{eqnarray}
\label{eq:enbg}
\epsilon_{\rm{bg}}=\frac{\hbar^2}{m a_{\rm{bg}}^2}.
\end{eqnarray}
Interestingly, 
it can be shown that the 
solution of Eq.~(\ref{eq:analroot}) is
equivalent to the solution,
in the limit when the effective range $r_0$ can be neglected,
of the equation for the binding energy given by Eq.(3) in Ref.~\cite{Marcelis05}.
Therefore, in the case of a very large background scattering length,
our theory reproduces correctly
the non-trivial energy dependence 
due to the interplay between the resonant background interaction and the Feshbach resonance
~\cite{Marcelis04,VanKempen04}.

In the limit 
of a vanishing background 
scattering length $a_{\rm{bg}}\rightarrow 0$, the solution of Eq.~(\ref{eq:analroot})
reduces to \cite{Duine04}
\begin{eqnarray}
\label{eq:Rembbind}
\epsilon_{\rm m}
=\delta(B)+\frac{\eta^2}{2}\left(
\sqrt{1-\frac{4\delta(B)}{\eta^2}}-1\right).
\end{eqnarray}

\begin{figure*}
  \begin{center}
    \begin{tabular}{cc}
      \resizebox{70mm}{!}{\includegraphics{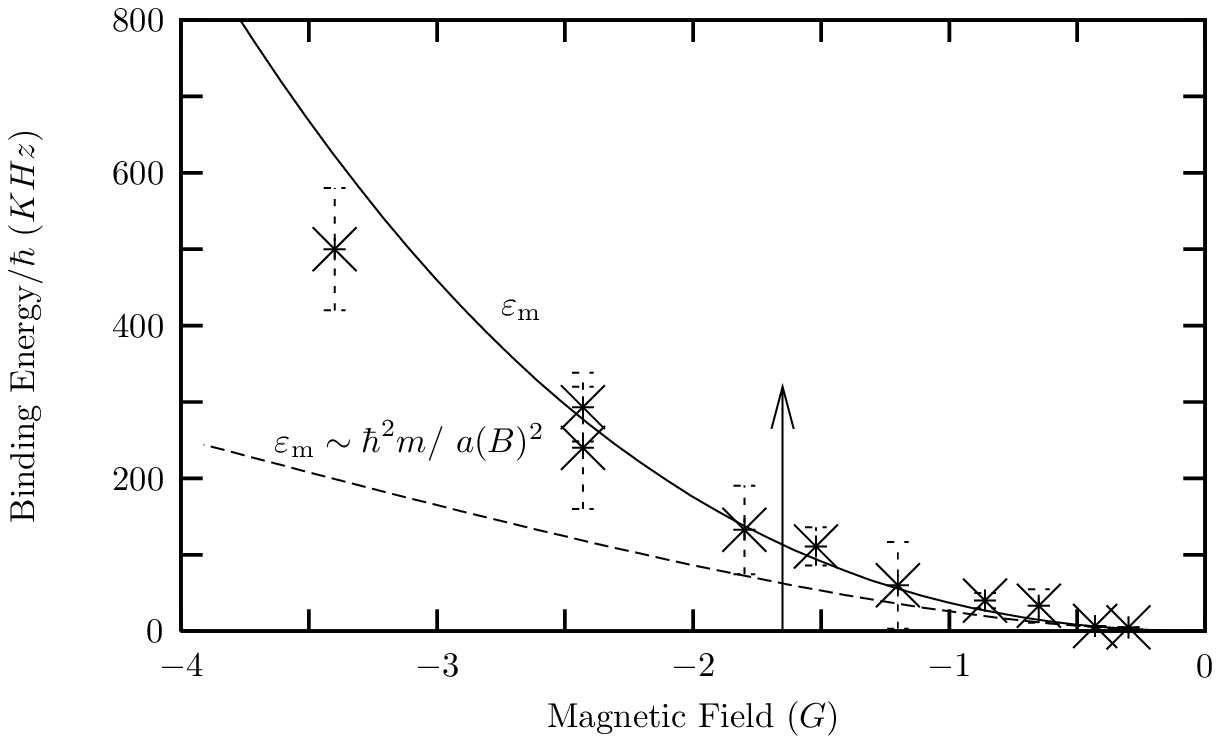}} & \hspace{1 cm}
      \resizebox{62mm}{!}{\includegraphics{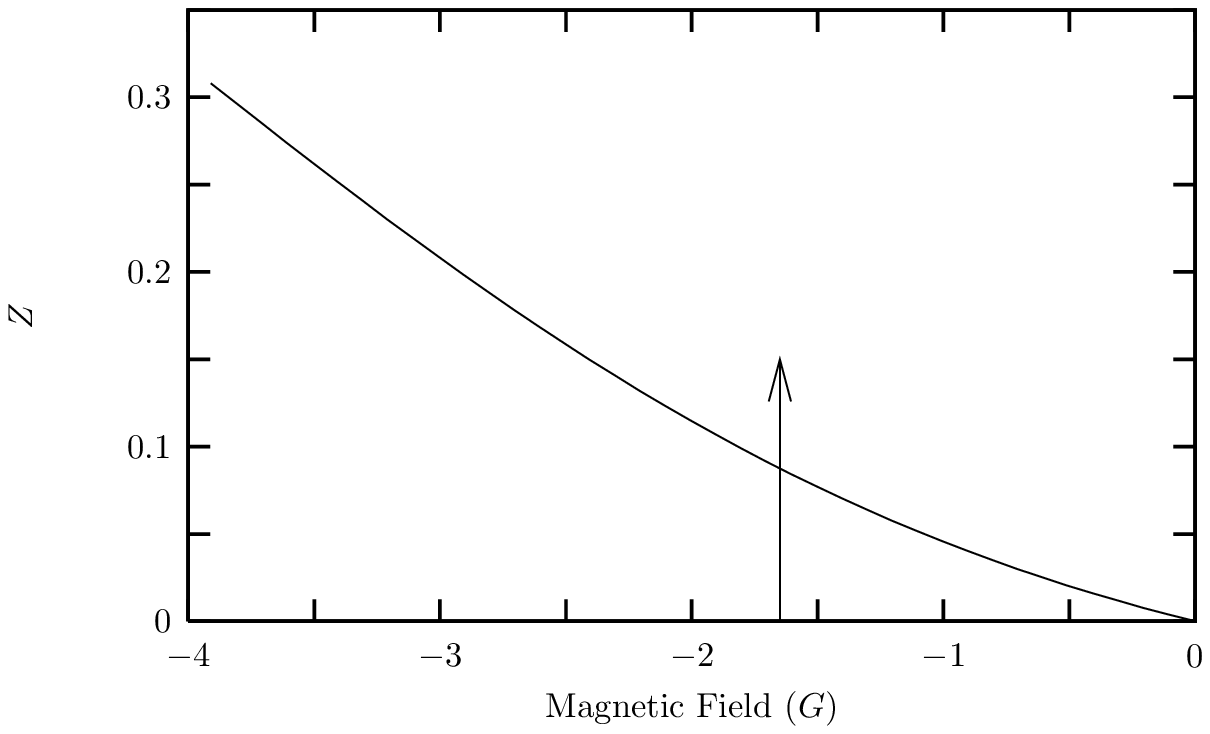}} 
    \end{tabular}
    \caption{Binding energy and the probability
    Z as a function of the magnetic
    field near the resonance at $224~G$ for a mixture of $^{40}$K atoms
    in the hyperfine states $|f,m_f\rangle=|9/2,-9/2\rangle$ and $|f,m_f\rangle=|9/2,-5/2\rangle$.
    The dashed line on the left picture represents the binding energy 
    approximated according to the Wigner formula in Eq.~(\ref{eq:bind})
    while the stars represent the experimental data taken from Ref. \cite{Ticknor03}.
    The arrow at about $B-B_0=-1.6~G$
    indicates the magnetic field associated with
    the energy $\epsilon_{\rm{bg}}$.
    The curves are calculated using the the same experimental
    parameters $a_{\rm{bg}}=174~a_0$, $\Delta\mu=1.27 \mu_B$ 
    and $\Delta B=9.7~G$ as in Ref. \cite{Drummond04}.}
    \label{Bind2}
  \end{center}
\end{figure*}

\begin{figure*}
  \begin{center}
    \begin{tabular}{cc}
      \resizebox{70mm}{!}{\includegraphics{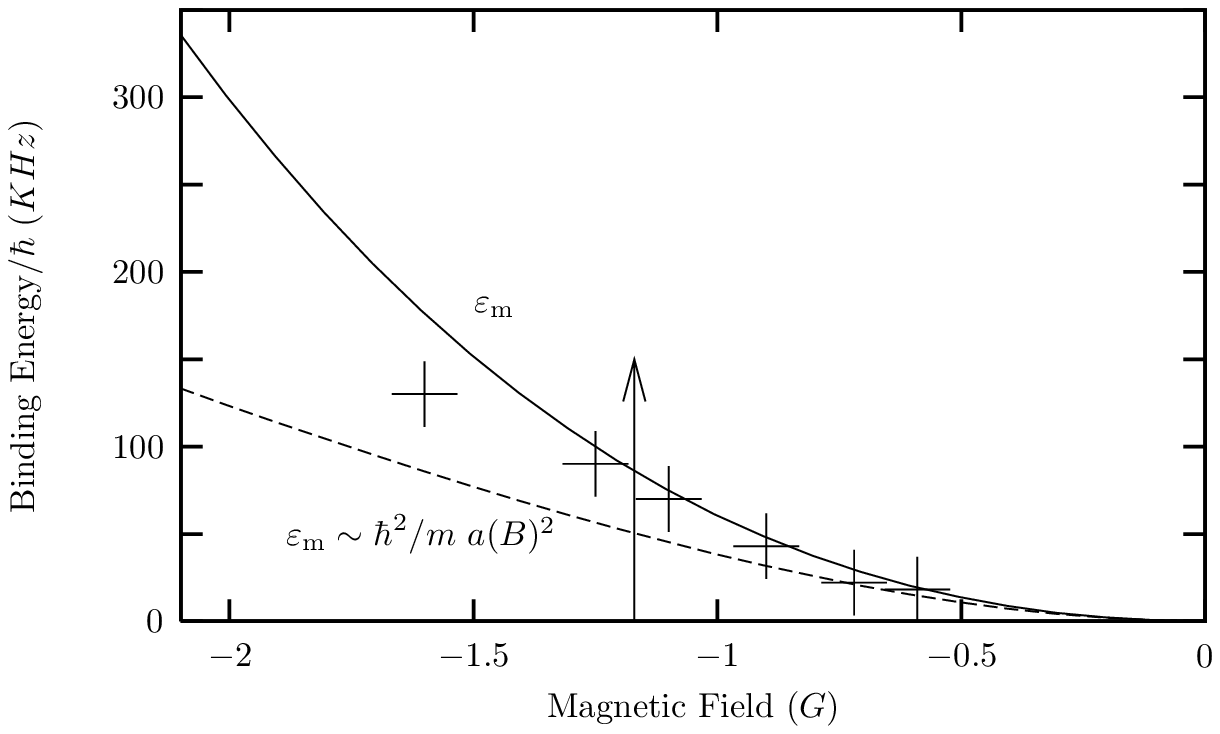}} &  \hspace{1 cm}
      \resizebox{62mm}{!}{\includegraphics{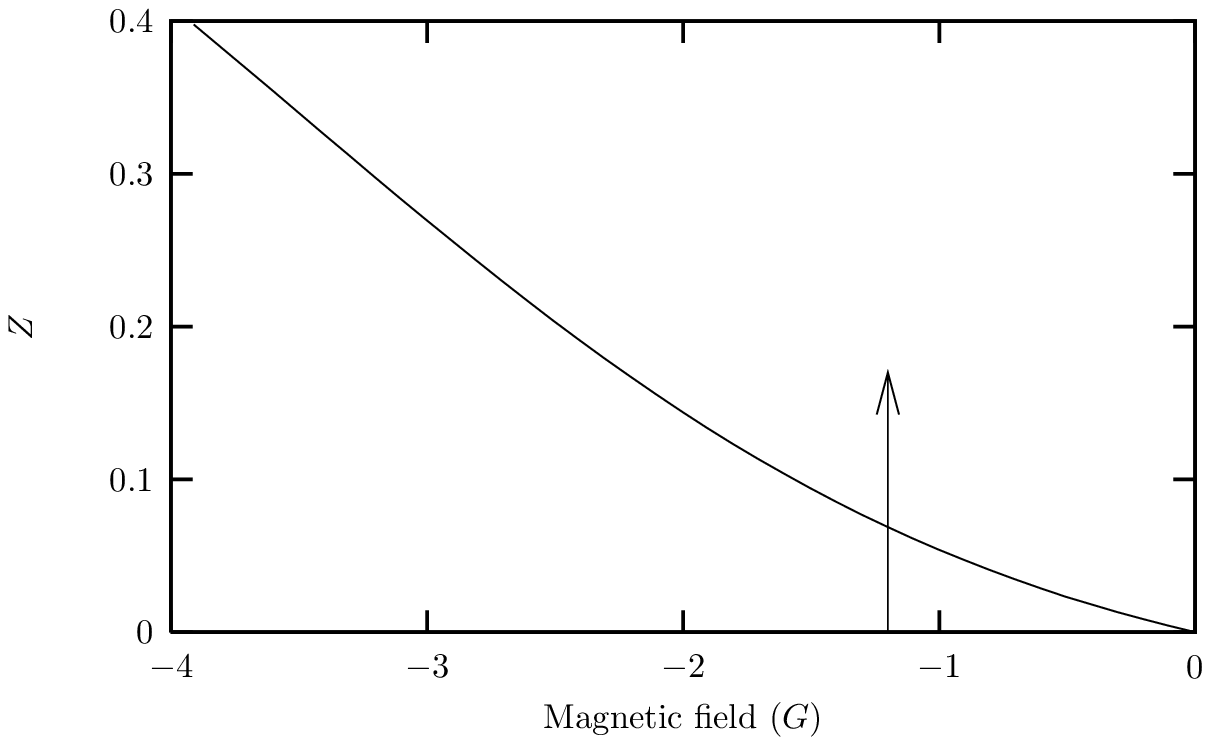}} 
    \end{tabular}
    \caption{Binding energy and the probability
    Z as a function of the magnetic
    field near the resonance at $202.1~G$ for a mixture of $^{40}$K atoms
    in the hyperfine states $|f,m_f\rangle=|9/2,-9/2\rangle$ and $|f,m_f\rangle=|9/2,-7/2\rangle$.
    The dashed line on the left picture represents the binding energy 
    approximated according to the Wigner formula in Eq.~(\ref{eq:bind})
    while the crosses represent the experimental data taken from Ref.~\cite{Ticknor03}.
    The arrow at about $B-B_0=-1.2~G$
    indicates the magnetic field associated with
    the energy $\epsilon_{\rm{bg}}$.
    The curves are calculated using the experimental
    parameters $a_{\rm{bg}}=174~a_0$, $\Delta\mu=(16/9) \mu_B$ 
    and $\Delta B=7.8~G$ as in Ref. \cite{Dennis05}.}
    \label{Bind3}
  \end{center}
\end{figure*}

\begin{figure*}
  \begin{center}
    \begin{tabular}{cc}
      \resizebox{75mm}{!}{\includegraphics{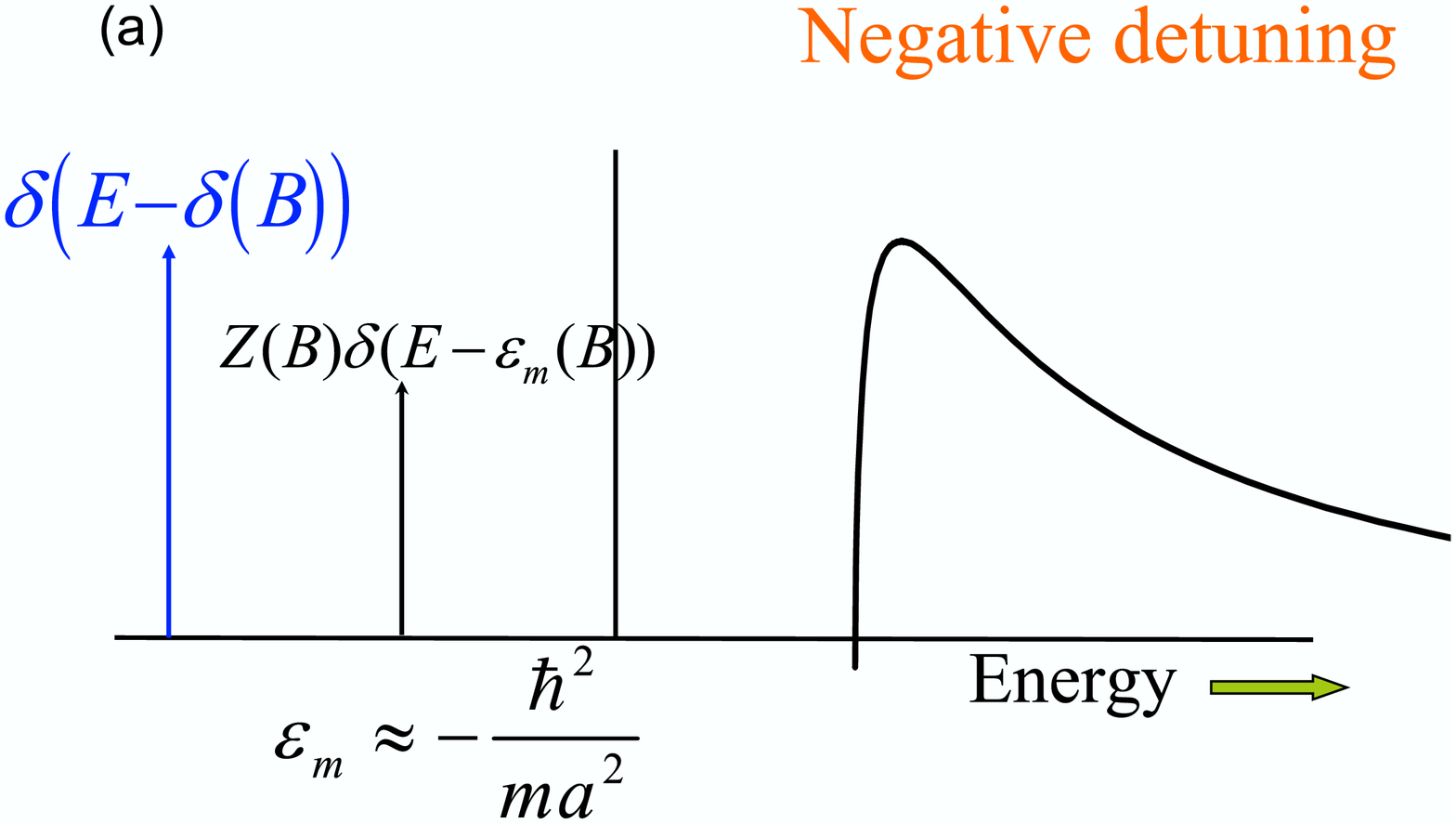}} &
      \resizebox{75mm}{!}{\includegraphics{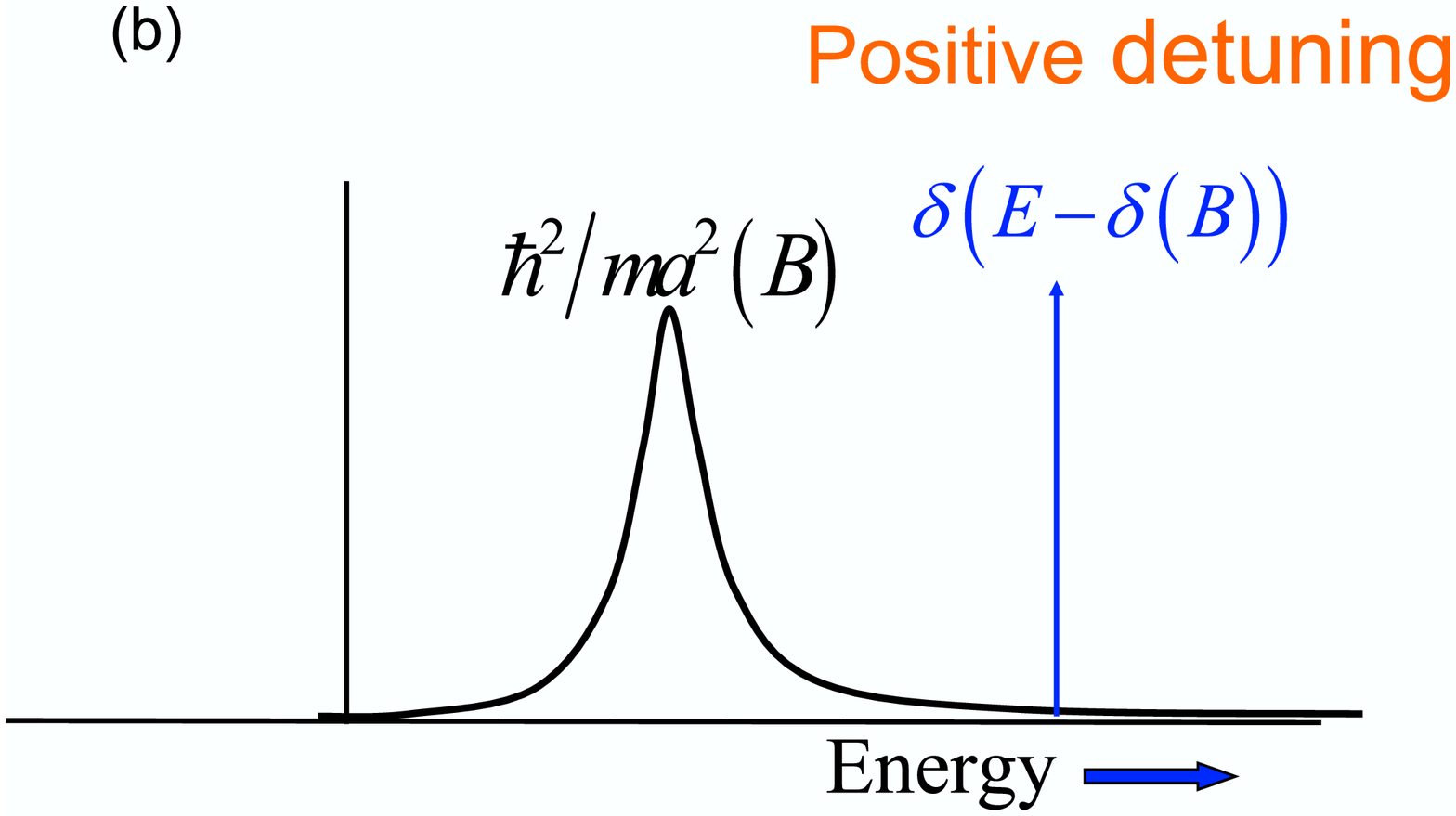}} 
    \end{tabular}
    \caption{Molecular density of states in the two-body limit 
    at negative detuning (a) and positive detuning (b).}
    \label{densityofstates}
  \end{center}
\end{figure*}

The bound-state energy  $ \epsilon_{\rm{m}}(B) $
of the dressed molecule as a function of the magnetic 
field is shown on the
left sides of Figs.~\ref{Bind1},~\ref{Bind2}
and~\ref{Bind3}
for three different resonances.
Near resonance the binding energy goes to zero 
and the square-root term in Eq.~(\ref{eq:root}),
which describes the dressing effects of the open channel, dominates
the linear term. In this approximation the location of the pole is at
\begin{eqnarray}
\label{eq:bind}
\epsilon_{\rm{m}}(B)=-\frac{\hbar^2}{m a^2(B)}.
\end{eqnarray}   
according to the Wigner's formula in Eq.~(\ref{eq:WignLaw}) for the binding energy in resonant scattering.
At large negative detuning, far away from resonance, the dressing effects of the
coupling with the open channel represent only a small perturbation. 
In this limit,
the energy of the Feshbach molecule
approaches the binding energy of the bare molecule which 
scales linearly with the magnetic field as 
\begin{align}
\label{eq:linearbinding}
\epsilon_{\rm{m}}(B)\simeq \delta(B).
\end{align}
In general the residue of the pole is given by
\begin{align}
\label{eq:zeta}
Z(B)&=\left[1-\frac{\partial\hbar\Sigma^{\rm{2B}}\left(z\right) }{\partial z}
\right]_{z=\epsilon_{\rm{m}}(B)}^{-1}=\nonumber\\
&=\left[1+\frac{\eta}{2 \sqrt{|\epsilon_{\rm{m}}|}\left(
1+\sqrt{\frac{|\epsilon_{\rm{m}}|}{\epsilon_{\rm{bg}}}}\right)^2}\right]^{-1}
\end{align}
and is always smaller than one, because physically the 
wavefunction of the dressed molecular state near
the Feshbach resonance is given by the linear superposition
\begin{align}
\label{eq:wavefctmol}
  \langle {\bf r}| \chi_{\rm m}; {\rm dressed} \rangle&\simeq
                 \sqrt{Z(B)} \chi_{\rm m}({\bf r}) |{\rm closed} \rangle
                 \nonumber \\
& + \sqrt{1-Z(B)}\frac{1}{\sqrt{2 \pi a(B)}}
\frac{e^{-r/a(B)}}{r}
                           |{\rm open} \rangle ~,
\end{align}
where $\chi_{\rm m}({\bf r})$ denotes the wavefunction of the bare
molecular state in the closed channel of the Feshbach problem. The
dressed molecular state therefore only contains with an amplitude $\sqrt{Z(B)}$ the
bare molecular state $|\chi_{\rm m};{\rm bare} \rangle$ of the closed channel.
The remaining $\sqrt{1-Z(B)}$ component is carried by the continuum of the scattering
states in the open channel. The density of states of the dressed
molecule at negative detuning is shown in Fig.~\ref{densityofstates}(a).
The probability $Z$ corresponds to the wavefunction
renormalization constant of the
bare molecule in the closed channel. It goes 
linearly in $|B-B_0|$ to zero near resonance, where the
dressing due to the open channel is maximum. It goes to one monotonically
at large negative detunings where the eigenstate of the Feshbach problem approaches
the bare molecule of the closed channel. (See also the right sides of
Figs.~\ref{Bind1}
,~\ref{Bind2}
and~\ref{Bind3}).
For a broad resonance, we usually have that $\eta^2 >\epsilon_{\rm{bg}}$. In this case
the "crossover" between these two regimes takes place at 
detunings such that 
\begin{align}
\label{eq:condition1}
|\epsilon_{\rm{m}}|\simeq\epsilon_{\rm{bg}}.
\end{align} 
When $a_{\rm{bg}}$ is small 
the turning point between the quadratic and the linear regime is given
by the condition $|\epsilon_{\rm{m}}|\simeq\eta^2$.
The quantity $Z$ is also related to the magnetic moment $\mu_{\rm magn}$ of the dressed molecules.
Taking the derivative on both sides respect to the magnetic field $B$ in Eq.~(\ref{eq:root})
we have that
\begin{align}
\label{eq:derbinden}
\frac{\partial \epsilon_{\rm{m}}}{\partial B}-\Delta \mu-\frac{\partial \hbar\Sigma^{\rm{2B}}}{\partial \epsilon_{\rm{m}}}
\frac{\partial \epsilon_{\rm{m}}}{\partial B}=0,
\end{align}
where we recall again that $\Delta \mu$ is the difference in the magnetic moment between the atomic pair
and the bare molecule
and closed channel in absence of coupling. 
Using the definition of $Z$, Eq.~(\ref{eq:derbinden}) can be rewritten as
\begin{align}
\label{eq:magnmom0}
Z(B)\Delta \mu=\frac{\partial \epsilon_{\rm{m}}}{\partial B}\equiv\Delta \mu-\mu_{\rm magn}(B).
\end{align}
Therefore the change of the difference in magnetic
moments due to the dressing of the molecule near resonance
given by the relation
\begin{align}
\label{eq:magnmom}
\mu_{\rm magn}(B)=\Delta \mu(1-Z(B)),
\end{align}
is automatically included in our theory
\cite{Duine04}. The curve of Eq.~(\ref{eq:magnmom}) is in very good agreement \cite{Falco05}
with the measurement of the dressed-molecule magnetic moment in 
$^6$Li gas taken if Ref. \cite{Jochim03a} and
discussed earlier in Sec~\ref{atomic}.

For positive detuning $\delta(B)>0$  the solution of Eq.~(\ref{eq:root})
has a negative imaginary part, in agreement with the fact that the molecule decays when
its energy is above the two-atom continuum threshold.
The molecular density of states at positive detuning does no longer have a delta function
at negative energy but is given by a Lorentzian-like curve 
depicted in Fig.~\ref{densityofstates}(b).
The analytic expression which gives the maximum in the density of states 
$\epsilon_{\rm{m}}$ at positive detuning
can be easily calculated in the limit of $a_{\rm{bg}}=0$
and reads \cite{Romans03}
\begin{align}
\label{eq:posdetmax}
\epsilon_{\rm m}
=\frac{1}{3}\left(\delta-\frac{\eta^2}{2}+\sqrt{\frac{\eta^4}{4}
-\eta^2\delta+4\delta^2}\right).~ 
\end{align}
At large positive detuning, the peak of the Lorentzian distribution is located
at the detuning.  
Near resonance, it is shifted from the detuning by the nonzero
coupling, and located at $\hbar^2/m a^2(B)$ exactly as the real pole at negative detuning.

\subsection{Many-body $\rm{T}-$matrix revisited: functional integral approach.}
\label{sec:many-body}

The approximation developed 
to solve the two-body scattering problem, based on Eq.~(\ref{eq:theo}),
can be generalized rather straightforwardly
to include the Pauli-blocking effects on the many-body  $\rm{T}-$matrix and 
on the molecular propagator.
Nevertheless, for completeness we prefer here
to rederive the many-body $\rm{T}-$matrix approach also by means 
of functional-integral methods \cite{Kleinert78,Nagele88,Stoof01}. 

The path integral expression for the partition function of the effective atom-molecule
model in Eq.~(\ref{eq:Ham}) is
\begin{align}
\label{eq:partfunct}
Z=\int
d[\phi^{*}_{\rm{m}}]d[\phi_{\rm{m}}]
\left(\prod_{\sigma}
d[\psi^{*}_{\sigma}]d[\psi_{\sigma}]\right)
e^{-S[\phi^{*}_{\rm{m}},\phi_{\rm{m}},\psi^{*}_{\sigma},\psi_{\sigma}]/\hbar}
\end{align}
with the Euclidean action  given by
\begin{widetext}
{\small{
\begin{align}
\label{eq:action}
S[\phi^{*}_{\rm{m}},\phi_{\rm{m}},\psi^{*}_{\sigma},\psi_{\sigma}
]=&
\int_0^{\hbar\beta}\!\! d\tau \!\int\! d\textbf{x}
\sum_{\sigma}
 \psi^{*}_{\sigma
}(\textbf{x},\tau)\left[\hbar\frac{\partial}{\partial
\tau}-\frac{\hbar^{2}\nabla^{2}}{2m}-\mu\right]\psi_{\sigma
}(\textbf{x},\tau)\nonumber
\\ 
+
&
\int_0^{\hbar\beta}\! \!d\tau \!\int \!d\textbf{x}
\;\phi_{\rm{m}}^{*}(\textbf{x},\tau)\left[\hbar\frac{\partial}{\partial
\tau}-\frac{\hbar^{2}\nabla^{2}}{2(2m)}+\delta_{\rm{bare}}(B)-2\mu\right]\phi_{m}(\textbf{x},\tau)
\\ 
+&\int_0^{\hbar\beta}\!\! d\tau \!\int\! d\textbf{x}\int\! d\textbf{x}'~
V_{\rm{bg}}\left(\textbf{x}-\textbf{x}'
\right)\psi^{*}_{\uparrow}(\textbf{x},\tau)
\psi^{*}_{\downarrow}(\textbf{x}',\tau)
\psi_{\downarrow}(\textbf{x}',\tau)\psi_{\uparrow}(\textbf{x},\tau)\nonumber 
\\ 
+&\int_0^{\hbar\beta}\! \!d\tau \!\int\! d\textbf{x}\int \! d\textbf{x}'~
\left[g^*\left(\textbf{x}-\textbf{x}'
\right)
\phi_{\rm{m}}^*\left((\textbf{x}+\textbf{x}')/2,\tau\right)
\psi_{\downarrow}
(\textbf{x}',\tau)\psi_{\uparrow}(\textbf{x},\tau)+
{\rm{c}}.{\rm{c}}.\right],\nonumber 
\end{align}
}}

\noindent where $\phi_{\rm{m}}$ denotes the molecular bosonic field and the Grassmann fields
$\psi_{\sigma}$ represent the fermionic atoms in the two
atomic hyperfine states. 
By means of the Hubbard-Stratonovich transformation
\begin{eqnarray}
\label{eq:hubbstrat}
e^{-(\psi_{\uparrow}\psi_{\downarrow}|V_{\rm{bg}}|\psi_{
\downarrow}\psi_{\uparrow})}\,= \,\int d[\Delta]d[\Delta^{*}] \,\,e^{+(\Delta|\psi_{
\downarrow}\psi_{\uparrow})+(\psi_{
\uparrow}\psi_{\downarrow}|\Delta)+(\Delta|V^{-1}_{\rm{bg}}|\Delta)}
\end{eqnarray}
we can introduce the auxiliary field $\Delta(\textbf{x},\tau)$,
which describes the quantum fluctuations in the particle-particle channel
according to the ladder approximation. 
Using this identity, we can rewrite the full action as

{\small{
\begin{align}
\label{eq:actionbis}
S[\phi^{*}_{\rm{m}},\phi_{\rm{m}},\psi^{*}_{\sigma},\psi_{\sigma},
\Delta^{*},\Delta]&= 
\int_0^{\hbar\beta}\!\! d\tau \!\int\! d\textbf{x}
\sum_{\sigma} \psi^{*}_{\sigma
}(\textbf{x},\tau)\left[\hbar\frac{\partial}{\partial
\tau}-\frac{\hbar^{2}\nabla^{2}}{2m}-\mu\right]\psi_{\sigma
}(\textbf{x},\tau)
\\&+\int_0^{\hbar\beta}\! \!d\tau \!\int \!d\textbf{x}
\;\phi_{\rm{m}}^{*}(\textbf{x},\tau)\left[\hbar\frac{\partial}{\partial
\tau}-\frac{\hbar^{2}\nabla^{2}}{2(2m)}+\delta_{\rm{bare}}(B)-2\mu\right]\phi_{m}(\textbf{x},\tau) 
- \frac{|\Delta(\textbf{x},\tau)|^{2}}{V_{\rm{bg}}}\nonumber\\&
+\int_0^{\hbar\beta} d\tau \int d\textbf{x} \;
\left(g_{\rm{bare}}\,\phi_{\rm{m}}(\textbf{x},\tau)-\Delta(\textbf{x},\tau)\right)\psi^{*}_{\uparrow}
(\textbf{x},\tau)\psi^{*}_{\downarrow}(\textbf{x},\tau)
\nonumber\\& +\int_0^{\hbar\beta} d\tau \int
d\textbf{x} \;
\left(g_{\rm{bare}}\,\phi^{*}_{\rm{m}}(\textbf{x},\tau)-\Delta^{*}(\textbf{x},\tau)\right)
\psi_{\downarrow}(\textbf{x},\tau)\psi_{\uparrow}(\textbf{x},\tau)\nonumber.
\end{align}
}}

\noindent Note that we have made the replacements 
$V_{\rm{bg}}(\textbf{x}-\textbf{x}')=V_{\rm{bg}}\delta(\textbf{x}-\textbf{x}')$
and $g(\textbf{x}-\textbf{x}')=
g_{\rm{bare}}\delta(\textbf{x}-\textbf{x}')$
as in Eq.~(\ref{eq:effpotbare}) above.

Recalling the expression of the non interacting atomic Green's function 
\begin{eqnarray}
\label{eq:nonintGr}
 G^{-1}_{\sigma,0}(\textbf{x} ,\tau;\textbf{x}',\tau')=-\frac{1}{\hbar} \left\{
\hbar\frac{\partial}{\partial
\tau}-\frac{\hbar^{2}\nabla^{2}}{2m}-\mu
\right\}\delta(\textbf{x}-\textbf{x}')\delta(\tau-\tau')
\end{eqnarray}
and using a $2\times2$ matrix (Nambu-space) notation, the dressed atomic Green's function can be written as
 \begin{align}
\label{eq:intatoGr}
&{\textbf{G}_{\rm f}}^{-1}
(\textbf{x},\tau;\textbf{x}',\tau') =
\left[%
\begin{array}{cc}
  G^{-1}_{\uparrow,0}(\textbf{x},\tau;\textbf{x}',\tau') & 0 \\
  0 & -G^{-1}_{\downarrow,0}(\textbf{x}',\tau';\textbf{x},\tau) \\
\end{array}%
\right]+\\
&-\frac{1}{\hbar}\left[%
\begin{array}{cc}
  0 & g_{\rm{bare}}\,\phi_{\rm m}(\textbf{x},\tau)-\Delta(\textbf{x},\tau) \\
  g_{\rm{bare}}^{*}\,\phi^{*}_{\rm m}(\textbf{x},\tau)-\Delta^{*}(\textbf{x},\tau) & 0 \\
\end{array}%
\right]\times \delta\left(\textbf{x}-\textbf{x}'\right)\delta\left(\tau-\tau'\right)
\nonumber\\
&\,\,\,\,\,\,\,
\,\,\,\,\,\,\,\,
\,\,\,\,\,\,\,\,\,
\,\,\,\,\,\,\,\,\,\,\,\,
\,\,\,\,\,\,\equiv  {\mathcal{\bf{G}}}^{-1}_{{\rm f},0}(\textbf{x},\tau;\textbf{x}',\tau')
-\bf{\Sigma}_{\rm{f}}(\textbf{x},\tau;\textbf{x}',\tau')\nonumber.
\end{align}
The integration over the atomic fields now involves a Gaussian integral
which adds the formal result $-\hbar \rm{Tr}[\ln(-{\mathcal{\bf{G}_{\rm f}}})]$
to the effective action, that
can be treated perturbatively.
The effective Gaussian action is obtained
by keeping terms up to the quadratic level in the $\Delta$ and $\phi_{\rm{m}}$ fields.
It reads
\begin{align}
\label{eq:actiontris}
S[\Delta^{*},\Delta,\phi^{*}_{\rm{m}},\phi_{\rm{m}}]=&\int^{\hbar\beta}_{0} d
\tau \int d\textbf{x}\;
\Delta(\textbf{x},\tau)^{*}\left[-\frac{1}{V_{\rm{bg}}}\right]\Delta(\textbf{x},\tau)
+ \left[\frac{\hbar}{2}\rm{Tr}
(\mathcal{\textbf{G}}_{f,0} {\bf{\Sigma}}_{\rm{f}})^2 \right]\\
&\!\!\!\!\!\!\!\!+\int^{\hbar\beta}_{0} d \tau \int
d\textbf{x}\;\phi_{\rm{m}}^{*}(\textbf{x},\tau)
\left[\hbar\frac{\partial}{\partial
\tau}-\frac{\hbar^{2}\nabla^{2}}{2(2m)}-\delta_{\rm{bare}}
(B)-2\mu\right]\phi_{\rm{m}}(\textbf{x},\tau),
\nonumber
\end{align}
\end{widetext}
where the trace $\rm{Tr}(\mathcal{\textbf{G}}_{{\rm f},0} {\bf{\Sigma}}_{\rm{f}})^2$
can be rewritten as
\begin{align}
\label{eq:tracebis}
\rm{Tr}(\mathcal{\textbf{G}}_{{\rm f},0} {\bf{\Sigma}}_{\rm{f}})^2
&=
\frac{2}{\hbar^2}\int^{\hbar\beta}_{0}\ \!\!d \tau \int
d\textbf{x}\int^{\hbar\beta}_{0}\ \!\! d \tau'\int
d\textbf{x}'\\ \nonumber\\&\left[-g_{\rm{bare}}\,\phi_{\rm m}(\textbf{x}',\tau')+\Delta(\textbf{x}',\tau')\right]
\times\nonumber\\
\nonumber\\
&\hbar\Pi_{0}(\textbf{x},\tau ;\textbf{x}',\tau')
\left[-g_{\rm{bare}}^{*}\,\phi^{*}_{\rm m}(\textbf{x},\tau)+\Delta^{*}(\textbf{x},\tau)\right]\nonumber
\end{align}
if we introduce the free pair propagator 
\begin{align}
\label{eq:freepartprop}
\hbar\Pi_{0} \left(\textbf{x},\tau ;
\textbf{x}',\tau'\right)=-G_{\uparrow,0} \left(\textbf{x},\tau ;
\textbf{x}',\tau'\right) G_{\downarrow0} \left(\textbf{x},\tau ;
\textbf{x}',\tau'\right)
\end{align}
in coordinate space. 
Using as a definition of the pairing-field propagator  
\begin{align}
\label{eq:Deltafield}
-\hbar 
G_{\Delta_{\rm{bg}}}^{-1}(\textbf{x},\tau;\textbf{x}',\tau')&=\\
-&\frac{\delta(\textbf{x}-\textbf{x}')\delta(\tau-\tau')}{V_{\rm{bg}}}+
\Pi_{0} (\textbf{x},\tau ;
\textbf{x}',\tau')\nonumber
\end{align}
the effective Gaussian action in Eq.~(\ref{eq:actiontris})
can be put finally in the compact form
\begin{widetext}
\begin{align}
\label{eq:actionquadris}
&S[\Delta^{*},\Delta,\phi^{*}_{\rm{m}},\phi_{\rm{m}}]
=\int d\tau \int d\textbf{x}\;\int d\tau' \int
d\textbf{x}'{\Big{\{}}
\Delta^{*}(\textbf{x}',\tau')
\left(-\hbar
 G_{\Delta_{\rm{bg}}}^{-1}
(\textbf{x},\tau;\textbf{x}',\tau')\right)
\Delta(\textbf{x},\tau)+
\nonumber \\
&\,\,\,\,\,\,\,\,\,\,\,\,\,\,\,\,\,\, 
\,\,\,\,\,\,\,\,\,\,
\,\,\,\,\,\,\,\,\,+ \phi_{\rm{m}}^{*}(\textbf{x}',\tau')\left[-\hbar G_{0
}^{-1}(\textbf{x},\tau;\textbf{x}',\tau')+\frac{1}{\hbar}g_{\rm{bare}}
\hbar\Pi_{0}(\textbf{x},\tau ;
\textbf{x}',\tau') g_{\rm{bare}} \right]\phi_{\rm{m}}(\textbf{x},\tau)+ \nonumber\\
&-\frac{1}{\hbar}\Delta^{*}(\textbf{x},\tau)
\hbar\Pi_{0}  \left(\textbf{x},\tau ;\textbf{x}',\tau'\right)  g_{\rm{bare}}
\phi_{\rm m}(\textbf{x}',\tau')
-\frac{1}{\hbar}
\phi^{*}_{\rm m}\left(\textbf{x},\tau\right)g_{\rm{bare}}
\hbar\Pi_{0}\left(\textbf{x},\tau ;\textbf{x}',\tau'\right) 
\Delta(\textbf{x}',\tau'){\Big{\}}},
\end{align}
\end{widetext}
\noindent which represents a generalization to the 
two-channel atom-molecule model
of the Gaussian fluctuation theory developed in Ref.~\cite{SadeMelo93}
for a single-channel resonantly-interacting Fermi gas.
From Eqs.~(\ref{eq:Deltafield}) and~(\ref{eq:TMBGg}) we see that 
the Fourier transform of the pair propagator $G_{\Delta_{\rm{bg}}}$
can be identified with the many-body $\rm{T}_{\rm{bg}}-$matrix. 
It can be  
easily calculated by performing the
sum over the  Matsubara frequencies 
in $\Pi_0$ and using
the Lippmann-Schwinger equation in Eq.~(\ref{eq:Lipp})
at $E=0$.
We obtain \cite{SadeMelo93,Stoof96}

{\small{
\begin{align}
\label{eq:Randeria}
&\hbar G_{\Delta_{\rm{bg}}}^{-1}\left(\mathbf{K},\Omega_n\right)\!=\!
{{\rm T}_{\rm{bg}}^{\rm{MB}}}^{-1}\left(\mathbf{K},\Omega_n\right)\!=\!\\\nonumber 
&-\frac{1}{V}\sum_{\mathbf{k}}\left[
\frac{1-N_{\mathbf{K}/2+\mathbf{K}}-N_{\mathbf{K}/2-\mathbf{k}}}{i\Omega_n+2\mu-
2\epsilon_\mathbf{k}-\epsilon_{\mathbf{K}}/2}+\frac{1}{2 \epsilon_\mathbf{k}}
\right]+\frac{m}{4\pi\hbar^2 a_{\rm{bg}}}.
\end{align}
}}

\noindent The Fermi distribution factors $N_\mathbf{k}$ 
describe the effects of Pauli blocking in the medium.

The renormalized pair propagator $G_{\Delta}$,
which includes the effects of the resonant
scattering, is calculated by integrating out also the 
molecular field in Eq.~(\ref{eq:actionquadris}).
After some algebra, which involves the definitions in Eqs.~(\ref{eq:effpotbare}-\ref{eq:TMBGg}),  
we recover the expected result
\begin{align}
\label{eq:genRanderia}
G_{\Delta}
\left(\mathbf{K},\Omega_n\right)/\hbar\!\!&=
{\rm T}^{\rm{MB}}\left(\mathbf{K},\Omega_n\right)={\rm T}_{\rm{bg}}^{\rm{MB}}
\left(\mathbf{K},\Omega_n\right)\\&
+g^{\rm{MB}*}\left(\mathbf{K},\Omega_n\right)
\frac{G\left(\mathbf{K},\Omega_n\right)}{\hbar}
g^{\rm{MB}}\left(\mathbf{K},\Omega_n\right)\nonumber.
\end{align}
Note that in this derivation we have 
assumed $\Pi=\Pi_{0}$
hich corresponds to neglecting 
many-body effects in fermionic atomic propagators. 
Nevertheless, 
this restriction would disappear if
we consider the complete series in the expansion
of the $-\hbar {\rm{Tr}}[\ln(-{\mathcal{\bf{G}}})]$ term.
At the saddle point $\Delta\left(\mathbf{x},\tau\right)=0$,
there are no contributions
coming from the $\hbar {\rm{Tr}}[\ln(-{\mathbf{G}})]$ term, 
and the solution of
the pair propagator reduces to  
\begin{align}
\label{eq:saddlegenRanderia}
\hbar G_{\Delta}^{-1}&\left(\mathbf{K},\Omega_n\right)\!\!=\\
&\left[\frac{4\pi\hbar^2 a_{\rm{bg}}}{m}+\frac{g^2_{\rm{bare}}}
{i\hbar\Omega_n+2\mu-\delta_{\rm{bare}}
-\epsilon_{\mathbf{K}}/2}\right]^{-1}\nonumber.
\end{align}
Thus the Feshbach molecule makes the pair field dynamical, even before 
including the fluctuations around the saddle point, in contrast with the usual
BCS theory, where the dynamics of the pair field is generated only via the
$\hbar \rm{Tr}[\ln(-{\mathbf{G}})]$ term.  
As we will see in two next section,
this can have a large effect on the bare molecular 
component above the resonance
in the case of a narrow resonance $\eta^2\leq \epsilon_F$.

\subsection{Many-body $\rm{T}-$matrix for a very broad resonance.}
\label{sec:many-body-broad}

For a very broad resonance,
the dynamics of the molecular boson
close to resonance, 
is dominated by the selfenergy in the molecular
propagator in Eq.~(\ref{eq:DysMol}). The dynamics of the bare
boson, given by the term $i\hbar\omega$ in the free propagator, is 
then effectively screened by dressing effects.
This can be explicitly verified by carefully studying
the low-energy behaviour of the exact 
many-body $\rm{T}-$matrix in Eq.~(\ref{eq:genRanderia}).

At fixed density the width of the BEC-BCS crossover region, delimited by 
\begin{eqnarray}
\label{eq:crossover}
|k_F a(B)| \simeq \sqrt{\epsilon_F}\frac{\eta}{|\delta(B)|}\leq 1,
\end{eqnarray}  
is
determined by the energy scale $\eta^2$, related to the width of the 
specific resonance under consideration. 
As the latter increases, so does the range of the magnetic
fields which spans the crossover region.
Within this range the detuning in general turns out,
except of course very close to resonance,
to be much larger than the other
energy scales
\begin{eqnarray}
\label{eq:broadrescons1}
|\delta(B)|\gg |2\mu|,\epsilon_
{\rm{m}}.
\end{eqnarray}
To put it more precisely, at positive detuning it is $\delta(B)\gg 2\mu\simeq 2\epsilon_F$. 
At negative detuning, 
the chemical potential 
approaches half the molecular binding energy 
$ 2|\mu|\simeq |\epsilon_
{\rm{m}}|\ll |\delta(B)|$. Furthermore, at 
low energies $\hbar\omega\simeq \epsilon_F$
and low momenta $\hbar|\mathbf{q}|\simeq \hbar k_F$,
we have $ \hbar\omega\ll\hbar\Sigma_{\rm{m}}
\left(\mathbf{q},\omega\right)\simeq\eta \Pi \left(\mathbf{q},\omega\right)$
due to the large coupling $\eta $.
Under these conditions, the full molecular propagator 
in Eq.~(\ref{eq:DysMol}) and~(\ref{eq:genRanderia}) can be approximated by 
\begin{eqnarray}
\label{eq:apprDysMol}
\hbar G\left(\mathbf{q},\omega\right)^{-1}\simeq-\delta
_{\rm{bare}}
-
\hbar\Sigma_{\rm{m}}
\left(\mathbf{q},\omega\right).
\end{eqnarray}
Introducing also
\begin{eqnarray}
\label{eq:bareresint}
{\rm{V}}_{\rm{res}}=-\frac{g_{\rm{bare}}^2}
{\delta
_{\rm{bare}}
}
\end{eqnarray}
as a new definition of the bare resonant part of the potential and  
using  the  set of relations in Eqs.~(\ref{eq:effpotbare}-\ref{eq:TMBGg}),
we obtain without any further approximation 
\begin{align}
\label{eq:apprTMB}
{\rm T}^{\rm{MB}}\left(\mathbf{K},\omega\right)\simeq &{\rm T}_{\rm{bg}}^{\rm{MB}}
\left(\mathbf{K},\omega\right)
+g^{\rm{MB}*}\left(\mathbf{K},\omega\right)\nonumber\\
&\times\frac{1}{-\delta_{\rm{bare}}-
\hbar\Sigma_{\rm{m}}
\left(\mathbf{K},\omega\right)}g^{\rm{MB}}\left(\mathbf{K},\omega\right)\nonumber\\
&=
\frac{
{\rm V}_{\rm{bg}}+{\rm{V}}_{\rm{res}}
}{
1-\left(
{\rm V}_{\rm{bg}}+{\rm{V}}_{\rm{res}}\right)\Pi\left(\mathbf{K},\omega\right)
}.
\end{align}
\begin{figure}
\begin{center}
 \resizebox{80mm}{!}{ \epsfig{figure=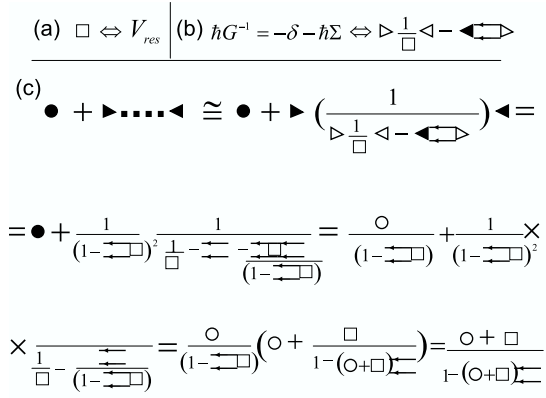,angle=-90}}
  \caption{{\footnotesize{(a) Diagrammatic representation of the bare resonant
  interaction introduced in Eq.~(\ref{eq:bareresint}). (b) Approximated form
  of the molecular propagator defined in Eq.~(\ref{eq:apprDysMol}).
  (c) Diagrammatic derivation of the formula in Eq.~(\ref{eq:apprTMB}).  
  }}} 
   \label{StoofC}
  \end{center}
\end{figure}
This is the formal solution of a Bethe-Salpeter equation for the many-body $\rm{T}-$matrix of the 
bare interaction ${\rm V}_{\rm{bg}}+{\rm{V}}_{\rm{res}}$
\cite{Beliaev57,Galitskii57,Stoof96}. A derivation of Eq.~(\ref{eq:apprTMB}) based on the diagrammatic calculus
developed in Sec.~\ref{sec:twobodyphys} is shown in Fig. \ref{StoofC}.
Renormalizing the bare interactions to the two-body $\rm{T}-$matrix 
as in Eq.~(\ref{eq:Randeria}) this equation can be rewritten as \cite{SadeMelo93}
\begin{align}
\label{eq:apprTMBren}
\hbar &G_{\Delta}^{-1}\left(\mathbf{K},\Omega_n\right)\!\!
={{\rm T}^{\rm{MB}}}^{-1}\left(\mathbf{K},\Omega_n\right)\simeq \\&
\frac{m}{4\pi\hbar^2 a(B)}-
\frac{1}{V}\sum_{\mathbf{k}}
\left[
\frac{1-N_{\mathbf{K}/2+\mathbf{k}}-N_{\mathbf{K}/2-\mathbf{k}}}{i\Omega_n+2\mu-
2\epsilon_\mathbf{k}-\epsilon_{\mathbf{K}}/2}+\frac{1}{2 \epsilon_\mathbf{k}}
\right]\nonumber.
\end{align}
In the limit of zero-energy scattering
in the vacuum it reduces to the atom-atom effective interaction in Eq.~(\ref{eq:sclen})
\begin{eqnarray}
\frac{4\pi\hbar^{2}a(B)}{m}&=\frac{4\pi\hbar^{2}a_{bg}(B)}{m}-\frac{g^2(B)}{\delta(B)}\\
&\equiv\frac{4\pi\hbar^{2}}{m}\left(a_{\rm{bg}}(B)+a_{\rm{res}}(B)\right)
\nonumber
\end{eqnarray}
that we obtain by integrating out the molecular field from the very beginning
in the original action in Eq.~(\ref{eq:action}).
This shows that for 
a very broad resonance,
the low-energy behaviour of the 
many-body $\rm{T}-$matrix~(\ref{eq:genRanderia})
in most of the BEC-BCS crossover
region~(\ref{eq:crossover}),
does not differ significantly
from the many-body
$\rm{T}-$matrix of a {\it{single-channel}} interaction with scattering length $a(B)$.
A detailed analysis of the analytic structure of the many-body 
$\rm{T}-$matrix~(\ref{eq:apprTMBren}) has been carried out by
Combescot in Ref.~\cite{Combescot03a,Combescot03b}.

However, even for a broad resonance, a complete approach to investigate
the physics across the Feshbach resonance, cannot 
neglect the closed channel from the beginning.
This is essentially for two reasons. 
First, at sufficiently negative detuning,
where the systems approaches a gas of weakly-interacting bare molecules
with binding energy linear in the magnetic field, the closed channel dominates. 
In that regime, the approximation introduced in Eq.~(\ref{eq:apprDysMol}) fails
because the analytic expression of the two-body scattering amplitude
contained in Eq.~(\ref{eq:apprTMB}) has the wrong pole structure in 
the complex plane in order to reproduce the correct binding energy.
The pole which denotes the binding energy of the molecular state 
follows Wigner's formula $\epsilon_{\rm{m}}(B)=-\hbar^2/m a^2(B)$,
which is quadratic in the detuning, at every magnetic field below resonance.
Therefore only a two-channel model can allow a full description 
at every magnetic field across the resonance.
Secondly, near the resonance the gas consists of a mixture of atoms and dressed
molecules which can 
in principle be measured separately. Dressed molecules have nonzero
components on two different spin states configurations, even above 
the Feshbach resonance, which can also be measured by experiments \cite{Partridge05}.
Thus, even when the dressing effect is so strong that the wavefunction
contains almost completely only atomic scattering states in the open channel,
the distinction between the two different components can be crucial for the
interpretation of the experiments. 

\subsection{Many-body molecular selfenergy.}
\label{sec:selfen}
The physics of the dressed molecules is described by the molecular propagator
in the presence of the medium.
The effective Gaussian action for the molecules is obtained 
by integrating out the pairing field $\Delta\left(\mathbf{x},\tau\right)$ 
in Eq.~(\ref{eq:actionquadris})
\begin{widetext}
\begin{align}
\label{eq:effactmol}
S[\phi^{*}_{\rm{m}},\phi_{\rm{m}}]
=\int d\tau \int d\textbf{x}\;\int d\tau' \int
d\textbf{x}'\;
  &\phi_{\rm{m}}^{*}(\textbf{x}',\tau')\left[-\hbar G_{0
}^{-1}\left(\textbf{x},\tau;\textbf{x}',\tau'\right)\right.\\
&\left.
+\frac{1}{\hbar}g_{\rm{bare}}
\hbar\Pi_{0}
\left(g_{\rm{bare}}+\frac{1}{\hbar^2} G_{\Delta_{\rm{bg}}}
\left(\textbf{x},\tau;\textbf{x}',\tau'\right)
\hbar\Pi_{0}
g_{\rm{bare}}\right)  
 \right]\phi_{\rm{m}}(\textbf{x},\tau). \nonumber
\end{align}
When we recall the definition of the pairing 
propagator~$G_{\Delta_{\rm{bg}}}$ in Eq.~(\ref{eq:Deltafield}) 
and Eqs.~(\ref{eq:gMBtris}-\ref{eq:TMBGg}),
we observe that the effect of the Gaussian integration over the
field $\Delta_{\rm{bg}}$
consists of dressing
the atom-molecule coupling constant
with ladder diagrams. Therefore, we obtain the 
renormalized molecular propagator
\begin{align}
S[\phi^{*}_{\rm{m}},\phi_{\rm{m}}]
=\int d\tau \int d\textbf{x}
\int d\tau' \int d\textbf{x}'\; \phi^{*}_{\rm{m}}(\textbf{x}',\tau') \left[
-\hbar G_{0}^{-1}
(\textbf{x},\tau;\textbf{x}',\tau')
+ \frac{1}{\hbar}~g~
\hbar\Pi_0(\textbf{x},\tau;\textbf{x}',\tau')
~g^{\rm{MB}} \right] \phi_{\rm{m}}(\textbf{x},\tau).
\end{align}  
\end{widetext}
By repeating the same renormalization procedure of the ultraviolet divergencies
described above in Eqs.~(\ref{eq:2Bpropren}) and (\ref{eq:rendet}), 
the many-body molecular propagator can be rewritten as
\begin{align}
\label{eq:MBprop}
\hbar{G^
{\rm{MB}}}^{-1}
\left(\mathbf{k},E\right)
=E+&2\mu-\delta(B)-\epsilon_{\mathbf{k}}/2\\
&-\hbar\Sigma^{\rm{MB}}_{\rm{m}}\left(\mathbf{k},E\right)
+\hbar\Sigma^{\rm{2B}}_{\rm{m}}\left(\mathbf{0},0\right)\nonumber,
\end{align}
where the many-body selfenergy in the ladder approximation 
is given by 
\begin{align}
\label{eq:MBSelfenergy}
\hbar&\Sigma^{\rm{MB}}_{\rm{m}}\left(\mathbf{q},E\right)
-\hbar\Sigma^{\rm{2B}}_{\rm{m}}\left(\mathbf{0},0\right)\!\!=
\\
&\frac{1}{V}
\sum_{\mathbf{k}}
|g^{\rm{MB}}\left(2\mathbf{k},2\epsilon_\mathbf{k}\right)|^2\left[
\frac{1-N_{\mathbf{q}/2+\mathbf{k}}-N_{\mathbf{q}/2-\mathbf{k}}}{E+2\mu-
2\epsilon_\mathbf{k}-\epsilon_{\mathbf{q}}/2}+\frac{1}{2 \epsilon_\mathbf{k}}
\right]\nonumber.
\end{align}

If we neglect the many-body corrections in $g^{\rm{MB}}$ and the Fermi
factors in the numerator of the integrand, we recover the two-body result in Eq.~(\ref{eq:self}).
The integral of Eq.~(\ref{eq:MBSelfenergy}) cannot be performed analytically
but some progress can be made by considering a special limit.
Neglecting the many-body effects in the dressed coupling
constant $g^{\rm{MB}}$, the expression reduces to
\begin{align}
\label{eq:MBSelfenergy1}
\hbar&\Sigma^{\rm{MB}}_{\rm{m}}\left(\mathbf{q},E\right)
-\hbar\Sigma^{\rm{2B}}_{\rm{m}}\left(\mathbf{0},0\right)=\\&\frac{1}{V}
\sum_{\mathbf{k}}
|g^{\rm{2B}}\left(\mathbf{0},2\epsilon_\mathbf{k}\right)|^2\left[
\frac{1-N_{\mathbf{q}/2+\mathbf{k}}-N_{\mathbf{q}/2-\mathbf{k}}}{E+2\mu-
2\epsilon_\mathbf{k}-\epsilon_{\mathbf{q}}/2}+\frac{1}{2 \epsilon_\mathbf{k}}
\right]\nonumber.
\end{align}
In this case, the integral can be integrated analytically at $T=0$ but 
the solution is rather cumbersome and we do not reproduce it here.
The solution at $T=0$~ in the limit $a_{\rm{bg}}\rightarrow 0$,
is given by \cite{Falco04b}
\begin{align}
\label{eq:selfenknd}
&\hbar\Sigma_{{\rm m}}^{(+)}({\bf q},\omega)=-\eta i
\sqrt{\hbar\omega+2\mu-\frac{\epsilon_{{\bf
q}}}{2}}\nonumber\\
&+2\eta\frac{\sqrt{2 \mu}}{\pi}+\eta \frac{\hbar\omega}{\pi\sqrt{2\epsilon_{\bf q}}}
{\rm{ln}}\left[\frac{\hbar\omega-\epsilon_{\bf q}+2\sqrt{\mu\epsilon_{\bf q} }}
{\hbar\omega-\epsilon_{\bf q}-2\sqrt{\mu\epsilon_{\bf q} }}
\right]+\nonumber\\
&+\frac{\eta}{\pi}\sqrt{\hbar\omega+2\mu-\frac{\epsilon_{{\bf q}}}{2}}\left[
{\rm{ln}}\frac{\sqrt{\hbar\omega+2\mu-\frac{\epsilon_{{\bf q}}}{2}}
-(\sqrt{2\mu}+\sqrt{\frac{\epsilon_{{\bf q}}}{2}})}
{\sqrt{\hbar\omega+2\mu-\frac{\epsilon_{{\bf q}}}{2}}
+(\sqrt{2\mu}+\sqrt{\frac{\epsilon_{{\bf q}}}{2}})}\right.\nonumber\\
&+\left.
\rm{ln}\frac{\sqrt{\hbar\omega+2\mu-\frac{\epsilon_{\bf q}}{2}}
-(\sqrt{2\mu}-\sqrt{\frac{\epsilon_{{\bf q}}}{2}})}
{\sqrt{\hbar\omega+2\mu-\frac{\epsilon_{{\bf q}}}{2}}
+(\sqrt{2\mu}-\sqrt{\frac{\epsilon_{{\bf q}}}{2}})}
\right],
\end{align}
\begin{figure*}
    \begin{tabular}{cc}
      \resizebox{70mm}{!}{\includegraphics{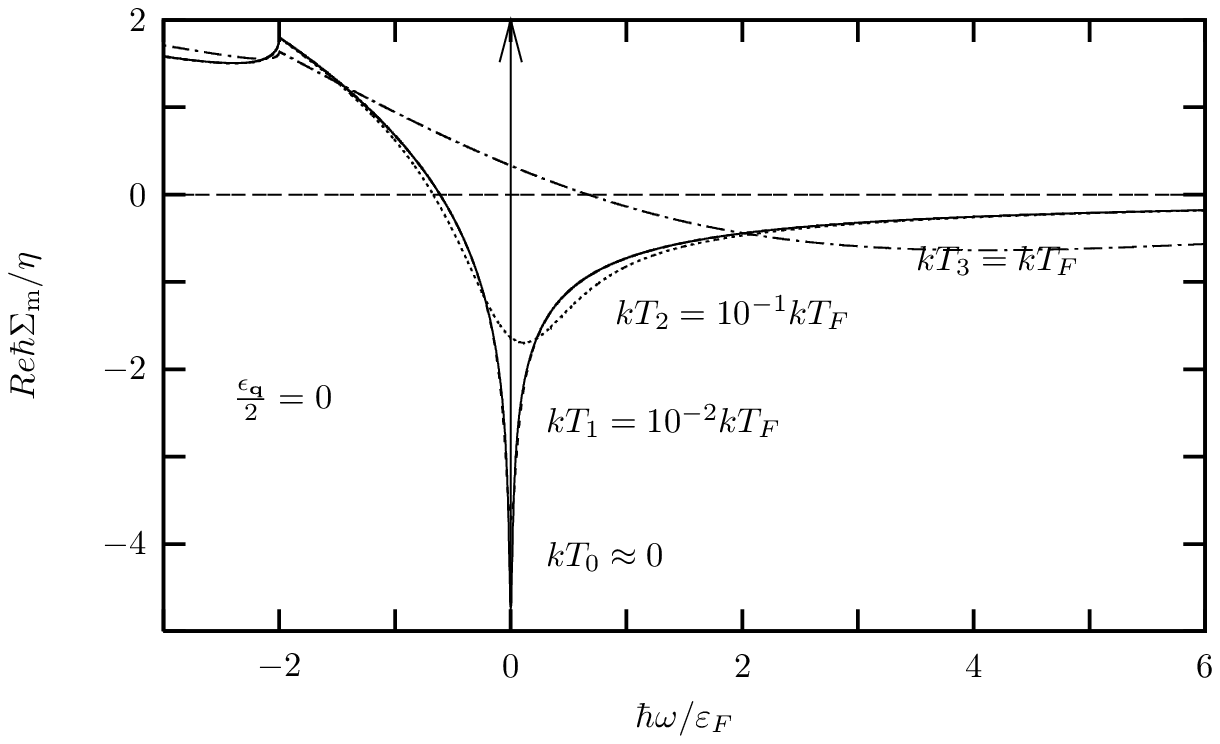}} &
      \resizebox{70mm}{!}{\includegraphics{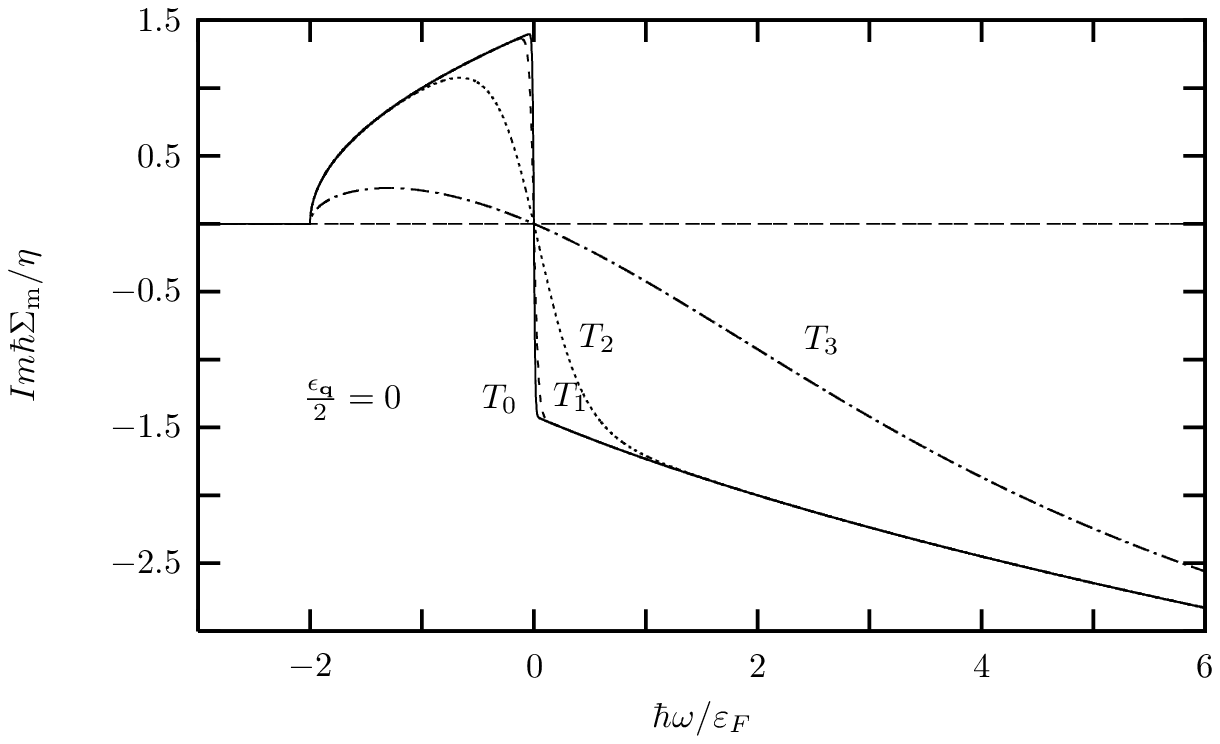}} \\
    \end{tabular}
    \caption{Real part (left) and imaginary part (right) of the molecular
    selfenergy in Eq.~(\ref{eq:MBSelfenergy1}), with $\mathbf{q}=0$ and 
    $|g^{\rm{2B}}\left(\mathbf{0},2\epsilon_\mathbf{k}\right)|^2\simeq 
     |g^{\rm{2B}}\left(\mathbf{0},\mathbf{0}\right)|^2=g^2$, and
    at different temperatures.
    }
    \label{Self1}
\end{figure*}

\begin{figure*}
    \begin{tabular}{cc}
      \resizebox{70mm}{!}{\includegraphics{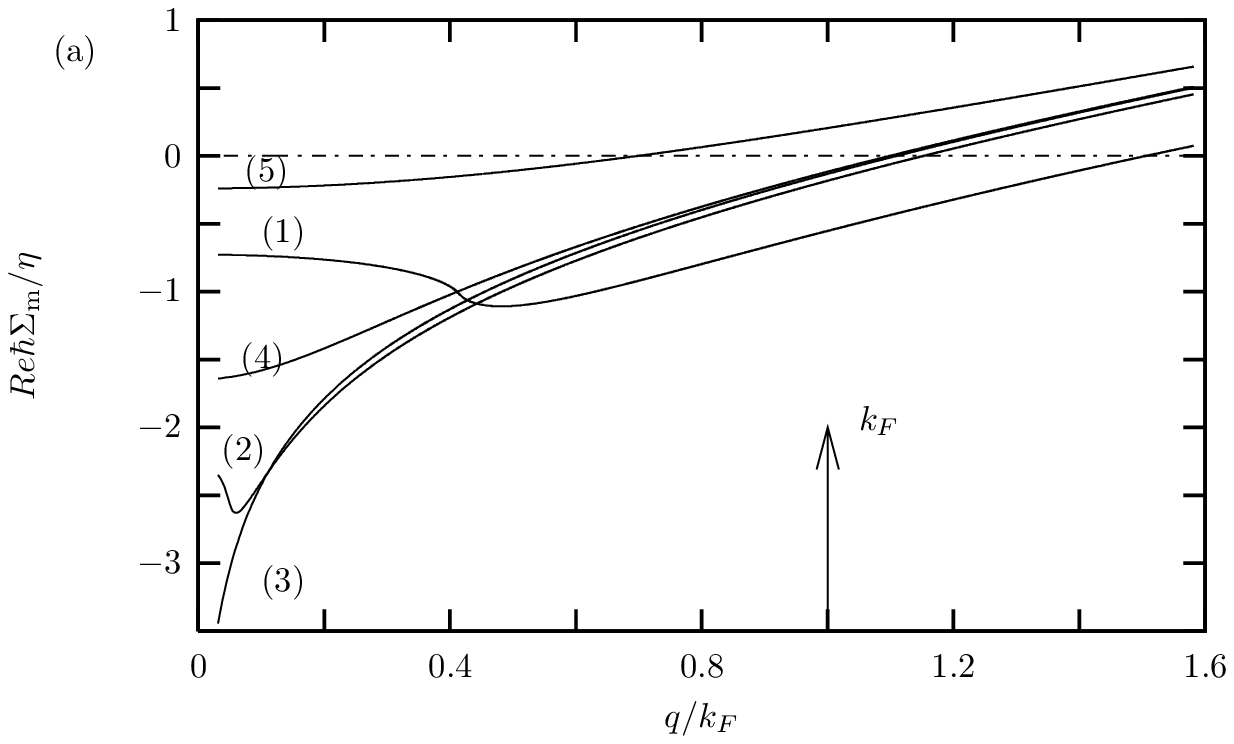}} &
      \resizebox{70mm}{!}{\includegraphics{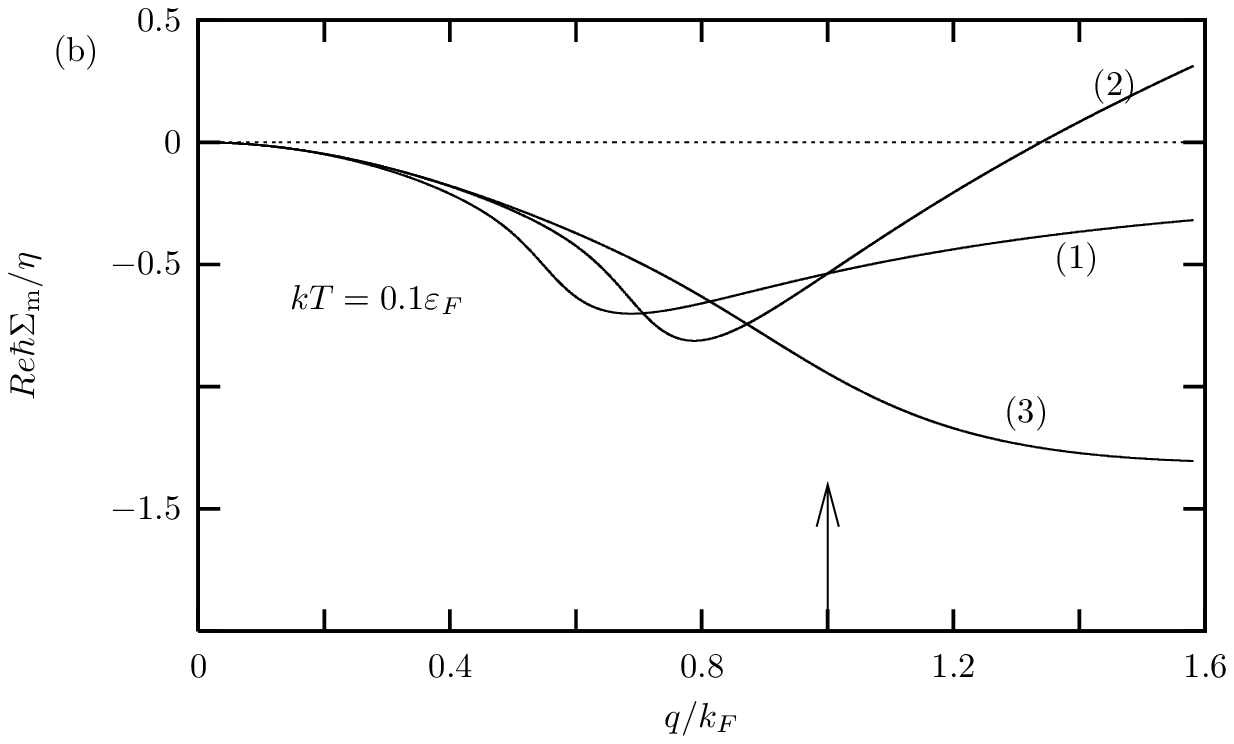}} \\
    \end{tabular}
    \caption{(a) Real part of the molecular
    selfenergy in Eq.~(\ref{eq:MBSelfenergy1})
    as a function of the momentum ${\mathbf{q}}$,
    at different energies and temperatures.
    The various curves are
     ~$(1)~ kT=0,~ \hbar\omega=\epsilon_F;~
(2)~ kT=0,~ \hbar\omega=0.1\cdot\epsilon_F;~
(3)~ kT=0,~ \hbar\omega=0.01\cdot\epsilon_F;~
(4)~ kT=0.1\cdot kT_F,~\hbar\omega=0.01\cdot\epsilon_F;~
(5)~ kT=1.\cdot kT_F,~\hbar\omega=0.01\cdot\epsilon_F$;
    (b) Real part of the molecular
    selfenergy in Eq.~(\ref{eq:MBSelfenergy1})
    as a function of the scaling variable $\hbar\omega/\epsilon_{\mathbf{q}}$.
    The curves are $(1)~ Re\hbar\Sigma_{\rm{m}}
(\omega/\epsilon_{\mathbf{q}},\mathbf{q})/\eta~,~
(2)~Re\hbar\Sigma_{\rm{m}}
(\omega/\epsilon_{\mathbf{q}},\mathbf{k}_F)/\eta~,~
(3)~Re\hbar\Sigma_{\rm{m}}
(\omega/\epsilon_{\mathbf{q}},0.3\cdot\mathbf{k}_F)/\eta$. }
    \label{Self4}
\end{figure*}

\noindent As an important point, we must mention
the fact that in Eq.~(\ref{eq:MBSelfenergy1})
the Galilean invariance is lost.
This is because the Fermi sea
introduces a preferred coordinate frame
in the description of the scattering.
Physical insight is gained by observing the  $T=0$ solution for 
$\mathbf{q}=0$. In that case the solution is
\begin{align}
\label{eq:MBSelfenergy2}
\hbar\Sigma^{\rm{MB}}_{\rm{m}}\left(\mathbf{0},E\right)&
-\hbar\Sigma^{\rm{2B}}_{\rm{m}}\left(\mathbf{0},0\right)=
\eta\frac{\sqrt{-E-2\mu}}{1+\sqrt{\frac{-E-2\mu}{\epsilon_{\rm{bg}}}}}
+\\&\frac{\eta}{1+\frac{E+2\mu}{\epsilon_{\rm{bg}}}}
\times\left(
 \frac{4}{\pi}\sqrt{\epsilon_{\rm{bg}}}
~{\rm{Arctan}}\left[\frac{\sqrt{2\epsilon_F}}
{\sqrt{\epsilon_{\rm{bg}}}}\right] \right.\nonumber\\
&\left.- 
\frac{4}{\pi}\sqrt{-E-2\mu}~{\rm{Arctan}}\left[\frac{\sqrt{2\epsilon_F}}
{-E-2\mu}\right]
\right)\nonumber
\end{align}
or   
\begin{align}
\label{eq:MBSelfenergy3}
&\hbar\Sigma^{\rm{MB}}_{\rm{m}}\left(\mathbf{0},E\right)
-\hbar\Sigma^{\rm{2B}}_{\rm{m}}\left(\mathbf{0},0\right)=
\eta
\sqrt{-2\mu-E} \\
&+ \frac{4}{\pi}\eta\sqrt{2\epsilon_F} - 
\frac{4}{\pi}\eta\sqrt{-E-2\mu}~{\rm{Arctan}}\left[\frac{\sqrt{2\epsilon_F}}
{-E-2\mu}\right]
\nonumber
\end{align}
in the limit of $a_{\rm{bg}}\rightarrow 0$ when 
it is possible to neglect the energy dependence
of the atom-molecule coupling constant.
\begin{figure*}
    \begin{tabular}{cc}
      \resizebox{65mm}{!}{\includegraphics{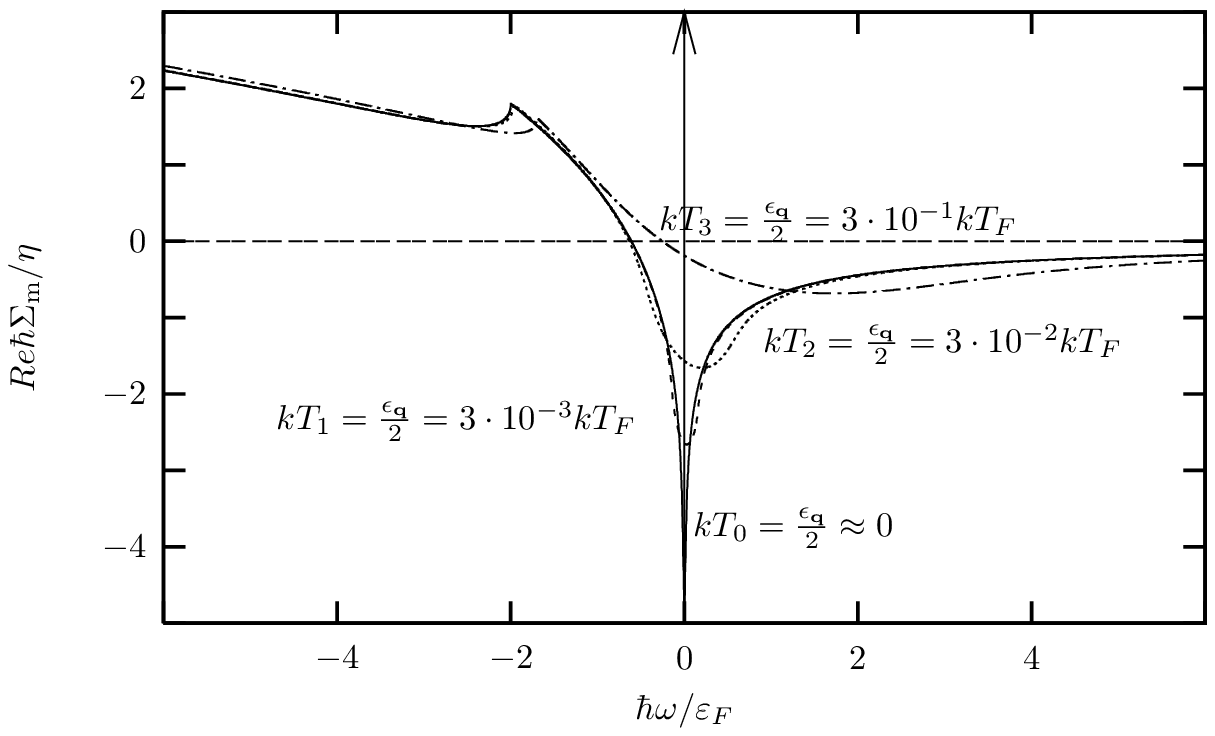}} & \hspace{0.7 cm}
      \resizebox{70mm}{!}{\includegraphics{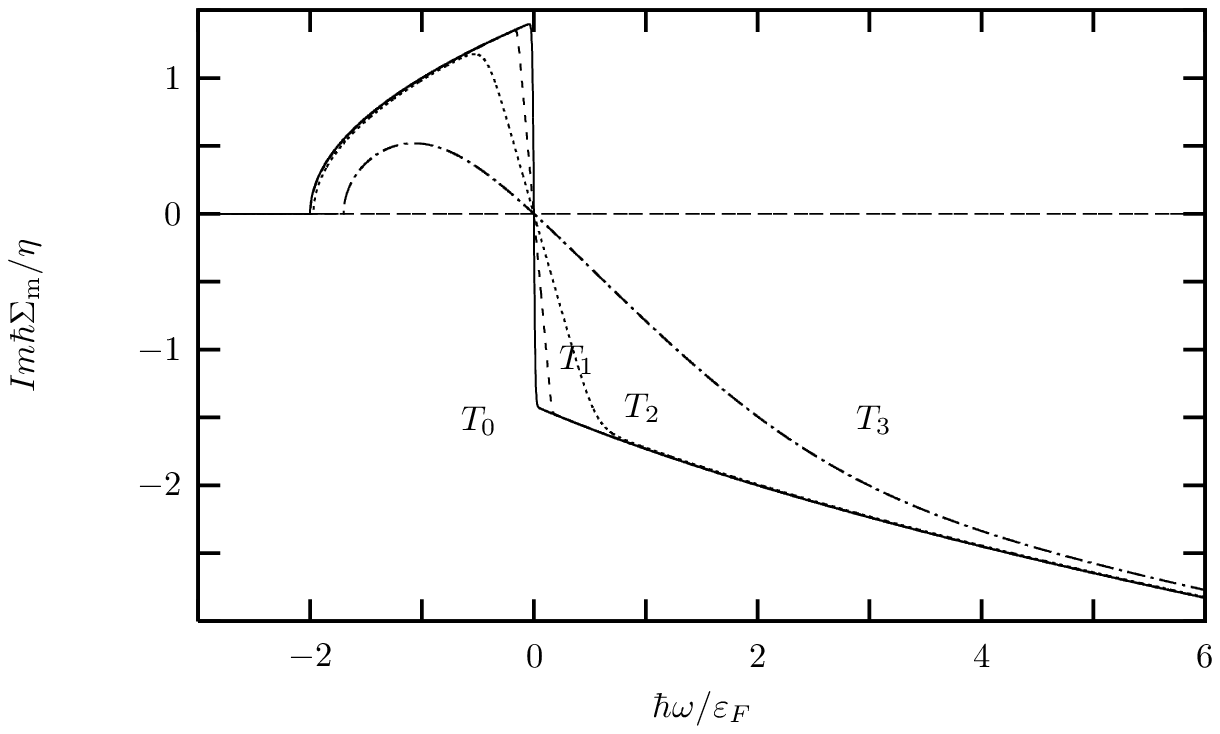}} \\
    \end{tabular}
    \caption{Real part (left) and imaginary part (right) of the molecular
    selfenergy in Eq.~(\ref{eq:MBSelfenergy1}), with  
    $|g^{\rm{2B}}\left(\mathbf{0},2\epsilon_\mathbf{k}\right)|^2\simeq 
     |g^{\rm{2B}}\left(\mathbf{0},\mathbf{0}\right)|^2=g^2$, and
    at different momenta and temperatures.
    }
    \label{Self2}
\end{figure*} 
\begin{figure*}
    \begin{tabular}{cc}
      \resizebox{70mm}{!}{\includegraphics{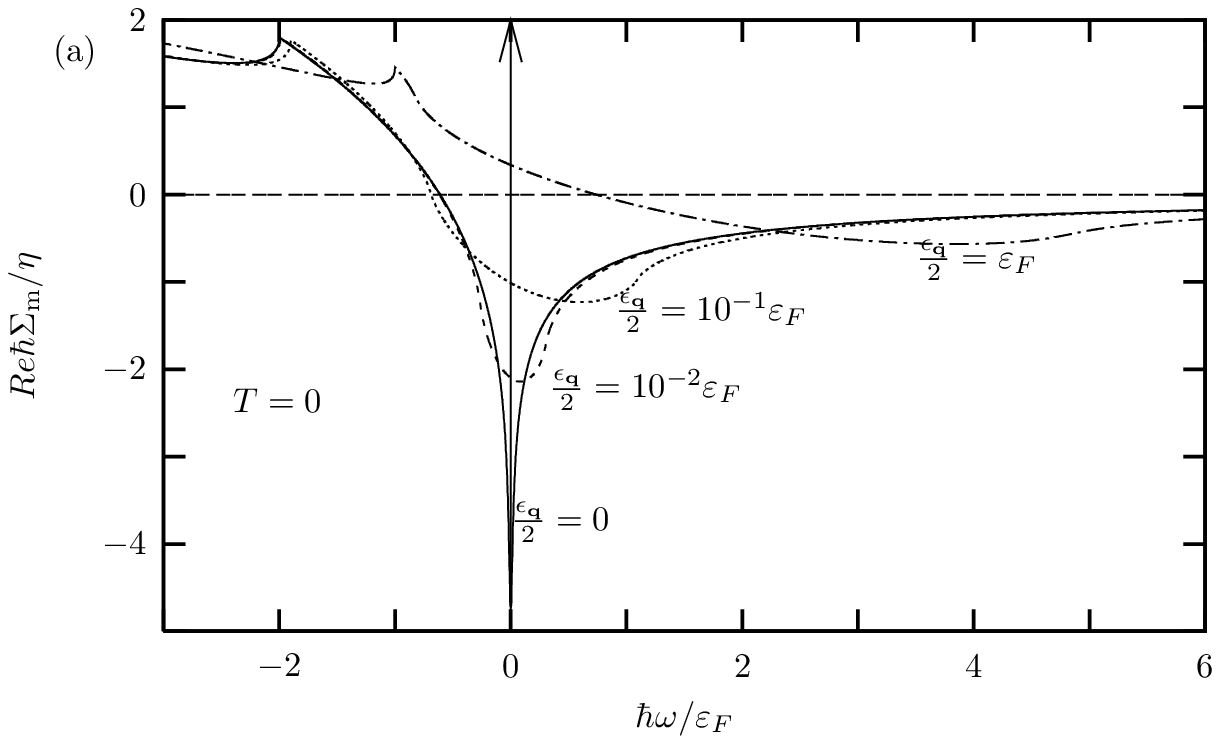}} & \hspace{0.6 cm}
      \resizebox{66mm}{!}{\includegraphics{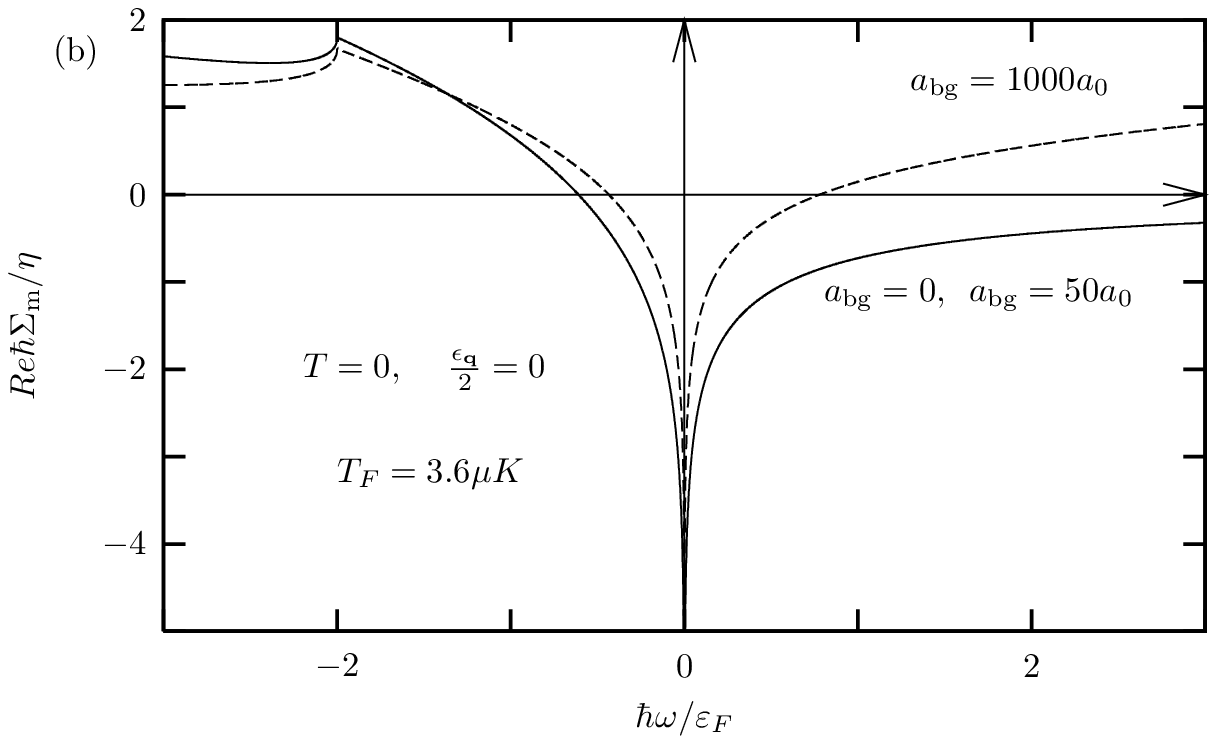}} \\
    \end{tabular}
    \caption{(a) Real part of the molecular
    selfenergy~(\ref{eq:selfenknd}),
    at different momenta.
    (b) Real part of the molecular
    selfenergy in Eq.~(\ref{eq:MBSelfenergy2}),
    for different value of the background scattering length $a_{\rm bg}$.}
    \label{Self3}
\end{figure*}
Only the 
imaginary part of the selfenergy in Eq.~(\ref{eq:MBSelfenergy1}) can be calculated by simple
analytic methods. In the limit $a_{\rm{bg}}\rightarrow 0$ we find \cite{Ohashi0203}
\begin{align}
\label{eq:imselfen}
{\rm{Im}}&\{\hbar\Sigma^{\rm{MB}}_{\rm{m}}\left(\mathbf{q},E^{+}\right)\}
=~{\eta \over \beta \sqrt{\frac{\epsilon_{\mathbf{q}}}{2}}}
\Theta(E+2\mu-\frac{\epsilon_{\mathbf{q}}}{2})\nonumber\\
&\times {\rm{ln}}~\left[
{\cosh{\beta \over 2}({E^{+} \over 2}
+\sqrt{(E^{+}+2\mu-\frac{\epsilon_{\mathbf{q}}}{2})
\frac{\epsilon_{\mathbf{q}}}{2}})
\over
\cosh{\beta \over 2}({E^{+} \over 2}
-\sqrt{(E^{+}+2\mu-\frac{\epsilon_{\mathbf{q}}}{2})
\frac{\epsilon_{\mathbf{q}}}{2}
})}\right]. 
\end{align}
The real part at nonzero temperatures can only be calculated numerically.
The effects of the nonzero temperature corrections
are shown in Figs.~\ref{Self1} and~\ref{Self2}.
Fig.~\ref{Self3}(a) illustrates the real part of the zero-temperature
selfenergy in Eq.~(\ref{eq:selfenknd}) at different momenta ${\mathbf{q}}$.
Fig.~\ref{Self3}(b) shows the effect of 
the corrections introduced by the background scattering length $a_{\rm{bg}}$.
In Fig.~\ref{Self4}(a)
the real part of the 
selfenergy in Eq.~(\ref{eq:MBSelfenergy1}) is plotted
at fixed energy $\hbar\omega$ as a function of 
the momentum $\hbar q$ at different temperatures.
Finally in Fig.~\ref{Self4}(b)
the relative dependence between the two variables
$\omega$ and ${\mathbf{q}}$ is illustrated.

\subsection{Fermi edge effects for a narrow Feshbach resonance.}
\label{sec:Kondo}

A two-component Fermi gas with negative $s-$wave interactions exhibits a superconducting
instability when lowering the temperature to $T\ll T_F$. 
The instability is
signaled by a singularity at $\mathbf{K}=\Omega=0$ in the two-particle
propagator $ G_{\Delta}$ which describes the Cooper channel.
For a weak attractive background interaction $a_{\rm{bg}}<0$, for example, the
pair propagator in Eq.~(\ref{eq:Randeria})
of the traditional BCS theory 
\begin{align}
\label{eq:TCbgesssing}
\hbar G_{\Delta_{\rm{bg}}}^{-1}&\left(\mathbf{0},0\right)\!\!=\!\!
{{\rm T}_{\rm{bg}}^{\rm{MB}}}^{-1}\left(\mathbf{0},0\right)\!\!=\\&
\!\!-\frac{1}{V}\sum_{\mathbf{k}}\left[
\frac{1-2 N_\mathbf{k}}{2\mu-
2\epsilon_\mathbf{k}}+\frac{1}{2 \epsilon_\mathbf{k}}
\right]+\frac{m}{4\pi\hbar^2 a_{\rm{bg}}}\nonumber
\end{align}
has a pole at \cite{SadeMelo93,Gorkov60}
\begin{eqnarray}
\label{eq:TCbgesssing1}
T_{\rm{c}} \simeq \frac{8\epsilon_F}{k_{\rm B} \pi} e^{\gamma-2}\ 
             \exp \left\{ -\frac{\pi}{2k_F|a_{\rm{bg}}|}  \right\}~,
\end{eqnarray}
where $\gamma=0.5772$\ is Euler's constant.

From a mathematical point of view, the logarithmic singularity is generated by
the rather sharp Fermi surface in Eq.~(\ref{eq:TCbgesssing}).
The same structure occurs in the selfenergy in Eq.~(\ref{eq:MBSelfenergy1}).
This is because many-body effects on the
propagation of the molecule and the Cooper instability
are described by the same ladder diagrams.
As a result, the pairing mechanism must be enhanced by the presence 
of the molecular bound state. 
The effects of the resonant interaction on 
the Cooper instability, 
are considered in the next chapter,
by focusing on the singularities of
the total many-body $\rm{T}-$matrix in Eq.~(\ref{eq:genRanderia}).
Here we 
follow the opposite logic and 
concentrate on the effects of 
the logarithmic singularity, introduced by the atomic Fermi sea, on 
the molecular field.

The  bare molecular field  
is a true independent degree of freedom.
Even at the lowest order in the selfenergy, its propagator
\begin{eqnarray}
\label{eq:baremolprop1}
\hbar G_0^{-1}\left(\mathbf{K},\Omega_n\right)=
\left(i\hbar\Omega_n-\frac{\epsilon_{\mathbf{K}}}{2}+2\mu-\delta_{\rm{bare}}
\right)
\end{eqnarray}
is dynamical.
This is in contrast with
the auxiliary pairing field $\Delta_{\rm{bg}}$
of BCS theory, which, at lowest order, is described by a static propagator 
\begin{eqnarray}
\label{eq:static}
\hbar G^{-1}_{\Delta_{\rm{bg}}}=\frac{1}{V_{\rm{bg}}}.
\end{eqnarray}
This explains why we expect the logarithmic singularity
to have nontrivial effects on the molecule thermodynamics.
It turns out that
it can lead to important consequences,
expecially
in the case of a narrow resonance $\eta^2\leq \epsilon_F$.
Surprisingly, the bare
molecular component of the gas, becomes 
strongly enhanced by the presence of a sharp Fermi surface
when approaching the resonance 
from positive detuning.
The mechanism underlying this effect can be better understood
when examining the analogy with other well-known phenomena
in condensed-matter physics.
In particular, the resonant molecular level embedded in the continuum
of the atoms, is reminescent of the Anderson model for a
quantum dot, where, a localized electron level is located just
below the Fermi energy of the metal leads.

\begin{figure}[h]
 \begin{center}
 \includegraphics[width=0.95 \columnwidth]{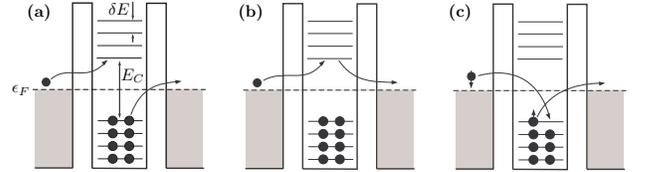}
 \caption{Examples of co-tunneling processes:
 (a) inelastic co-tunneling transferring an electron between the leads leaves 
 behind an electron-hole pair in the dot; (b) elastic co-tunneling; (c) elastic co-tunneling
 with a flip of the spin. This figure is taken from Ref. \cite{Glazman04}}.
 \label{DOS2}
 \end{center}
\end{figure}

Quantum dots are small solid-state devices in which the number of
electrons is a well defined integer N \cite{Gordon98,Glazman04}. 
In Fig.~\ref{DOS2}~ the quantum dot is sketched as
an electron box, separated from the leads by tunable tunnel
barriers 
and with a number of spin-degenerate energy states 
that can be 
single or doubly occupied by electrons of either spin up or down.
An electron on the leads cannot tunnel onto the dot because it would cost an energy
$E_C-\epsilon_F$, which is assumed to be much larger than the thermal energy $k_{\rm B} T$ of
the leads. 
The inhibition of such a transition, called the Coulomb blockade,
suppresses exponentially the conduction through the dot at low temperature.
This suppression occurs because the process of electron transport through
the dot involves a real transition to the state in which the charge of the dot differs by
one unit from the thermodynamically most probable value.
However, the Heisenberg uncertainty principle allows higher-order
processes for short durations, in which virtual states participate in the tunneling
process. The leading contributions to this activationless transport are provided
by the inelastic and elastic co-tunneling processes described in Fig.~\ref{DOS2}. 

\begin{figure*}
  \begin{center}
    \begin{tabular}{cc}
      \resizebox{68mm}{!}{\includegraphics{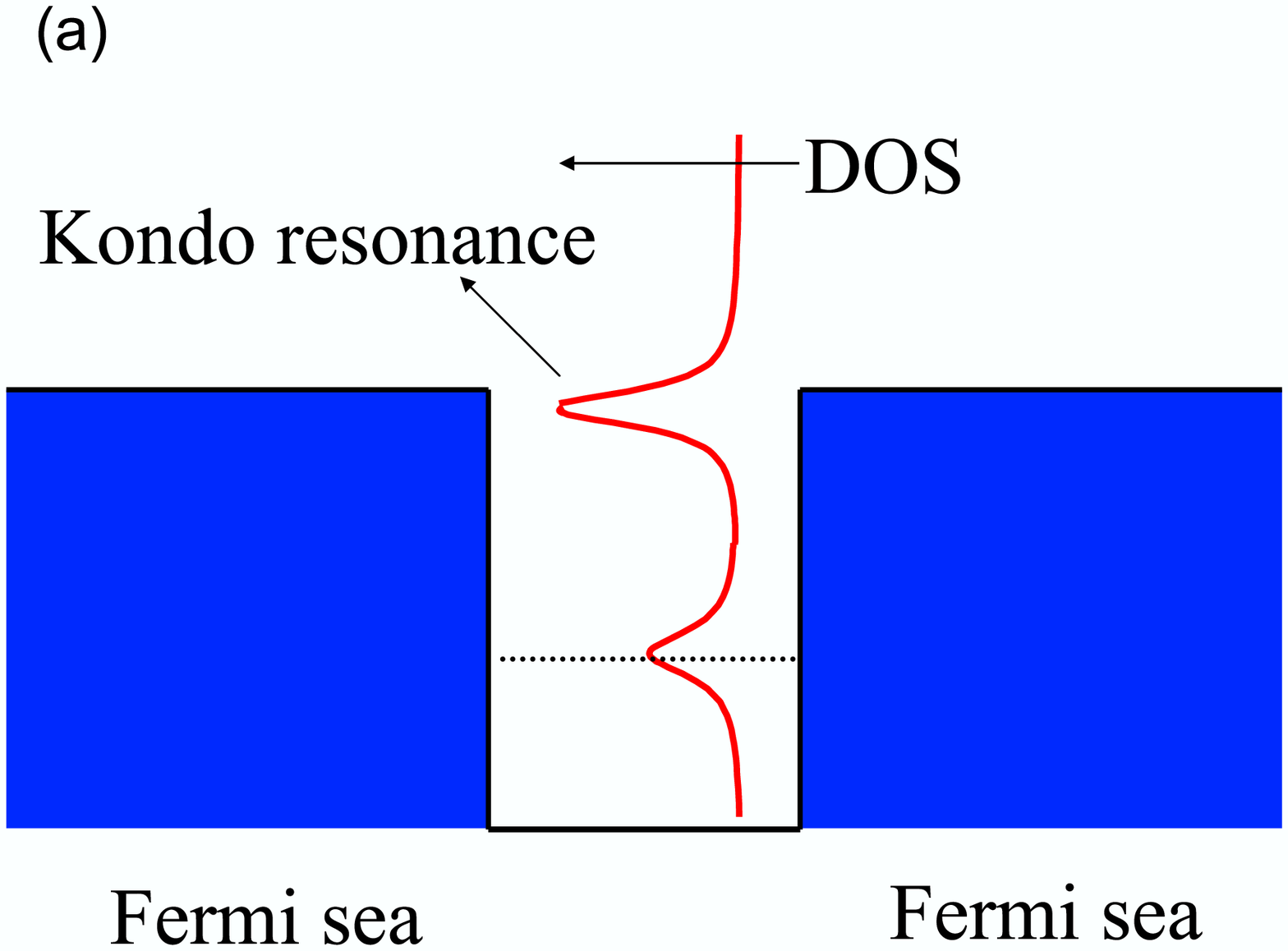}} & \hspace{3 cm}
      \resizebox{68mm}{!}{\includegraphics{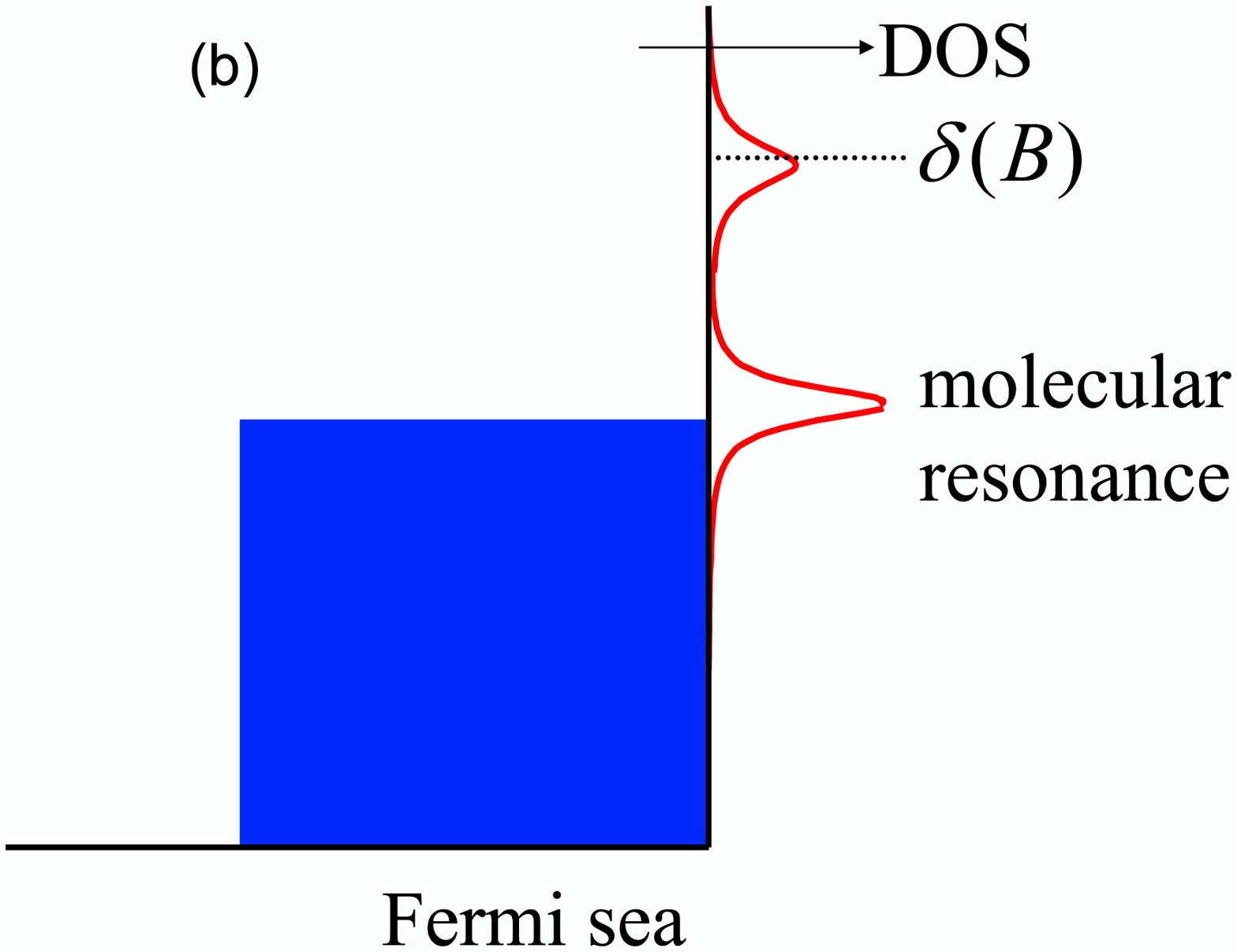}} \\       
    \end{tabular}
    \caption{(a) Kondo resonance in quantum dots;
    (b) Many-body resonance in the molecular density of states
    near the Feshbach resonance}
    \label{dos}
  \end{center}
\end{figure*}

It turns out that in a quantum dot the amplitude of the elastic
co-tunneling process with a singly-occupied level of Fig~\ref{DOS2} (c),
diverges logarithmically when the energy $\hbar\omega\simeq k_{\rm B} T$ of an incoming electron
approaches zero.
The singularity in the transition amplitude gives a dramatic enhancement of
the conductance through the dot. 
In the elastic co-tunneling process (c) the electron on the dot is
quickly replaced by another
electron,
when the electron on the dot tunnels to
one of the leads. 
The characteristic time scale for such a co-tunneling
is about $\hbar/E_C$. Events of type (c) also effectively
flip the spin on the dot. Successive spin-flip processes
screen the local spin on the dot
until a spin singlet is formed by the electrons 
in the leads and on the dot.
Macroscopically, the system enters in a many-body correlated quantum state.
The formation of this entangled state, which is a pure many-body effect, 
represents the Kondo effect in quantum dots \cite{Gordon98,Cronen98}. 
If we interprete the tunneling as an effective 
magnetic-exchange coupling, 
the physics
of a quantum dot between two leads becomes analogous to the 
original Kondo effect \cite{Hewson93}
for magnetic impurities coupled to conduction electrons in
a metal host.
As shown in  
Fig.~\ref{dos}(a),
the Kondo effect is signaled by
a narrow resonance at the Fermi energy in the density of
states of the dot. 

Returning back to our model, we expect something similar
to happen when  
the molecular level 
is located at
somewhat more than
twice the Fermi energy above the threshold
of the atomic continuum. 
Differently than in 
quantum dots, the molecular state has to lie above twice the Fermi energy of the atoms,
because otherwise the ground state of the gas 
would contain
a Bose-Einstein condensate of molecules.
\begin{figure}[h]
 \begin{center}
 \includegraphics[width=0.8 \columnwidth]{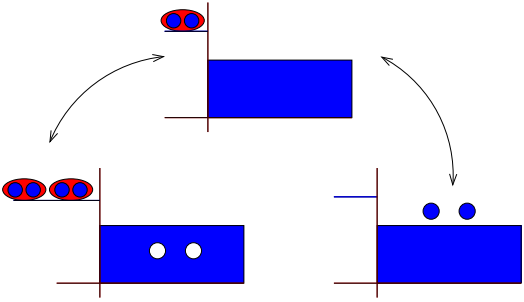}
 \caption{ Virtual tunneling of pairs of atoms 
 to the molecular level in the Fermi gas near a Feshbach resonance.}
 \label{VarGrState}
 \end{center}
\end{figure} 
The atom-molecule coupling produces virtual transitions
between the molecules and
two atoms above and two holes below the Fermi sea,
as it is shown in Fig.~\ref{VarGrState}. In this case the virtual co-tunneling
of pairs of atoms to the molecular level is indeed reminiscent 
of the inelastic co-tunneling of Fig.~\ref{DOS2} (c).
However, in our case, no Coulomb blockade is required.
The Fermi sea removes the symmetry between 
scattering to and from the virtual molecular level,
because atoms can tunnel only in pairs.
This 
leads to the logarithmic singularity in the selfenergy in Eq.~(\ref{eq:MBSelfenergy1})
as mentioned above. The onset of a singularity 
in the selfenergy signales the formation of a new many-body collective state.
Taking the lowest order precesses shown in Fig.~\ref{VarGrState} this many-body resonant 
state can be written as a linear superposition
\begin{eqnarray}
\label{eq:ansatz}
|\Psi_{\rm K}\rangle\simeq\sqrt{Z} \left[b_{\bf 0}^{\dagger}|\Phi_{\rm
F}\rangle+\sum_{|\bk|> k_{\rm F}} \alpha_{\bf k}a_{\bf
k,\uparrow}^{\dagger}a_{\bf
-k,\downarrow}^{\dagger}|\Phi_F\rangle+\right.\nonumber\\
\left.\sum_{|\bk|< k_{\rm F}}\beta_{\bf k}b_{\bf 0}^{\dagger}b_{\bf
0}^{\dagger} a_{\bf k,\uparrow}a_{-\bf k,\downarrow}|\Phi_{\rm F}
\rangle\right],
\end{eqnarray} 
where $|\Phi_{\rm F}\rangle$ indicates the filled Fermi sea
and the molecular field is treated in the zero-mode approximation that 
considers only the lowest molecular state. This is justified at very low temperatures.
In the formation of the many-body coherent state the system gains an energy
\begin{equation}
\label{eq:shiftzero}
\Delta E_K =
\langle\Psi_{\rm K}|H|\Psi_{\rm K}\rangle
-E_G, 
\end{equation}
where $E_G=2\sum_{|\bk|< k_{\rm F}}(\epsilon_{\bk}-\mu)$
is the ground state of the Fermi mixture in absence of Feshbach resonace. This 
shift can be calculated variationally \cite{Mahan90,Gunnarson83} as follows. 
First we evaluate $H|\Psi_{\rm K}\rangle$
considering the Hamiltonian
given in Eq.~(\ref{eq:Ham1}). Then we project this state on the three different components $|\Psi_{\rm K,j}\rangle$
of the ground-state ansatz of Eq.~(\ref{eq:ansatz}). As a result Eq.~(\ref{eq:shiftzero}) is turned in three
equations for the variables $\Delta E$, $\alpha_{\bf k}$, $\beta_{\bf k}$. The variable $Z$ cancels out because
it is the normalization constant. Eliminating $\alpha_{\bf k}$ and $\beta_{\bf k}$ we obtain 
at $T=0$ a self-consistent
equation for $\Delta E$. This is 
\begin{align}
\label{eq:nonpertshift}
\Delta&E_K+2\mu-\delta(B) \simeq\\&\eta \frac{4\sqrt{2\mu}}{\pi}+\eta \frac{2}{\pi}
\sqrt{\Delta E_K+2\mu} \ln\frac{\sqrt{\Delta E_K+2\mu}-\sqrt{2\mu}}{\sqrt{\Delta E_K+2\mu}+\sqrt{2\mu}},\nonumber
\end{align}
where we have used that $\mu=\epsilon_F=\hbar^2 k_F^2/2m$.
At large detunings $\delta\gg2\mu$ this equation has only one solution at $ \Delta E_K\simeq\delta(B)$
as it is expected when resonance effects can be neglected.
However, when
$\delta\gtrsim2\mu$,
we have three solutions.
The lowest energy solution defines the ground state of the systems and occurs 
when $\Delta E_K\simeq 0$.  
The logarithm in the last term 
diverges when $\Delta E_K\rightarrow 0$. Therefore, we have
that $2\mu\gg\Delta E_K$, $|\ln\Delta E_K|\gg\Delta E_K$ and 
we can approximate Eq.~(\ref{eq:nonpertshift}) as
\begin{align}
\label{eq:nonpertshift1}
2\mu-\delta(B) \simeq\eta \frac{2\sqrt{2\mu}}{\pi}
\left[2+\ln\left(\frac{|\Delta E_K|}{8\mu}\right)\right].
\end{align}
This leads to a non-perturbative expression for the energy shift of the 
ground state given by
\begin{equation}
\label{eq:shift}
|\Delta E_K| \simeq\frac{8\mu}{e^2}e^{\frac{\pi}{2}
\frac{\sqrt{2\mu}}{\eta}}e^{-\frac{\pi}{2}\frac{\delta(B)}{\eta\sqrt{2\mu}}}.
\end{equation}
The change in the ground state energy defines a temperature, 
i.e.,
the analogous of the Kondo temperature. 

\begin{figure*}
  \begin{center}
    \begin{tabular}{cc}
      \resizebox{68mm}{!}{\includegraphics{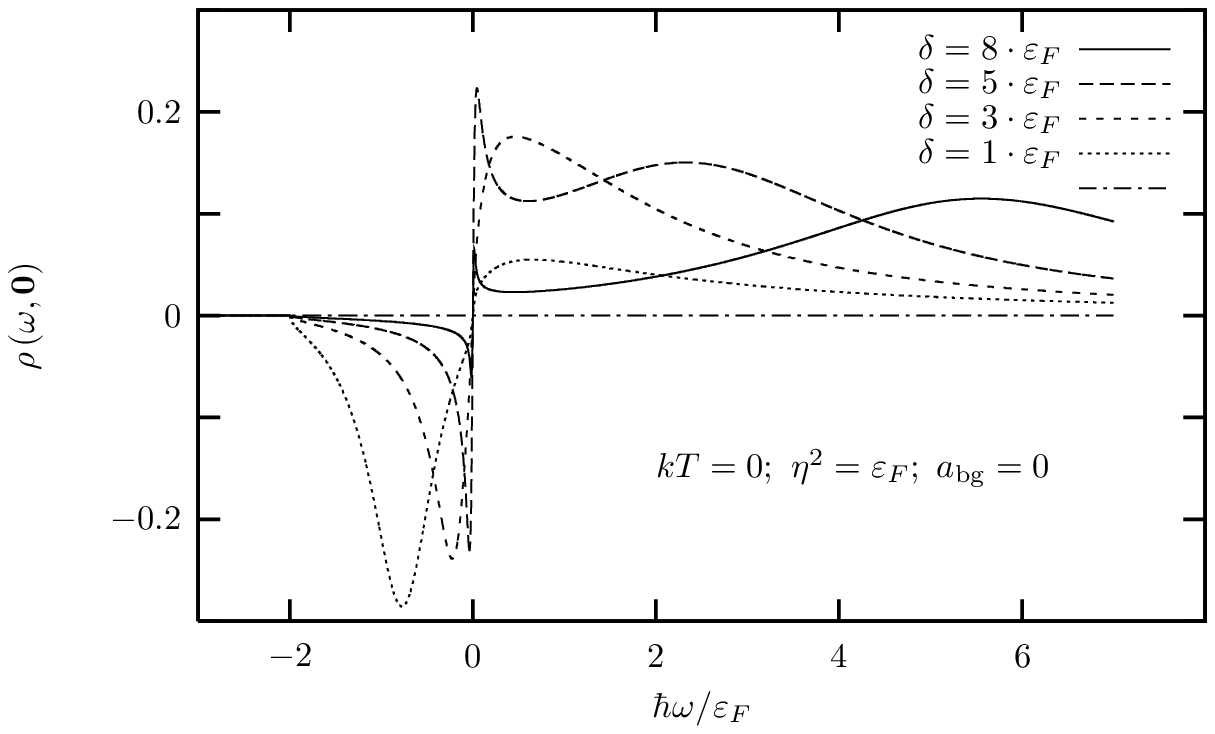}} &
   \hspace{2 cm}   \resizebox{70mm}{!}{\includegraphics{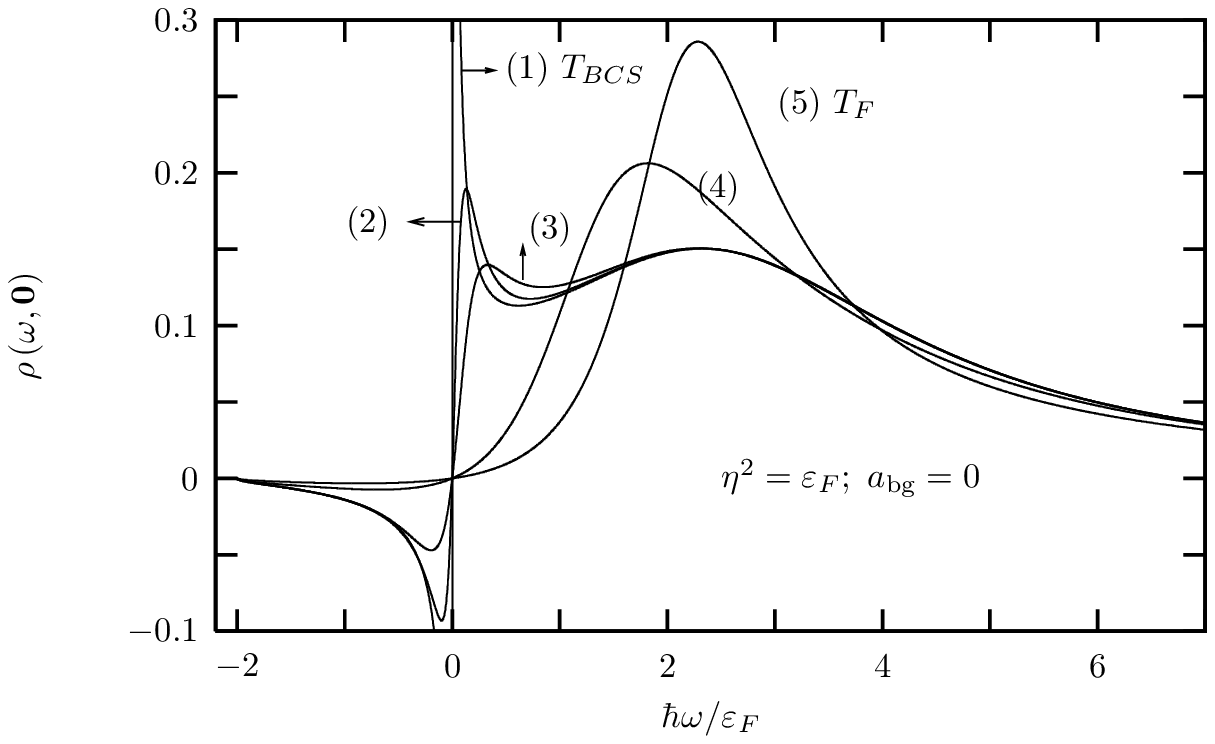}} \\
    \end{tabular}
    \caption{Molecular spectral density (left) calculated 
    from the selfenergy in Eq.~(\ref{eq:MBSelfenergy3}) at different detunings.    
   Molecular spectral density (right) calculated 
    from the selfenergy in Eq.~(\ref{eq:MBSelfenergy1}) at $\delta=5\epsilon_F$
    for different temperatures: (1) $T= T_F$, (2) $ T=0.5  T_F$,
    (3) $ T=0.1 T_F$, (4) $ T=0.06 T_F$, (5) $ T_{BCS}\simeq 0.02 k T_F$.}
    \label{rho1}
  \end{center}
\end{figure*}

\begin{figure*}
  \begin{center}
    \begin{tabular}{cc}
      \resizebox{68mm}{!}{\includegraphics{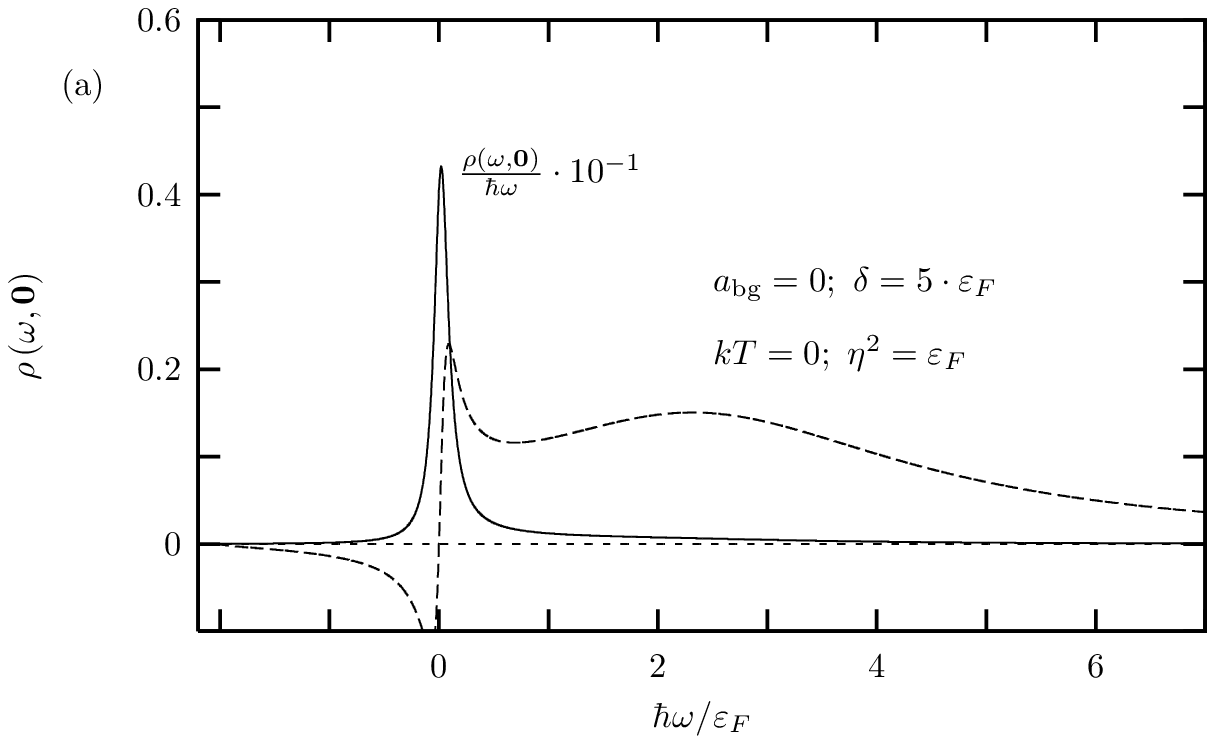}} & \hspace{2 cm}
      \resizebox{70mm}{!}{\includegraphics{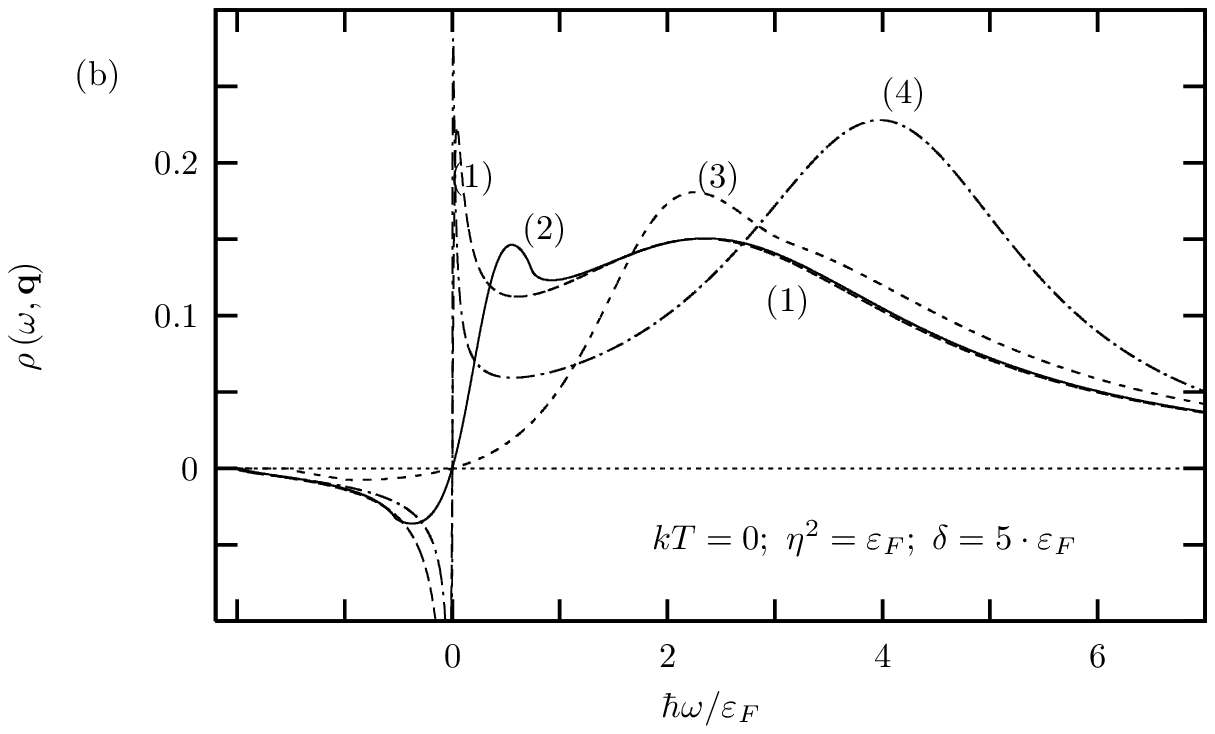}} \\
    \end{tabular}
    \caption{(a) The spectral density is plotted
    together with $\rho(\omega)/\omega$. The double structure
    of the molecular resonance merges then into a single Kondo-like resonance.
    (b) Effects on the molecular spectral density,
    calculated from the selfenergy in Eq.~(\ref{eq:MBSelfenergy3}),
    of the kinetic 
    energy of the molecules: (1) $q=0$, (2) $q\sim 0.3\cdot k_F$
    and (3) $q=k_F$. The curve (4) 
    is calculated from the selfenergy~(\ref{eq:MBSelfenergy2})
    using $q=0,~ a_{\rm{bg}}\sim 10^3 a_0,~
     \epsilon_F \sim 3.6 \mu K$. It
    shows the corrections
    given on the density of states by a large background scattering length.    
    }

    \label{rho2}
  \end{center} 
\end{figure*}

The singular behaviour of the selfenergy is associated with
the occurence of a resonance in the spectral density $\rho(\mathbf{q},\omega)$
of the molecular bosons
given by
\begin{equation}
\rho_{\rm m}({\bf
q},\omega)=-{\rm Im}[G_{\rm m}^{(+)}({\bf
q},\omega)]/\pi\hbar.
\end{equation}  
This feature is clearly illustrated in Fig.~\ref{dos}(b).
Fig.~\ref{rho1}~ shows the function  $\rho_{\rm m}(\mathbf{q},\omega)$
for different detunings and momenta $\mathbf{q}$
both at nonzero and zero temperature.
At large positive detuning, the spectral function shows just a single
broad peak centered around the detuning. This is the expected
situation for a single molecular state with a finite lifetime.
As the detuning gets closer to twice the Fermi energy,
the spectral density shows two other sharp peaks slightly above
and below the zero frequency. When approaching the resonance, 
the spectral density of the molecular field at low momenta and temperatures
does no longer satisfy the unitarity sum rule \cite{Farid01}
\begin{equation}
\int_{-\infty} ^{\infty}
d(\hbar \omega)\,\,\rho_{\rm m}({\bf q},\omega)=1, 
\label{eq:sumrule}
\end{equation} 
despite the fact that the selfenergy in Eq.~(\ref{eq:MBSelfenergy1}) is analytic
in the upper-half complex plane.
This happens already above the BCS critical temperature
and is due to the appearance of 
a complex pole in the upper half plane in the propagator. 
This feature is 
an artifact of the weak-coupling approximation, which 
neglects self-consistency effects on the Fermi propagators
in the calculation of the molecular selfenergy when the gas enters
the strong-coupling regime $k_F a(B)<-1$.
For fermions, Luttinger's theorem \cite{Luttinger58} shows that the causality
of the selfenergy is a sufficient condition to have also a causal propagator.
We are not aware of an analogous theorem in the case of bosons 
and our findings indeed present a counter example \cite{Coleman05}.
The double-peak feature 
arises because the spectral density must be negative at negative frequencies
for a bosonic field.
By plotting
the quantity $\rho_{\rm m}(\omega)/\omega$ \cite{HewsonComm},
as shown in Fig.~\ref{rho2}(a),
a single peak at zero frequency is obtained.
We call this a molecular Kondo resonance.
The effects of nonzero
momentum and of corrections due to a nonzero background scattering lenght $a_{\rm bg}$ 
are shown in Fig.~\ref{rho2}(b). Both tend to 
suppress the resonance.
Nonzero momentum corrections spoil the coherence of the many-body virtual 
cotunneling events similarly as BCS pairing is suppressed at nonzero
total momentum. 
A large background scattering length means that there is a bound state
in the open channel potential \cite{VanKempen04}. This induces some density of states
at large energy. 
Because the total density of states is normalized to $1$, this results
in an effective reduction of the resonance peak.

The additional spectral weight at low frequencies,
induced by the above mentioned many-body effects,
leads to an increase of the bare molecular component
in the gas, calculated by means of 
\begin{equation}
\label{eq:baremolnum}
n_{\rm B}(T)=\int_{-\infty}^{+\infty}\,d(\hbar\omega)\int \frac{{\rm d}
{\bf q}}{(2\pi)^3}\, \rho_{\rm m}({\bf q}
,\omega)N\left(\frac{\hbar\omega}{k_{\rm B} T}\right).
\end{equation} 
with respect to that estimated from two-body physics. 
The 
enhancement of the bare molecular components
for realistic resonance parameters 
is discussed in detail in Ref. \cite{Falco04b}.  

\subsection{Exact mapping to the anisotropic Kondo Hamiltonian.}
\label{mapping}
The analogy with the Kondo effect in a quantum 
dot, introduced in the previous section,
requires a more formal analysis.
It is very desirable to search
for an exact mapping from the atom-molecule Hamiltonian 
to the Hamiltonian of the Kondo problem. 

The physics of the Kondo effect is described by a model Hamiltonian based
on the assumption that the magnetic moment of 
a local impurity is coupled
via an antiferromagnetic exchange interaction $J$ with the conduction
electrons. This is known as the Kondo Hamiltonian and reads
\begin{align}
\label{eq:Kondomodel}
 {H}=\sum_{\textbf{k},\sigma}&
 \left(\epsilon_{\textbf{k}}\!-\!\mu\right)c^{+}_{\textbf{k},\sigma}c_{\textbf{k},\sigma}
+\sum_{\textbf{k},\textbf{k}'}\left[
J_{+}S^{+}c^{+}_{\textbf{k},\downarrow}c_{\textbf{k}',\uparrow}
\right.\\&\left.+J_{-}S^{-}c^{+}_{\textbf{k},\uparrow}c_{\textbf{k}',\downarrow} +
J_{z}S_{z}(c^{+}_{\textbf{k},\uparrow}c_{\textbf{k}',\uparrow}\!-\!
c^{+}_{\textbf{k},\downarrow}c_{\textbf{k}',\downarrow})\right]\nonumber, 
\end{align}
where $\epsilon_{\textbf{k}}$ is the energy of the conduction
electrons and 
$S_{z}$ and $S^{\pm}(\equiv S_{x}\pm i S_{y})$ are the spin
operators for the impurity with spin $1/2$. 
The operators
$c_{\textbf{k},\alpha}$ and $c^{+}_{\textbf{k},\alpha}$
correspond to the creation and annihilation operators of
conduction electrons with momenta $\textbf{k}$ and one of the
possible spin states $\alpha=|\uparrow\rangle$, or $|\alpha=\downarrow\rangle$ which scatters
of the impurity. The coupling constants $J_{\pm}$
and $J_{z}$ describe spin-flip
scattering. 

The Hamiltonian  of our model in Eq.~(\ref{eq:Ham}) appears
rather different than the Hamiltonian in Eq.~(\ref{eq:Kondomodel}),
but it becomes
more similar when a restricted many-body Fock space is considered \cite{Tsvelik04}.
For our present purpose, we can in first instance
consider only the lowest-energy molecular state \cite{Barankov04}. 
We can also neglect the effects of the background interaction.
Within these approximations,
the atom-molecule Hamiltonian in Eq.~(\ref{eq:Ham})
becomes
\begin{align}
\label{eq:Ham1}
{ H}=&\!\!\!\sum_{\mathbf{k},\sigma 
\in \{\uparrow,\downarrow\}}\!\!\! 
\left(\epsilon_{\mathbf{k}}-\mu\right)a^{\dagger}_{\mathbf{k},\sigma} a_{\mathbf{k},\sigma}+
(\delta(B)-2\mu)
 b^{\dagger}_{0}
b_{0}+
\nonumber\\
&+\frac{1}{\sqrt{V}}
\sum_{\mathbf{k}}
(g^*~ b^{\dagger}_{0}
a_{{\mathbf k},\downarrow}a_{-{\mathbf k},\uparrow}+{\rm h.c.})~.
\end{align}
The bosonic number operator $ b^{\dagger}_{0}
b_{0}$ acts on the 
occupied (unoccupied) lowest energy molecular state $|1\rangle$ ($|0\rangle$) according to
$b^{\dagger}_{0}b_{0}|1\rangle = 1|1\rangle$
and
$b^{\dagger}_{0}b_{0}|0 \rangle =0$.
Hence, in this reduced Hilbert space, we can formally identify the molecular operators
$b^{\dagger}_{0}$ and $b_{0}$ with spin $1/2$ raising and lowering operators
$S^+=S^{x}+ iS^{y}$ and
$S^-=S^{x}- iS^{y}$,
which satisfy
analogous relations
\begin{align}
&S^{+}S^{-}|+1/2 \rangle=\left(\frac{1}{2}+S^{z}\right)|+1/2 \rangle=1|+1/2\rangle \\
&S^{+}S^{-}|-1/2 \rangle=\left(\frac{1}{2}+S^{z}\right)|-1/2 \rangle = 0
\end{align}
in the two-dimensional Hilbert space $\{|+1/2 \rangle,|-1/2 \rangle\}$.

More progress can be made by rewriting the atom-molecular coupling 
in terms of the symmetrized atom-pair operators
${d}_{\textbf{k}}=a_{-\textbf{k},\downarrow}a_{\textbf{k},\uparrow}+
a_{\textbf{k},\downarrow}a_{-\textbf{k},\uparrow}/\sqrt{2}$ and
$d_{\textbf{k}}^{+}=a^{+}_{\textbf{k},\uparrow}a^{+}_{-\textbf{k},\downarrow}+
 a^{+}_{-\textbf{k},\uparrow}a^{+}_{\textbf{k},\downarrow}/\sqrt{2}$,
which satisfy the commutation relation
\begin{align}
\label{eq:commut}
[d_{\textbf{k}}^{+} ,
d_{\textbf{k}}]=
\frac{1}{2}\;\left(+N_{\textbf{k},\uparrow}+N_{-\textbf{k},\downarrow}+N_{-\textbf{k},\uparrow}
+ N_{\textbf{k},\downarrow}-2\right).
\end{align}
In the restricted two-dimensional Hilbert space, 
defined by the condition that the
eigenvalues of 
$N_{\textbf{k},\uparrow}+N_{-\textbf{k},\downarrow}$
and $N_{-\textbf{k},\uparrow} +
N_{\textbf{k},\downarrow}$ are either $2$ or $0$,
the commutator in Eq.~(\ref{eq:commut}) can be put \cite{Anderson58} into the matrix form 
$[d_{\textbf{k}}^{+},d_{\textbf{k}}]=\sigma^{z}_{\textbf{k}}$,
by means of the $z$-component Pauli matrix 
\begin{eqnarray}
\left( 
\begin{array}{cc}
1&0\\
0&-1\end{array}\right)
&=& \sigma^{z}_{\textbf{k}}.
\label{eq:Pauliz}
\end{eqnarray}
\begin{figure}
\begin{center}
\resizebox{40mm}{!}{\includegraphics{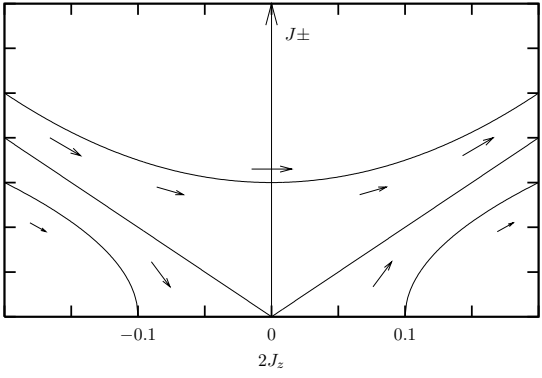}}
\caption{\footnotesize{Scaling trajectories 
for the anisotropic
Kondo model in Eq.~(\ref{eq:Kondomodel}) \cite{Anderson70}.    }}
 \label{J_flow}
 \end{center}
\end{figure}

\noindent This suggests the correspondence
${d}_{\textbf{k}}\rightarrow\sigma^{+}_{\textbf{k}}
\equiv\frac{1}{2}(\sigma^{x}_{\textbf{k}}+i\sigma^{y}_{\textbf{k}})$ and
$d_{\textbf{k}}^{+}\rightarrow
\sigma^{-}_{\textbf{k}}\equiv\frac{1}{2}(\sigma^{x}_{\textbf{k}}-i\sigma^{y}_{\textbf{k}})$.
Finally, by rewriting the spin ladder operators $\sigma^{\pm}_{\textbf{k}}$,
by means of the Abrikosov representation \cite{Abrikosov65}, in terms of anticommuting fermionic operators
$\sigma_{\textbf{k}}^{+}=c^{+}_{\textbf{k},\uparrow}c_{\textbf{k},\downarrow}$ and
$\sigma_{\textbf{k}}^{-}=c^{+}_{\textbf{k},\downarrow}c_{ \textbf{k},\uparrow}$
the atom-molecule Hamiltonian in Eq.~(\ref{eq:Ham}) can finally be transformed into 
\begin{align}
\label{eq:effKon}
{H}&=\sum_{\textbf{k},\alpha}\left(
\epsilon_{\textbf{k}}-\mu\right)c^{+}_{\textbf{k},\alpha}c_{\textbf{k},\alpha}+\left(\delta
-2\mu\right)
S_{z}\\&+\frac{g}{\sqrt{2V}}\sum_{\textbf{k}}[S^{+}c^{+}_{\textbf{k},\uparrow}c_{\textbf{k},\downarrow}
+S^{-}c^{+}_{\textbf{k},\downarrow}c_{\textbf{k},\uparrow}]\nonumber.
\end{align}
When $\delta=2\mu$,
this Hamiltonian is equivalent of the anisotropic Kondo Hamiltonian
in Eq.~(\ref{eq:Kondomodel})
if we state the correspondence $ g/\sqrt{2}=J_{+}=J_{-}$ and $J_{z}=0$.
At finite detuning
but larger than $2\mu$, the term $\left(\delta-2\mu\right) S_{z}$ 
represents an external magnetic
field applied on the localized spin, which removes its degeneracy
and ultimately destroys the Kondo effect.   
Moreover, it is important to realize that,
the lack of the nonflipping spin-interaction term, signaled by
the condition  $J_{z}=0$, does not 
rule out the possibility that such a term is generated by higher-order
processes.
As illustrated in the example of Fig.~\ref{J_flow}, 
the renormalization group flow   
in the solution of the Kondo
problem, can generate nonzero values
of $J_{z}$ starting from a point with $ J_{z}= 0 $ \cite{Anderson70}.
A renormalization group analysis of the coupling constants flow reveals that this 
occurs also in Eq.~(\ref{eq:effKon}).

\subsection{Fermi liquid description of the normal state in the weak coupling regime.}
\label{sec:Fermi liquid}
In absence of the resonance,  
the ultracold dilute fermionic mixture weakly-interacting in the 
background scattering channel above some certain critical temperature $T_c$,
is a Fermi liquid \cite{Migdal58}.
In this section, we consider the weak coupling regime $1/k_F a(B)\leq -1$
in presence of the Feshbach resonance from the point of view of the 
Fermi liquid theory \cite{Bruun03}. 
The thermodynamics of a normal-fluid Fermi liquid is determined by the fermionic Green's
function
\begin{eqnarray}
\label{eq:fermionicGreenFunct}
G_{\sigma}(\textbf{k},\omega_n)=\frac{\hbar}
{i\hbar\omega_n-(\epsilon_{\textbf{k}}-\mu)-\hbar\Sigma^f_{\sigma}(\textbf{k},\omega_n)}.
\end{eqnarray}
In a self-consistent approach, the fermionic 
selfenergy $\hbar\Sigma^f_{\sigma}(\textbf{k},\omega_n)$ is calculated 
in the ladder approximation from 
\begin{align}
\label{eq:fermionicSelfEnrNormst}
\hbar\Sigma^f_{\sigma}(\textbf{k},\omega_n)&=
-\frac{1}{(2\pi)^3}\int d^3\textbf{k}'\frac{1}{\beta\hbar}\sum_{\omega_{n'}}\\& 
\times\rm{T}^{\rm MB}(\textbf{k}+\textbf{k}',\omega_n+\omega_n')
~G_{-\sigma}(\textbf{k}',\omega_n')\nonumber,
\end{align}
where the many-body $\rm{T}$-matrix of Eq.~(\ref{eq:theo}) is determined   
by Eqs.~(\ref{eq:MBprop}), (\ref{eq:MBSelfenergy}) and (\ref{eq:Randeria}). 
From the selfenergy $\hbar\Sigma^f_{\sigma}(\textbf{k},\omega_n)$ we
can calculate the atomic spectrum given by the poles of the fermionic propagator
in Eq.~(\ref{eq:fermionicGreenFunct}) \cite{Migdal58,Galitskii57} according
to the equation
\begin{align}
\label{eq:fermionicSpectrum}
\omega_{\textbf{k}}=\frac{\hbar k^2}{2 m}+
\hbar\Sigma^f(\textbf{k},\omega_{\textbf{k}})=\epsilon_{\textbf{k}}+\hbar\Sigma^f(\textbf{k},\omega_{\textbf{k}}),
\end{align}
where we have suppressed the spin indices
because the two-species have the same mass and chemical potential. 
This equation can be simplified in the weak coupling limit as
\begin{align}
\label{eq:fermionicSpectrumB}
\omega_{\textbf{k}}\simeq\epsilon_{\textbf{k}}+\hbar\Sigma^f(\textbf{k},\epsilon_{\textbf{k}}),
\end{align}
because we expect 
a quasiparticle spectrum 
for the single particle excitations of the type $\omega_{\textbf{k}}=
\epsilon_{\textbf{k}}[1+O(k_F a(B))]$ and the selfenergy $\hbar\Sigma^f(\textbf{k},\omega_{\textbf{k}})$
to be of the order of $O(k_F a(B))$.
The real part ${\rm Re}[\hbar\Sigma^f(\textbf{k},\epsilon_{\textbf{k}})]$
gives a shift in the single particle energies of momentum $\textbf{k}$,
while the immaginary part ${\rm Im}[\omega_{\textbf{k}}]\equiv\hbar\gamma_{\textbf{k}}$
leads to a finite lifetime proportional to $1/\gamma_{\textbf{k}}$.
Since the lifetime becomes infinite at 
at $k_F$, the chemical potential is defined by
\begin{align}
\label{eq:fermionicMu}
\mu\equiv\omega_{k_F}=\epsilon_{k_F}+\hbar\Sigma^f(k_F,\epsilon_{k_F}).
\end{align}
Close to the Fermi surface the energy spectrum can be expanded into a Taylor
series
\begin{align}
\label{eq:quasipSpectrum}
{\rm Re}[\omega_{\textbf{k}}]=\epsilon_{k_F}+\frac{\hbar^2k_F}{m^*}(k-k_F)+O(k-k_F)^2,
\end{align}
which defines the effective mass
\begin{align}
\label{eq:effmass}
m^*=\hbar^2 k_F {\left(\frac{\partial {\rm Re}[\omega_{\textbf{k}}]}{\partial k}_{|k=k_F}\right)}^{-1}.
\end{align}
The effective mass determines the properties of the system in the zero-temperature limit.

In presence of the resonance it is
important to distinguish  \cite{Bruun03} between the weak-coupling regime
near a broad and a narrow resonance. 
For a broad resonance the condition $\eta^2\gg \epsilon_F$ implies that
in the weak-coupling regime $1/k_F a(B)\leq -1$ we have $\delta(B)\gg\epsilon_F$. Therefore,
we can neglect the retardation effects on the
atom-atom interaction induced by the dynamics of the bare molecular boson.
According to the discussion of Sec.~\ref{sec:many-body-broad},
near a broad resonance,
the many-body $\rm{T}$-matrix of Eq.~(\ref{eq:theo}) can be approximated
by the single-channel expression of Eq.~(\ref{eq:apprTMBren}).
Then the atomic selfenergy becomes 
\begin{widetext}
\begin{align}
\label{eq:fermionicSelfEnrNormstApprB}
\hbar\Sigma^f(\textbf{k},\omega_n)
=
-\frac{1}{(2\pi)^3}\int d^3\textbf{k}'\frac{1}{\beta\hbar}\sum_{\omega_{n'}}
\frac{4\pi\hbar^2 }{m}
\left[\frac{1}{a(B)}-\frac{4\pi\hbar^2 }{m}
\frac{1}{V} {\sum_{\mathbf{p}}} \left[
\frac{1-N_{(\mathbf{k}+\mathbf{k}')/2+\mathbf{p}}-
N_{\mathbf{(\mathbf{k}+\mathbf{k}')}/2-\mathbf{p}}}{i\omega_n+2\mu-
2\epsilon_\mathbf{p}-\epsilon_{\mathbf{(\mathbf{k}+\mathbf{k}')}}/2}+\frac{1}{2 \epsilon_\mathbf{p}}
\right]
\right]^{-1}\!\!\!\!G_{\sigma}(\textbf{k}',\omega_n')
\end{align}
\end{widetext}
In the weak-coupling limit the many-body $\rm{T}$-matrix in Eq.~(\ref{eq:fermionicSelfEnrNormstApprB}) 
can be expanded in powers of $a(B)$ and we obtain 
\begin{align}
\label{eq:fermionicSelfEnrNormstApprC}
\hbar\Sigma^f(\textbf{k},\omega_n)
=\frac{n}{2}\frac{4\pi\hbar^2 a(B)}{m}+\hbar\tilde{\Sigma}^f(\textbf{k},\omega_n),
\end{align}
where $n$ is the total atomic density of the gas and
$\hbar\tilde{\Sigma}^f(\textbf{k},\omega_n)$ 
represents a correction of order $(k_F a(B))^2$.
Calculating the selfenergy $\hbar\tilde{\Sigma}^f(\textbf{k},\omega_n)$
in the limit of zero temperature one finds 
the expression of $\omega_{\mathbf{k}}$, $\gamma_{\mathbf{k}}$ 
and $m^*$ up to order $(k_F a(B))^2$ of a dilute Fermi liquid theory
\cite{Galitskii57,Haussmann94,Heiselberg01,Bruun03,Combescot05}
\begin{align}
\label{eq:Galitskii}
\omega_{k_F}&=\frac{\hbar^2 k_F^2}{2m}\left[
1+\frac{4}{3\pi}(k_F a(B))\right.\nonumber\\
&+\left.\frac{4}{15\pi^2}(11-2{\rm ln}2)(k_F a(B))^2
\right]\nonumber\\
\frac{m^*}{m}&=1+\frac{8}{15\pi^2}(7 {\rm ln}-1)(k_F a(B))^2\nonumber\\
\gamma_{\mathbf{k}}&=-\frac{\hbar k_F^2}{2m}\frac{2}{\pi}\left(\frac{k_F-k}{k_F}\right)^2 
\end{align}
for $|k-k_F|/k_F\ll 1$.

In the case of a narrow resonance the bare dynamics of the molecular propagator 
cannot be neglected in the weak-coupling limit $1/k_F a(B)\leq -1$. Therefore, we 
have to retain the full expression 
of the molecular propagator in the many-body $\rm{T}$-matrix of
Eq.~(\ref{eq:theo}).
Using the formal definition of the effective mass in Eq.~(\ref{eq:effmass})  
and calculating the integral in Eq.~(\ref{eq:fermionicSelfEnrNormst}),
one recovers the additional contribution to the effective mass of Eq.~(\ref{eq:Galitskii})
\begin{align}
\label{eq:deltamasseff}
\frac{\delta m^*}{m}\simeq\frac{g^2}{2~\delta(B)^2} n,
\end{align}
calculated by Bruun and Pethick in Ref.~\cite{Bruun03}.
However, we have seen 
that, in the weak-coupling regime $1/k_F a(B)\leq -1$
on the positive side of the resonance, non-trivial correlations arise
in the molecular degree of freedom. These correlations
induce a resonance in the molecular spectral function. 
As a result the finite-lifetime molecule 
on the positive side of the resonance cannot be 
pictured as a well-defined quasiparticle
as one would naively think in the case of a narrow resonance. 
Therefore, it is reasonable to ask if it is not possible
that the Fermi-liquid description could partially
break down in the case of a narrow resonance in combination
with the onset of the molecular resonance at the Fermi surface.   
This requires a more detailed analysis of the fermionic Green's function
in Eq.~(\ref{eq:fermionicGreenFunct})
and will be addressed elsewhere.

\section{The equation for the critical temperature $T_C$ in the BEC-BCS crossover in resonant atomic Fermi gases.}
\label{tc}

The ability to control experimentally
the strength of the interactions between ultracold atoms,
offers the exciting possibility to study in detail the
crossover between 
the Bose-Einstein condensation (BEC) of diatomic
molecules and
the Bardeen-Cooper-Schrieffer (BCS)
transition of atomic pairs 
\cite{Eagles69,Leggett80,Nozieres85,Timmermans01,SadeMelo93,Ohashi0203,Milstein02,Mackey05}.

\subsection{Thouless criterion.}

According to the Thouless criterion
\cite{Thouless69}, the superfluid phase transition 
in a Fermi system  with attractive interactions,
occurs at the temperature for which 
the many-body $\rm{T}$-matrix in Eq.~(\ref{eq:theo})
develops a pole at $\mathbf{K}=\omega=0$.
At large positive detuning, i.e., in the weak-coupling BCS limit, 
the Thouless criterion yields the 
equation for the determination of the BCS critical temperature
of a Fermi gas with attractive interactions.
However, when approaching the resonance, the
attraction increases and the Fermi surface of the gas gets strongly
renormalized. Hence, the approximation $\mu\simeq\epsilon_F$, valid in 
the weak-coupling BCS limit, is no longer valid. Particle number conservation
gives the desired condition
on the chemical potential of the gas 
under equilibrium conditions
\cite{Eagles69,Leggett80,Nozieres85}.
This leads to
a set of two coupled equations for $T_C$ and $\mu$. 
In the opposite BEC limit 
the role of the two equations is inverted.
The equation of state determines the critical temperature
for the Bose-Einstein condensation of dressed molecules.
The pole of the 
many-body $\rm{T}$-matrix becomes two-body in nature.
This fixes the chemical potential to half the binding energy
of a dressed molecule. 

The equation of state can be determined in different approximations.
The simplest mean-field theory at nonzero temperature, neglects finite-lifetime effects of
the molecules and 
does not reproduce the correct binding energy of the dressed molecules.
The latter problem can be repaired by replacing in the equation of state the bare
energy of the molecular level with its dressed value 
calculated at Sec.~\ref{sec:twobodyphys} in the two-body scattering approximation.
This leads to the modified mean-field description \cite{Falco04a}
discusssed in section \ref{sec:meanfield}. 
However, this mean-field approach still neglects 
finite-lifetime effects of the molecules.
As a result, it exhibits a phase transition between 
the BCS phase and the BEC phase. 
Since we expect, from symmetry principles, a smooth curve for $T_C$ and $\mu$, this must be considered 
to be an artefact of the mean-field theory.
The inclusion of Gaussian fluctuations on top of the mean-field solution, 
according to the Nozieres-Schmitt-Rink approach \cite{Nozieres85,SadeMelo93},
is sufficient to obtain a smooth crossover in the two-channel model
\cite{Ohashi0203,Milstein02,Stajic03}. 
This approximation is described in Sec.~\ref{sec:GaussianF}.
A comparison of the critical temperature as a function of the magnetic field 
for a broad resonance 
reveals that
our modified mean-field approach agrees quantitatively fairly well with the Nozieres-Schmitt-Rink result.
The Gaussian approximation exhibits a maximum in the curve for ${ T}_C$
\cite{Milstein02,Stajic03,GriffinComm}. 
It has been shown that,
in a self-consistent approach, this maximum in the curve for ${T}_C$ disappears both in 
a single-channel \cite{Haussmann94} and in a two channels model \cite{Xia-Ji05}.
In Sec.~\ref{sec:selfcons}, we discuss the self-consistent equations in connection with our dressed-molecule
picture through a systematic analysis of the BEC limit where analytical calculation are possible.

\subsection{Modified mean-field equations.}
\label{sec:meanfield}
In order to study the pole structure of the 
many-body $\rm{T}$-matrix in Eq.~(\ref{eq:theo}),
it is crucial to consider explicitly the many-body corrections
of the atom-molecule coupling constant in the calculation
of the selfenergy of the dressed molecular propagator in Eq.~(\ref{eq:MBSelfenergy}).
However, a more transparent derivation of the equation for the critical
temperature can be obtained by rewriting the 
many-body $\rm{T}$-matrix in Eq.~(\ref{eq:theo}) in terms of the simpler
two-body $\rm{T}$-matrix in Eq.~(\ref{eq:twobody}).
We start by considering the definition of the many-body $\rm{T}$-matrix in Eq.~(\ref{eq:totMBbis})
\begin{align}
{\rm T}^{\rm{MB}}\left(\mathbf{K},E\right)=
{\rm V}_{\rm{eff}}
+
{\rm V}_{\rm{eff}}
\Pi\left(\mathbf{K},E\right)
{\rm T}^{\rm{MB}}\left(\mathbf{K},E\right),
\nonumber
\end{align}
with ${\rm V}_{\rm{eff}}$ given in Eq.~(\ref{eq:effpotbare}).
Then we replace the kernel in Eq.~(\ref{eq:kernel})
\begin{align}
\hbar\Pi\left(\mathbf{K},E\right)=&-
\frac{k_{\rm B} T}{\hbar V}\sum_{\mathbf{k},\omega_n}
G_{\uparrow}
\left(\frac{\mathbf{K}}{2}+\mathbf{k},\frac{E}{2} +\omega_n\right)\nonumber\\
&\times
G_{\downarrow}
\left(\frac{\mathbf{K}}{2}-\mathbf{k},\frac{E}{2}-\omega_n\right)
\end{align}
with the result of Eq.~(\ref{eq:freepartprop})
\begin{align}
\label{eq:kernelTmatr}
\hbar\Pi_{0}\left(\mathbf{K},E\right)=&
-\frac{k_{\rm B} T}{\hbar V}\sum_{\mathbf{k},\omega_n}
G_{0,\uparrow}
\left(\frac{\mathbf{K}}{2}+\mathbf{k},\frac{E}{2} +\omega_n\right)
\nonumber\\&\times G_{0,\downarrow}
\left(\frac{\mathbf{K}}{2}-\mathbf{k},\frac{E}{2}-\omega_n\right)
\\
=&\frac{\hbar}{V}\sum_{\mathbf{k}}\left[
\frac{1-N_{\mathbf{K}/2+\mathbf{k}}-N_{\mathbf{K}/2-\mathbf{k}}}{E +2\mu-
2\epsilon_{\mathbf{k}}-\epsilon_{\mathbf{K}}/2}\right]\nonumber,
\end{align}
as we have done in the calculation of the molecular selfenergy in Eq.~(\ref{eq:MBSelfenergy}).
Subtracting from the many-body $\rm{T}$-matrix equation in Eq.~(\ref{eq:totMBbis}) the two-body
limit
\begin{eqnarray}
\label{eq:tot2B}
{\rm T}^{\rm{2B}}
\left(E\right)
=
{\rm V}_{\rm{eff}}
+{\rm V}_{\rm{eff}}
\frac{1}{V}\sum_{\mathbf{k}}\left[
\frac{1}{E -
2\epsilon_\mathbf{k}}\right]
{\rm T}^{\rm{2B}}\left(E\right)
\end{eqnarray}
at $E=2\mu$, the many-body $\rm{T}$-matrix can be put into the form
\begin{align}
\label{eq:manybodyTmatrix}
{\rm T}^{\rm{MB}}
&\left(\mathbf{K},E\right)\\&
=
\frac{{\rm T}^{\rm{2B}}
\left(2\mu^+\right)}{1-{\rm T}^{\rm{2B}}\left(2\mu^+\right)
\left[
\Pi_{0}
\left(\mathbf{K},E\right)
-\frac{1}{V}\sum_{\mathbf{k}}\left[
\frac{1}{2\mu^+ -
2\epsilon_\mathbf{k}}\right]\nonumber
\right]},
\end{align}
where the two-body $\rm{T}$-matrix ${\rm T}^{\rm{2B}}\left(E\right)$ 
is known from Eqs.~(\ref{eq:twobody}-\ref{eq:2Bproprenbis}).
At~$\mathbf{K}=\omega=0$ this becomes
\begin{eqnarray}
\label{eq:manybodyTmatrixzeroen}
{\rm T}^{\rm{MB}}
\left(\mathbf{0},0\right)
=
\frac{{\rm T}^{\rm{2B}}
\left(2\mu^+\right)}{1+{\rm T}^{\rm{2B}}\left(2\mu^+\right)
\Xi\left(\mathbf{0},0^+\right)
},
\end{eqnarray}
where 
\begin{eqnarray}
\label{eq:Xi}
\Xi
\left(\mathbf{K},E\right)
=
\frac{1}{V}\sum_{\mathbf{k}}
\left[
\frac{N_{\mathbf{K}/2+\mathbf{k}}+N_{\mathbf{K}/2-\mathbf{k}}}{E +2\mu-
2\epsilon_{\mathbf{k}}-\epsilon_{\mathbf{K}}/2}\right].
\end{eqnarray}
According to the Thouless criterion
the denominator of Eq.~(\ref{eq:manybodyTmatrixzeroen})
vanishes at the critical temperature.
To illustrate more clearly the physics involved we consider first a system when
the background interaction can be neglected, i.e., the limit $a_{\rm{bg}}\rightarrow 0$.
This is the case, for example, in $^{40}{\rm K}$ mixtures where the background
scattering length is small and does not contribute to the pairing
mechanism because is positive.
In that case the condition to have a pole in Eq.~(\ref{eq:manybodyTmatrixzeroen}) 
becomes
\begin{eqnarray}
\label{eq:Thoulsspole}
\delta(B)-2\mu+\eta \sqrt{-2\mu}-g^2\Xi
\left(\mathbf{0},0^+\right)
=0.
\end{eqnarray}
The critical temperature $T_c$ and the chemical potential $\mu$ are
determined at every  magnetic field  in the mean-field approximation 
by solving self-consistently
Eq.~(\ref{eq:Thoulsspole}) together with the equation of state.
At the mean-field level, neglecting the finite 
lifetime of the molecules, the latter is given by
\begin{align}
\label{eq:TcStEq}
n_{\rm tot}=&-\frac{1}{V}\frac{\partial{\Omega}}{\partial{\mu}}\\=-\frac{1}{V}&
\frac{\partial{}}{\partial{\mu}}\left(-
k_B T\sum_{\sigma}{\rm Tr}\left[{\rm ln}({G}_{\sigma,0}^{-1})\right]+
k_B T~{\rm Tr}\left[{\rm ln} (G_{0}^{-1})\right]
\right)\nonumber\\
=\frac{1}{V}&\sum_{\mathbf k}\left[2 N(\epsilon_{\mathbf k}-\mu)
+2 N_B(\epsilon_{\mathbf k}/2+\delta(B)-2\mu)\right]\nonumber
\end{align}
where the free-particle atomic and molecular propagators are
given by $G_{0}/\hbar=(i\hbar\omega-\epsilon_{\mathbf k}/2+2\mu-\delta(B))^{-1}$
and $G_{\sigma,0}/\hbar=(i\hbar\omega-\epsilon_{\mathbf k}+\mu)^{-1}$,
and $N_B$ is the Bose distribution function.
However, to neglect completely selfenergy effects in the molecular
propagator is not consistent. 
This can be repaired in a first 
approximation, by replacing the detuning $\delta$ in the free molecular
propagator occurring in Eq.~(\ref{eq:TcStEq}) with
the two-body curve of Eq.~(\ref{eq:posdetmax})
\begin{align}
\label{eq:posdetmaxB}
\epsilon_{\rm m}
=\frac{1}{3}\left(\delta-\frac{\eta^2}{2}+\sqrt{\frac{\eta^4}{4}
-\eta^2\delta+4\delta^2}\right) 
\end{align}
that we have discussed in Sec.~2.3. This replacement is necessary
because the energy of the dressed molecule in the equation of state
must be the same as the binding energy resulting from the chemical
potential in the BEC limit from the gap equation.  
As we shall see, this is crucial in the case of 
a crossover near a broad resonance
where the binding energy is quadratic in the detuning in a large part of the crossover region.
For a very narrow resonance, when $\eta^2\ll\epsilon_F$, the binding
energy of the dressed molecule can be approximated as $\epsilon_{\rm m}\simeq\delta(B)$
except that on a 
vanishing small 
range of magnetic field close to resonance
and the equation of state in Eq.~(\ref{eq:TcStEq}) turns out to be a good approximation.
However, the resonances used in current experiments on the BEC-BCS crossover 
are broad in nature.  
Therefore, the mean-field equation of state has to be modified to
\begin{align}
\label{eq:TcStEq1}
n=
\frac{1}{V}\sum_{\mathbf k}\left[2 N(\epsilon_{\mathbf k}-\mu)
+2 N_B(\epsilon_{\mathbf k}/2+\epsilon_{\rm m}-2\mu)\right].
\end{align}
Note that the two-body quantity $\epsilon_{\rm m}$ does not account for a
possible many-body shift of the molecular bound-state. However, 
for a broad resonance, 
this seems to be a smaller correction when compared to the two-body 
shift.

At positive detuning, in the BCS limit, the chemical potential is positive
and the many-body term of Eq.~(\ref{eq:Thoulsspole}) is
\begin{eqnarray} 
g^2\Xi\left(\mathbf{0},0^+\right)=g^2 {\rm{Re}}\left[
\Xi\left(\mathbf{0},0^+\right)\right]+\eta \sqrt{-2\mu}.
\end{eqnarray}
Therefore we obtain the gap equation
\begin{eqnarray} 
\label{eq:resGapeq}
\delta(B)-2\mu=g^2{\rm{Re}}\left[\Xi\left(\mathbf{0},0^+\right)\right].
\end{eqnarray}
Note that our derivation is based on the many-body $\rm{T}$-matrix calculated in the ladder
approximation. Therefore, it neglects the Gorkov's corrections \cite{Gorkov60}
to the transition temperature introduced by the effects of the density fluctuations
of the medium on the effective two-body interactions.

It turns out that for a narrow resonance 
$\eta^2\simeq \epsilon_F$, we can be in the weak-coupling regime
$k_F|a(B)|\leq 1$ at magnetic fields such that 
$\delta\simeq 2\mu\simeq \epsilon_F$.
In this case the BCS critical temperature is given by 
\begin{eqnarray} 
\label{eq:resBCSTcnarw}
T_{\rm{c}} \simeq \frac{8\epsilon_F}{k_{\rm B} \pi} e^{\gamma-2}
e^{\frac{2}{\pi}\frac{\sqrt{2\mu}}{\eta}}\ 
             \exp \left\{ -\frac{\pi}{2k_F|a(B)|}  \right\}~.
\end{eqnarray} 
When the resonance is broad and $\eta^2\gg 2\epsilon_F$, 
the detuning is always $\delta\gg 2\mu\simeq \epsilon_F$
in the weak
coupling range $k_F|a(B)|\leq 1$, 
and the 
term $e^{\frac{2}{\pi}\frac{\sqrt{2\mu}}{\eta}}$ can be neglected.
In this limit the BCS critical temperature 
can be approximated by
\begin{eqnarray} 
\label{eq:resBCSTcbrd}
T_{\rm{c}} \simeq \frac{8\epsilon_F}{k_{\rm B} \pi} e^{\gamma-2}\ 
             \exp \left\{ -\frac{\pi}{2k_F|a(B)|}  \right\}~.
\end{eqnarray}
This is the same result that one would obtain  
by looking at the pole of the 
the {\it{single-channel}} many-body $\rm{T}$-matrix given by Eq.~(\ref{eq:apprTMB}).
Therefore, we conclude that the positive shift of the 
weak-coupling BCS critical temperature by the factor $e^{\frac{2}{\pi}\frac{\sqrt{2\mu}}{\eta}}$, 
characteristic of the resonant superfluidity, can be observed only
in narrow resonances.

At large negative detuning, 
the pole of the many-body $\rm{T}$-matrix in Eq.~(\ref{eq:totMBbis})
determines the chemical potential $2\mu$ of the dressed molecule,
i.e., the energy we need to make a dressed molecule from two atoms at zero momenta. 
In the BEC limit this is very large and negative and corresponds to the
binding energy of the dressed molecule.
Indeed the pole of the many-body $\rm{T}$-matrix becomes two-body in nature
because, for negative $\mu$, 
the many-body factor is
\begin{eqnarray} 
g^2\Xi\left(\mathbf{0},0^+\right)=g^2{\rm{Re}}\left[\Xi\left(\mathbf{0},0^+\right)\right]
\end{eqnarray}
and vanishes exponentially as $|2\mu|/k_{\rm B} T\rightarrow \infty$.
The condition~(\ref{eq:Thoulsspole}) becomes
\begin{eqnarray}
\label{eq:twobodymu}
\delta(B)-2\mu+\eta \sqrt{-2\mu}\simeq 0,
\end{eqnarray}
which is the same algebraic equation as in Eq.~(\ref{eq:root}), 
again in the limit of $a_{\rm{bg}}\rightarrow 0$,
satisfied by the binding energy in Eq.~(\ref{eq:Rembbind})
\begin{align}
\label{eq:RembbindB}
\epsilon_{\rm m}
=\delta(B)+\frac{\eta^2}{2}\left(
\sqrt{1-\frac{4\delta(B)}{\eta^2}}-1\right).
\end{align}
Notice that the curve $\epsilon_{\rm{m}}$ of Eq.~(\ref{eq:posdetmaxB}),
that we have replaced in the equation of state
coincides with the binding energy of Eq.~(\ref{eq:RembbindB}) at every negative detuning.

Once more, it is instructive to distinguish between the two 
limits of narrow $\eta^2\ll\epsilon_F$ and broad 
$\epsilon_F\ll\eta^2$ resonances.
On the BEC side of a narrow resonance, when the chemical potential
approaches half the binding energy we have $\mu\simeq\epsilon_{\rm{m}}/2\simeq \delta /2$, and
the dressed molecular wavefunction has only a small
amplitude in the open channel, that is 
$Z\simeq 1$.
This can be checked easily if we assume that the chemical potential
approaches half the molecular binding energy $\mu\simeq\epsilon_{\rm{m}}/2$
approximatively 
at some value $B_{\rm BEC}$ when $k_F |a(B_{\rm BEC})|\simeq 1$. From the definition of
the resonant scattering length in Eq.~(\ref{eq:sclen}),
we have that $|\delta(B_{\rm BEC})|\simeq \eta \sqrt{\epsilon_F}$.
Then we can evaluate with the help of the formula in Eq.~(\ref{eq:RembbindB})
the binding energy
for this value of the magnetic field. This gives 
\begin{align}
\label{eq:idea1}
{\epsilon_{\rm m}}(B_{\rm BEC})
\simeq -\eta \sqrt{\epsilon_F}+\frac{\eta^2}{2}\left(
\sqrt{1+\frac{4\sqrt{\epsilon_F}}{\eta}}-1\right).
\end{align}
Substituting this expression in the definition of $Z$ in Eq.~(\ref{eq:zeta}) 
and expanding the result in the small parameter $\eta^2/\epsilon_F$,
we obtain
\begin{align}
\label{eq:idea2}
Z(B_{\rm BEC})\simeq 1-\frac{1}{2}\left(\frac{\eta^2}{\epsilon_F}\right)^{1/4}\simeq 1.
\end{align}
Therefore, for a narrow resonance the bare component of the dressed molecule
cannot be neglected.
It turns out that also in the so-called universal region,
where $1/k_F|a(B)|\gg 1$,
we have $0<Z<1$ and the thermodynamics is actually not universal.

For a very broad resonance, 
the bare component $Z$ 
in the universal region $1/k_F|a(B)|\gg 1$,
is not strictly zero but it is rather small.
When the
chemical potential
approaches half the binding energy, 
the bare component weight is still $Z\simeq 0$.
Infact, expanding Eq.~(\ref{eq:idea1}) in $\epsilon_F/\eta^2$
we find
\begin{align}
\label{eq:idea3}
Z(B_{\rm BEC})\simeq 2\frac{\epsilon_F}{\eta^2}\ll 1.
\end{align}
This implies that 
at magnetic fields where 
the BEC-BCS crossover takes place,
the dressed molecule is 
almost completely in the open channel component.  Furthermore, the binding 
energy in Eq.~(\ref{eq:RembbindB}) can be approximated like in Eq.~(\ref{eq:bind}) by 
\begin{eqnarray}
\label{eq:bindres}
\epsilon_{\rm{m}}(B)=-\frac{\hbar^2}{m a^2(B)}\propto\delta^2\,
\end{eqnarray} 
which is equivalent to neglecting the term $2\mu$ in Eq.~(\ref{eq:twobodymu}). 
Note that this is the same solution
determined by the pole 
of the the many-body $\rm{T}$-matrix in the 
{\it{single-channel}} approximation of Eq.~(\ref{eq:apprTMB}).
It is also clear now why,
in the case of a broad resonance, it is essential
to replace in the equation of state the detuning with the curve of Eq.~(\ref{eq:posdetmaxB}) 
in order to avoid inconsistency in the BEC limit between the two crossover equations.
In conclusion, in the case of a broad resonance, we can 
roughly speaking distinguish between
two different crossovers. 
At detunings where the many-body crossover 
to a 
Bose-Einstein condensate of
dressed molecules takes place, the 
dressed molecules live essentially in the states of the atomic continuum.
At larger negative detuning, the wavefunction of the dressed molecules
undergoes a second two-body crossover to the bare closed channel 
eigenstate.
The energy scale of this second crossover is $\eta^2$ as mentioned above.

In general, 
as far as the background interaction satisfies
the weak-coupling condition $k_F|a_{\rm{bg}}|< 1$,
the above discussion does not change qualitatively,  
if we extend the analysis 
of the many-body $\rm{T}$-matrix in Eq.~(\ref{eq:manybodyTmatrixzeroen})
without taking the limit $a_{\rm{bg}}=0$.
The gap equation in Eq.~(\ref{eq:resGapeq}) 
in the BCS weak coupling regime modifies to
\begin{eqnarray} 
\label{eq:totGapeq}
1+\left({\rm T}^{\rm{2B}}_{\rm{bg}}(0)+g^2\frac{1}{2\mu-\delta(B)}
\right)\Xi\left(\mathbf{0},0^+\right)\simeq 0.
\end{eqnarray}
In the case of a broad resonance the weak-coupling solution 
is again given by Eq.~(\ref{eq:resBCSTcbrd}) and can now be written as
\begin{eqnarray} 
\label{eq:totBCSTcbrd}
T_{\rm{c}} \simeq \frac{8\epsilon_F}{k_{\rm B} \pi} e^{\gamma-2}\ 
             \exp \left\{ -\frac{\pi}{2k_F
	     \left(|a_{\rm{res}}|+|a_{\rm{bg}}|\right)}  \right\}~.
\end{eqnarray}
At large negative detuning
the pole of the many-body 
$\rm{T}$-matrix~(\ref{eq:manybodyTmatrixzeroen})
reproduces the correct two-body physics of the dressed molecules.
The equation for the chemical potential in Eq.~(\ref{eq:twobodymu})
assumes the general form as in Eq.~(\ref{eq:root})
\begin{eqnarray}
\label{eq:tottwobodymu}
\delta(B)-2\mu+\frac{\eta}{1+|a_{\rm{bg}}|\sqrt{\frac{-2\mu m}{\hbar^{2}}}} 
\sqrt{-2\mu}
\simeq 0.
\end{eqnarray}
Therefore the chemical potential goes to half the binding
energy of the dressed molecules
$\epsilon_{\rm{m}}$ given by the more general solution~(\ref{eq:analroot}). 
The analysis concerning the differencies between the 
broad and narrow resonances remains unchanged,
except that the energy of the second two-body crossover defined above
is now near $\hbar^2/m a_{\rm{bg}}^2$.
This analysis holds in general for the resonances
currently used in experiments. One exception, however, concerns 
the region at large positive detuning of the 
the broad resonance 
at $834~G$ in a gas of of $^6$Li atoms.
In that case the background interactions turn out to be 
so large that $\hbar^2/m a_{\rm{bg}}^2\simeq \epsilon_F$.
In other words, the gas never enters a weak-coupling regime
at positive detuning and the gap equation in Eq.~(\ref{eq:totGapeq})
is not expected to be valid.

\begin{figure}
\vspace{0.4 cm}
\begin{center}
\resizebox{80mm}{!}{ \epsfig{figure=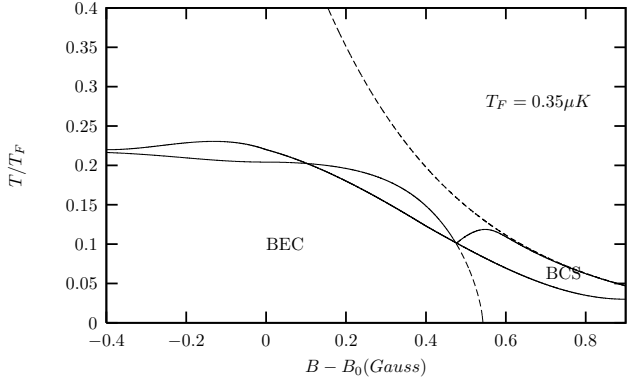}}
 \caption{
Phase diagram
of atomic $^{40}$K 
near the broad resonance near 201.2 G,
as a function of magnetic field and
temperature for a Fermi temperature of the gas of $T_{\rm F} =
0.35~\mu$K. The solid line exhibiting a kink at about
$0.5~G$ gives the critical temperature for
either a Bose-Einstein condensation of dressed molecules or a
Bose-Einstein condensation of atomic Cooper pairs. The critical
temperature for the latter is calculated by simultaneously
solving the BCS gap equation of Eq.~(\ref{eq:Thoulsspole}) and the equation of state of an ideal
mixture of atoms and dressed molecules as in Eq.~(\ref{eq:TcStEq1}). For
comparison, the upper dashed curve is the analytical BCS result of Eq.~(\ref{eq:resBCSTcbrd}). 
The lower dashed line is the crossover between the two Bose-Einstein
condensed phases. The other solid smooth line describes the result of the Nozieres-Schmitt Rink
approximation.}
\label{phasediagramA}
\end{center} 
\end{figure}

Figure~\ref{phasediagramA} shows the mean-field curve for the critical temperature
for a gas of $^{40}$K atoms near the broad resonance near 201.2 G.
The mean-field approach based on Eqs.~(\ref{eq:Thoulsspole}) and~(\ref{eq:TcStEq1})
does not describe a smooth BEC-BCS crossover because it neglects finite lifetime
effects of the molecules. The curve for 
$T_c$ exibits a sharp
phase transition between a Bose-Einstein condensation phase of dressed molecules with binding energy 
$\epsilon_{\rm{m}}$ to a Bardeen-Cooper-Schriffer superfluid.
This phase transition is only an artifact of the mean-field aproximation because the two phases break the same 
$U(1)/Z_2$ symmetry
\cite{Leggett80}. However, the dressed-molecules are the real physical components of the system. Therefore, 
the location of the kink in curve for $T_c$ of Fig.~\ref{phasediagramA} assumes an interesting
interpretation. According to the poor man's approach of Ref. \cite{Falco04a}, it fixes a scale in the magnetic
field below which, the superfluid transition temperature is determined by the 
condensation temperature of the dressed molecules.
Note that this value of the magnetic field
coincides also with the magnetic field value at which the size of the dressed-molecule,
that is equal to $a(B)$,
approaches the interparticle separation $k_F^{-1}$ \cite{Adenkov04}.
It is important to stress that, without the replacement in the equation of state of the detuning
with the true location of the molecular level in presence of coupling described by Eq.~(\ref{eq:posdetmaxB}),
this scale would have been orders of magnitudes smaller.
Interestingly, the location of the crossover line in our mean-field picture 
that is shown in the dashed curve of Fig.~\ref{phasediagramA},
coincides with the threshold for the non zero condensate
fraction in the $\left(T/T_C, \Delta B\right)$ plane 
found by Regal {\it{et al.}} in the experiment of Ref.\cite{Jin04} that we have discussed in Sec.~\ref{atomic}.

\subsection{Theory of Gaussian fluctuations.}
\label{sec:GaussianF} 

A smooth curve for the critical temperature $T_c$ can be obtained by including
finite lifetime effects of the molecular field in the theory. 
This can be achieved by considering the quadratic fluctuations
on top of mean-field, according to the Nozieres-Schmitt Rink approach
\cite{Nozieres85,SadeMelo93,Ohashi0203,Milstein02,Stajic03}.
In the Nozieres-Schmitt Rink approximation, one equation is still determined by 
the Thouless condition expressed by Eq.~(\ref{eq:totGapeq}). 
The equation of state, however, is modified because it has to incorporate the effects
of Gaussian fluctuations around the saddle point. The new equation of state reads
\cite{Ohashi0203}
\begin{align}
\label{eq:TcStEqGausFl}
n=-\frac{1}{V}\frac{\partial{\Omega}}{\partial{\mu}}=
-\frac{\partial{}}{\partial{\mu}}&\left(
-\frac{1}{\beta V}\sum_{\sigma}{\rm Tr}\left[{\rm ln}~(\beta~{G}_{\sigma,0}^{-1})\right]\right.\\&\left.+
\frac{1}{\beta V}~{\rm Tr}\left[{\rm ln}~(\beta~{\rm G}^{-1})\right]
\right),\nonumber
\end{align}
where the full molecular propagator $G$ is given by Eqs.~(\ref{eq:MBprop})
and~(\ref{eq:MBSelfenergy})
and the symbol
${\rm Tr}[~]$ denotes the sum over the Matsubara frequencies and momenta.
This equation describes a gas of atoms and dressed
molecules \cite{note}. 
Using the definition of the full molecular propagator Eq.~(\ref{eq:TcStEqGausFl})
can be rewritten as
\begin{align}
\label{eq:TcStEqGausF2}
&n=\\&
\frac{1}{\hbar \beta V}\sum_{\sigma}{\rm{Tr}}\left[{{G}_{\sigma,0}}\right]
-2\frac{1}{\hbar \beta V}{\rm{Tr}}\left[{{{\rm G}}}\right]+
\frac{1}{\hbar \beta V}{\rm{Tr}}\left[{\rm G}\cdot\frac{\partial{}}{\partial{\mu}}
\hbar \Sigma_{\rm m}\right]\nonumber.
\end{align}
The partition of the different contributions in this equation of state suggests a possible
interpretation of each term on the right-hand-side of Eq.~(\ref{eq:TcStEqGausF2}).
The first term denotes the number of unbound atoms. The second term describes 
the contribution of the bare molecular component while the third one defines
the open-channel component of the dressed molecules. The analysis of the BEC extreme of the crossover
reveals that this 
interpretation is correct in that regime. In the BEC limit the chemical potential
approaches half the binding energy of the dressed molecules, i.e., $\mu\simeq\epsilon_{\rm{m}}/2$.
Therefore, the chemical potential becomes large and negative and the 
contribution of the unbound atoms can be neglected. In first approximation the gas 
constituted mainly of dressed molecules
\begin{align}
\label{eq:TcStEqGausF3}
n\simeq2~n_{\rm m}.
\end{align}
We want to show that in that limit
\begin{align}
\label{eq:TcStEqGausF4}
-2\frac{1}{\hbar \beta V}{\rm{Tr}}\left[{{{\rm G}}}\right]+
\frac{1}{\hbar \beta V}{\rm{Tr}}\left[{\rm G}\cdot\frac{\partial{}}{\partial{\mu}}
\hbar \Sigma_{\rm m}\right]= n_{\rm m}.
\end{align}
Taking the contribution of the pole of the dressed molecular propagator we have
\begin{eqnarray}
\label{eq:TcStEqGausF5}
&-2&\frac{1}{\hbar \beta V}{\rm{Tr}}\left[{{{\rm G}}}\right]+
\frac{1}{\hbar \beta V}{\rm{Tr}}\left[{\rm G}\cdot\frac{\partial{}}{\partial{\mu}}
\hbar \Sigma_{\rm m}\right]=\nonumber\\
&=&\frac{1}{V}\sum_{\mathbf k}\left[2~Z~N\left(\epsilon_{\mathbf k}/2+\epsilon_{\rm m}-2\mu)\right)
\right.\nonumber\\ & &\left.
+ Z~N_B(\epsilon_{\mathbf k}/2+\epsilon_{\rm m}-2\mu)\cdot\frac{\partial{}}{\partial{\mu}}
\hbar {\Sigma_{\rm m}}_{|E=\epsilon_{\rm m}}\right]\nonumber\\
&=&\frac{1}{V}\sum_{\mathbf k}\left[2~Z~N\left(\epsilon_{\mathbf k}/2+\epsilon_{\rm m}-2\mu)\right)
\right.\nonumber\\ & &\left.
+2~(1-Z)~N_B(\epsilon_{\mathbf k}/2+\epsilon_{\rm m}-2\mu)\right]\nonumber\\ \nonumber\\
&=&2~Z~n_{\rm m}+2~(1-Z)~n_{\rm m}=2~n_{\rm m},
\end{eqnarray}
where we have used the definition of the residue $Z$ given in Eq.~(\ref{eq:zeta}), 
and that in the BEC limit we have $\partial{}/\partial{\mu}\simeq2\partial{}/\partial{E}$
because the many-body molecular selfenergy approaches its two-body limit given by 
Eq.~(\ref{eq:2Bproprenbis}). 

In the remaining part of this section we want to show that in the limit of a very 
broad resonance our two-channel model reproduces the results for $T_c$ 
found by Sa De Mel{\'o} {\it{et al.}} in 
Ref. \cite{SadeMelo93} based on a {\it{single-channel}} model \cite{Nozieres85}.
As we have shown in Sec.~\ref{sec:selfen}, 
in the case of a broad resonance, the 
general expression for the pairing field propagator $\hbar{\rm G}_\Delta$ given
in Eq.~(\ref{eq:genRanderia}) reduces to that given in Eq.~(\ref{eq:apprTMBren}).
The latter corresponds exactly to that defined by Sa De Mel{\'o} {\it{et al.}} in 
Ref. \cite{SadeMelo93}. 
This argument is sufficient to proof that our two-channel model is able to reproduce the results
of the single-channel model in the limit of a very broad resonance when $\eta^2\gg \epsilon_F$.
However, 
a detailed analysis of our crossover equations for a broad
resonance, gives us more insight about the two different descriptions.  
When $\eta^2\gg \epsilon_F$, 
the gap equation in Eq.~(\ref{eq:totGapeq}) 
can be approximated with   
\begin{eqnarray} 
\label{eq:totGapeqB}
1+\left({\rm T}^{\rm{2B}}_{\rm{bg}}(0)+g^2\frac{1}{-\delta(B)}
\right)\Xi\left(\mathbf{0},0^+\right)\simeq 0,
\end{eqnarray} 
along the all range of magnetic fields where the BCS-BEC crossover 
takes place. 
This is possible because, as we have explain in Sec.~\ref{sec:selfen},
in the case of a broad resonance we have $|\delta(B)|\gg|2\mu|,\epsilon_{\rm m}$ 
except of course very close to resonance. However,
in that region the gas becomes essentially universal and it can be shown that the
non universal correction
term $2\mu$ can also be neglected \cite{Ho03}. 
By using Eq.~(\ref{eq:sclen})
Eq.~(\ref{eq:totGapeqB}) can be rewritten as 
\begin{eqnarray} 
\label{eq:totGapeqC}
1+\left(\frac{4\pi \hbar^2 a(B)}{m}
\right)\Xi\left(\mathbf{0},0^+\right)\simeq 0,
\end{eqnarray} 
which corresponds to the single-channel gap equation of Sa De Mel{\'o} {\it{et al.}} in 
Ref. \cite{SadeMelo93}.
A similar analysis concerns the equation of state in Eq.~(\ref{eq:TcStEqGausF2}).
In the case of a broad resonance the weight of the bare molecular component 
remains quite small such that $Z\simeq0$ 
even when the chemical potential
approaches half the binding energy $\mu\simeq\epsilon_{\rm m}\simeq \hbar^2/m a^2$.
Therefore we can neglect completely the bare molecular contribution
to the number of particles and the equation of state reduces to
\begin{align}
\label{eq:TcStEqGausF7}
n\simeq
+\frac{1}{\hbar \beta V}\sum_{\sigma}{\rm{Tr}}\left[{{G}_{\sigma,0}}\right]
+
\frac{1}{\hbar \beta V}{\rm{Tr}}\left[{\rm G}\cdot\frac{\partial{}}{\partial{\mu}}
\hbar \Sigma_{\rm m}\right].
\end{align}
Furthermore, in the crossover near a broad resonance the bare dynamic
of the molecular field can be neglected and the molecular propagator
in Eq.~(\ref{eq:TcStEqGausF7}) can be approximated as in Eq.~(\ref{eq:apprDysMol}).
Therefore Eq.~(\ref{eq:TcStEqGausF7}) can be rewitten as
\begin{align}
\label{eq:TcStEqGausF8}
n\simeq\!\!
\frac{1}{\hbar \beta V}\sum_{\sigma}{\rm{Tr}}\left[{{G}_{\sigma,0}}\right]\!\!
+\!\!
\frac{1}{\hbar \beta V}{\rm{Tr}}\left[\frac{\hbar}{-\delta(B)-\hbar \Sigma_{\rm m}}
\cdot\frac{\partial{}}{\partial{\mu}}
\hbar \Sigma_{\rm m}\right],
\end{align}
or
\begin{align}
\label{eq:TcStEqGausF8a}
n\simeq\frac{\partial{}}{\partial{\mu}}&\left(
2~\frac{1}{\beta V}\sum_{\sigma}{\rm Tr}\left[{\rm ln}~({G}_{\sigma,0}^{-1})\right]-\right.\\ &\left.
\frac{1}{\beta V}~{\rm Tr}\left[{\rm ln}~(\frac{-\delta(B)-\hbar \Sigma_{\rm m}}
{\hbar})\right]
\right)\nonumber.
\end{align}
At this point, to make this part of the derivation more transparent we consider the special
case when the $a_{\rm bg}\rightarrow0$. In that case the molecular selfenergy
of Eq.~(\ref{eq:MBSelfenergy}) is
\begin{align}
\label{eq:MBSelfenergy1abg=0}
\hbar\Sigma_{\rm{m}}\left(\mathbf{q},\omega_n\right)
=g^2 \frac{1}{V}
\sum_{\mathbf{k}}
\left[
\frac{1-N_{\mathbf{q}/2+\mathbf{k}}-N_{\mathbf{q}/2-\mathbf{k}}}{i\hbar\omega_n+2\mu-
2\epsilon_\mathbf{k}-\epsilon_{\mathbf{q}}/2}+\frac{1}{2 \epsilon_\mathbf{k}}
\right].
\end{align}
Using 
the 
expression of the total scattering length in Eq.~(\ref{eq:sclen}) this can be rewritten
as in Ref. \cite{SadeMelo93}
\begin{align}
\label{eq:TcStEqGausF9}
n=\frac{\partial{}}{\partial{\mu}}\left(
\frac{1}{\beta V}\sum_{\sigma}{\rm Tr}\left[{\rm ln}~({G}_{\sigma,0}^{-1})\right]+
\frac{1}{\beta V}~{\rm Tr}\left[{\rm ln}~({{\rm G}_{\Delta}}^{-1}\right]
\right),
\end{align}
where ${\rm G}_{\Delta}$ is the pairing field propagator 
in the Gaussian approximation defined in 
Eq.~(\ref{eq:apprTMBren}).
In the general case when $a_{\rm bg}$ is finite, the derivation is 
analogous but only more involved because the energy dependence 
of the atom-molecule coupling \cite{explanation}. 

By comparing Eq.~(\ref{eq:TcStEqGausF7}) and Eq.~(\ref{eq:TcStEqGausF9})
we argue that the propagator ${\rm G}_{\Delta}$ of the composite boson 
in the single-channel picture contains the open-channel contribution
of the dressed-molecules.
The precise mathematical relation between the two entities will be discussed
further in one of the following sections where we a detailed analysis of
the BEC limit in the superfluid state. 

The curve of the critical temperature $T_c$ for the broad resonance in $^{40}{\rm K}$ atoms
based on the crossover Eqs.~(\ref{eq:totGapeqC}) and~(\ref{eq:TcStEqGausF9})
is shown also in Fig.~\ref{phasediagramA}. This result shows that 
our modified mean-field approach, based on the two-body physics of the dressed molecules,
can be considered a rather good approximation in view of its simplicity.

\subsection{Self-consistent equations.}
\label{sec:selfcons}

The curve of $T_c$ in Fig.~\ref{phasediagramA} calculated according the 
Nozieres-Schmitt Rink approximation exhibits a maximum. 
This maximum disappears in a
self consistent approach \cite{Haussmann94}. The latter
includes 
some fluctuations that lead to an 
increase of the effective mass of the dressed molecules
and therefore to a
descrease of
the transition temperature.
A self-consistent approach for the calculation of $T_c$ in a dilute 
gas of fermions interacting with strong attractive interactions
has been developed by Haussmann in Ref.~\cite{Haussmann94} in relation with a single-channel model and
has been extended by Xiang and Ji in Ref.~\cite{Xia-Ji05} to the two-channel Feshbach model.
Their numerical results shows clearly that 
the curve for $T_c$
in the BEC-BCS crossover evolves monotonically from one limit to the other
when selfenergy effects between atoms and 
molecules are treated self-consistently.
In a self-consistent approach, 
the conservation of the number of total particle in the BEC-BCS crossover
is given by the equation 
\begin{align}
\label{eq:TcStEqSelfCons}
n=
\frac{1}{\hbar \beta V}\sum_{\sigma}{\rm{Tr}}\left[{G_{\sigma}}\right]
-2
\frac{1}{\hbar \beta V}{\rm{Tr}}\left[{\rm G}
\right],
\end{align}
where the full atomic Green's function
is calculated from the many-body ${\rm T}-$matrix
as in Eq.~(\ref{eq:fermionicSelfEnrNormst}) but including
to all order the selfenergy effects in the fermionic and molecular propagators. 
In this section we want to discuss the 
two coupled Eqs.~(\ref{eq:TcStEqSelfCons})
and~(\ref{eq:totGapeq}) in connection with our dressed-molecules picture.
The analysis of the BEC limit is particularly illuminating to this purpose.
We consider here the case when the background interaction
can be neglected. However, as we have seen, it is always very easy to generalize our result including 
the effect of the non resonant interactions.
In the BEC limit we have $2\mu\simeq\epsilon_{\rm m}$ and $\epsilon_{\rm m}\gg k_B T,\epsilon_F$
and the fermionic selfenergy  
\begin{align}
\label{eq:fermionicSelfEnrNormstD}
\hbar\Sigma^f_{\sigma}(\textbf{k},\omega_n)=&
-\frac{1}{(2\pi)^3}\int d^3\textbf{k}'\frac{1}{\beta\hbar^2}\sum_{\omega_{n}'} 
\\
&\times g^2{\rm G}(\textbf{k}+\textbf{k}',\omega_n+\omega_n')
~G_{-\sigma}(\textbf{k}',\omega_n')\nonumber,
\end{align}
can be approximated to the lowest order in the fermionic propagator
as \cite{Haussmann94} 
\begin{align}
\label{eq:fermionicSelfEnrNormstE}
\hbar\Sigma^f_{\sigma}(\textbf{k},\omega_n)&\simeq\!\!
-G_{-\sigma,0}(\textbf{k},\omega_n)~g^2\!\!
\frac{1}{(2\pi)^3}\!\!\int \!\!d^3\textbf{k}'\frac{1}{\beta\hbar^2}\sum_{\omega_{n'}} 
{\rm G}(\textbf{k}',\omega_n'),\\
\label{eq:fermionicSelfEnrNormstEb}
&\simeq\frac{G_{-\sigma,0}(\textbf{k},\omega_n)}{\hbar}~g^2~Z~n_{\rm m}(T),
\end{align}
where $Z$ in the BEC limit is given by taking 
the limit $a_{\rm bg}\rightarrow 0$ in Eq.~(\ref{eq:zeta}), that is \cite{Duine04}
\begin{align}
\label{eq:Zduine}
Z=\left[1+\frac{\eta}{2 \sqrt{|\epsilon_{\rm{m}}|}}\right]^{-1}.
\end{align}
This means that the fermionic propagator assumes a double fraction structure as
\begin{align}
\label{eq:ferprdoppia}
G_{\sigma}(\textbf{k},\omega_n)=\frac{\hbar}{i\hbar\omega_n-(
\epsilon_{\textbf{k}}-\mu)-\frac{g^2~Z~n_{\rm m}(T)}{i\hbar\omega_n-(
\epsilon_{\textbf{k}}-\mu)}},
\end{align}
that is equivalent to a fermionic BCS propagator characterized by a "gap" 
in the spectrum defined as
\begin{align}
\label{eq:pseudogap}
|{\Delta_{\rm pg}}|^2\equiv g^2~Z~n_{\rm m}(T).
\end{align}
Therefore the equation of state in Eq.~(\ref{eq:TcStEqSelfCons}) can be rewritten as
\begin{align}
\label{eq:TcStEqSelfConsB}
n&=-2
\frac{1}{\hbar \beta V}{\rm{Tr}}\left[{\rm G}
\right]\\ 
-&2\frac{1}{(2\pi)^3}\int\!\! d^3\textbf{k}'\frac{1}{\beta\hbar}\sum_{\omega_{n'}}
\frac{\hbar(i\hbar\omega_n+(\epsilon_{\textbf{k}}-\mu))}
{(\hbar\omega_n)^2+(
\epsilon_{\textbf{k}}-\mu)^2+{|\Delta_{\rm pg}|}^2
}
\nonumber\\&=2~Z~n_{\rm m}(T)\nonumber\\
-&2\frac{1}{(2\pi)^3}\int d^3\textbf{k}'\frac{1}{\beta\hbar}\sum_{\omega_{n'}}
\frac{\hbar(i\hbar\omega_n+(\epsilon_{\textbf{k}}-\mu))}
{(\hbar\omega_n)^2+(
\epsilon_{\textbf{k}}-\mu)^2+{|\Delta_{\rm pg}|}^2
}\nonumber.
\end{align}
Far enough in the BEC limit, it is always possible to expand the first term on the right hand side
in powers of $\Delta_{\rm pg}/|\mu|\simeq\Delta_{\rm pg}/\epsilon_{\rm m}$ and 
we get  
\begin{align}
\label{eq:TcStEqSelfConsS}
n
=|\Delta_{\rm pg}|^2\frac{m^2}{4\pi \hbar^3 \sqrt{2 m |\mu|}}
+2~Z~n_{\rm m}(T),
\end{align}
that can be rewritten as
\begin{align}
\label{eq:TcStEqSelfConsT}
n
=2~(1-Z)~n_{\rm m}(T)
+2~Z~n_{\rm m}(T)=2~n_{\rm m}(T).
\end{align}
Therefore, at sufficiently negative detuning we find a gas of thermal dressed
molecules. However, we have not specified the approximation of 
the molecular selfenergy
in the molecular propagator ${\rm G}$. 
To the lowest order, the molecular selfenergy $\hbar\Sigma_{\rm m}$ is given by the
expression in Eq.~(\ref{eq:MBSelfenergy}) 
calculated taking two free fermionic propagators $G_{\sigma,0}G_{-\sigma,0}$.
In this approximation, 
Eq.~(\ref{eq:TcStEqSelfConsT}) reproduces 
the gas of non interacting dressed molecules
as in the Nozieres-Schmitt Rink approach found in Eq.~(\ref{eq:TcStEqGausF4}).
The interaction between the dressed molecules at $T\geq T_c$ is introduced 
calculating the molecular selfenergy to the lowest order in the 
fermionic selfenergy.
This requires to take one free and one dressed 
fermionic propagators 
($G_{\sigma,0}G_{-\sigma}$) 
in the diagram of the molecular selfenergy.
Taking, in the lowest order, the selfenergy in 
the fermionic propagator $G_{\sigma}$ given by Eq.~(\ref{eq:fermionicSelfEnrNormstD})
\begin{align}
\label{eq:fermionicSelfEnrNormstF}
\hbar\Sigma^f_{\sigma}(\textbf{k},\omega_n)\simeq&
-\frac{1}{(2\pi)^3}\int d^3\textbf{k}'\frac{1}{\beta\hbar^2}\sum_{\omega_{n'}}\\&
g^2~{\rm G}(\textbf{k}+\textbf{k}',\omega_n+\omega_n')
~G_{-\sigma,0}(\textbf{k}',\omega_n')\nonumber,
\end{align}
the 
molecular propagator assumes the form
\begin{align}
\label{eq:moleprop2a}
&\hbar G^{-1}\left(\mathbf{K},\Omega_n\right)=\\
&i\hbar\Omega_n\!-\!\frac{\epsilon_{\mathbf{K}}}{2}\!+\!2\mu-\delta(B)
\!-\!\hbar\Sigma_{\rm m}\left(\mathbf{K},\Omega_n\right)\!-\!
\hbar\Delta\Sigma_{\rm m}\left(\mathbf{K},\Omega_n\right)\nonumber.
\end{align}
The first-order correction $\hbar\Delta\Sigma_{\rm m}$ is given by
\begin{align}
\label{eq:selfencorr}
\hbar\Delta&\Sigma_{\rm m}\left(\mathbf{K},\Omega_n\right)=
\frac{1}{(2\pi)^3}\int d^3\textbf{K}'\frac{1}{\beta\hbar}\sum_{\Omega_{n'}}
G\left(\mathbf{K}',\Omega_n'\right)\nonumber\\
&\times{\rm{\Gamma}}^{\rm Born}_{\rm m}\left(\textbf{K}+\textbf{K}',\textbf{K}+\textbf{K}',
(\textbf{K}+\textbf{K}')/2,(\Omega_n+\Omega_n')/2\right), 
\end{align}
with the many-body molecule-molecule vertex given by
\begin{align}
\label{eq:molmolTborn}
{\rm{\Gamma}}^{\rm Born}_{\rm m}&\left(\textbf{K}+\textbf{K}',\textbf{K}+\textbf{K}',
(\textbf{K}+\textbf{K}')/2,(\Omega_n+\Omega_n')/2\right)=\nonumber\\&
=2~g^4~\frac{1}{(2\pi)^3}\int d^3\textbf{k}\frac{1}{\beta\hbar^4}\sum_{\omega_{n}}
G_{\uparrow,0}(\textbf{K}-\textbf{k},\Omega_n-\omega_n)\nonumber\\
&
\times G_{\downarrow,0}(\textbf{k},\omega_n)~G_{\uparrow,0}(\textbf{k},\omega_n) 
~G_{\downarrow,0}(\textbf{K}'-\textbf{k},\Omega_n'-\omega_n). 
\end{align}
In the BEC limit at temperature near $T_c$, 
we have $|\epsilon_{\rm m}(B)|\gg \hbar\Omega, \hbar^2\textbf{K}^2/4m$ and 
$|\Delta\Sigma_{\rm m}|\ll|\Sigma_{\rm m}|$ and the
molecular propagator of Eq~(\ref{eq:moleprop2a}) can be expanded around the pole 
of the molecular propagator without the $\hbar\Delta\Sigma_{\rm m}$ correction. This gives 
\begin{align}
\label{eq:moleprop2b}
&\hbar G^{-1}\left(\mathbf{K},\Omega_n\right)=\nonumber\\&\simeq\frac{1}{Z}\left[
i\hbar\Omega_n-\frac{\epsilon_{\mathbf{K}}}{2}+2\mu-\epsilon_{\rm m}(B)
-Z~\hbar\Delta\Sigma_{\rm m}\left(\mathbf{K},\Omega_n\right)\right]\\
&=\frac{1}{Z}\left[
i\hbar\Omega_n-\frac{\epsilon_{\mathbf{K}}}{2}+\mu_{\rm m}
-Z~\hbar\Delta\Sigma_{\rm m}\left(\mathbf{K},\Omega_n\right)\right]
\label{eq:moleprop2f},
\end{align}
where we have defined the dressed-molecule chemical potential as
$\mu_{\rm m}\equiv2\mu-\epsilon_{\rm m}(B)$.
Furthermore, the zero-momentum and energy correction to the molecular selfenergy
in Eq.~(\ref{eq:moleprop2b}) is given by
\begin{align}
\label{eq:selfencorrzerozero}
Z\hbar\Delta&\Sigma_{\rm m}\left(\mathbf{0},0\right)=\\=&
Z~\rm{\Gamma}^{\rm Born}_{\rm m}\left(\textbf{0},\textbf{0},
\textbf{0},0\right)\frac{1}{(2\pi)^3} 
\int d^3\textbf{K}'\frac{1}{\beta\hbar}\sum_{\Omega_{n'}}
G\left(\mathbf{K}',\Omega_n'\right)\nonumber\\
=& {\rm{\Gamma}}^{\rm Born}_{\rm m}\left(\textbf{0},\textbf{0},
\textbf{0},0\right)~Z^2~n_{\rm m}(T)\nonumber.
\end{align}
Evaluating the Matsubara summation and the momentum integral 
of Eq.~(\ref{eq:molmolTborn})
in the BEC limit
where  $2\mu\simeq\epsilon_{\rm m}$ we find
\begin{align}
\label{eq:selfencorrzero}
Z~\hbar\Delta\Sigma_{\rm m}\left(\mathbf{0},0\right)
=&g^4 \frac{m^{3/2}}{8\pi\hbar^3|\epsilon_{\rm m}(B)|^{3/2}} ~Z^2~n_{\rm m}(T)\nonumber\\
=&g^2 \frac{1-Z}{Z|\epsilon_{\rm m}(B)|}
~Z^2~n_{\rm m}(T),
\end{align}
where we have used the definition of $Z$ in Eq.~(\ref{eq:Zduine}).
In the case of a very broad resonance at large negative detuning, 
when the gas of dressed molecules enters the weak-coupling regime $k_F a(B)\leq 1$ 
and the chemical potential is essentially equal to half the binding
energy of the two-body dressed molecules,
the wavefunction in Eq.~(\ref{eq:wavefctmol}) 
of the dressed molecule can still contain only a very small amplitude in the closed bare
molecular channel. This means that
the binding energy
is in the quadratic regime in the magnetic field given by Wigner's formula
$\epsilon_{\rm{m}}\simeq-\hbar^2/m a^2(B)$
and that $Z\simeq 0$. In this case Eq.~(\ref{eq:selfencorrzero}) can be 
written as
\begin{align}
\label{eq:selfencorrzerob}
Z~\hbar\Delta\Sigma_{\rm m}\left(\mathbf{0},0\right)&\simeq
2\frac{4\pi\hbar^2 2 a(B)}{2 m}n_{\rm m}(T)\\ &\equiv2
{\Gamma}^{\rm Born}_{\rm m}\left(\textbf{0},\textbf{0},
\textbf{0},0\right)n_{\rm m}(T)\nonumber.
\end{align}
This term in the dressed molecular propagator of Eq.~(\ref{eq:moleprop2b})
represents  a many-body mean-field shift of a dilute gas of molecules
interacting with positive scattering length $2 a(B)$.

The interaction energy 
of the dressed molecular gas has been measured
\cite{Bartenstein04} by fitting 
a Thomas-Fermi profiles to the spatial distributions of the trapped ultracold Fermi gas
at densities of about $10^{13}$ cm$^{-3}$
in the BEC limit.
These experiments consider the crossover across the broad resonance at about $834~G$
in atomic $^6$Li atoms. They assume, for magnetic fields between
$600~G$ and $780~G$, 
a dilute gas of tightly-bound interacting molecules.  
They find that the molecule-molecule scattering length
is in agreement with $a_{\rm{m}}(B)\simeq0.6 ~a(B)$.
The origin of the value of $0.6 ~a(B)$ has been explained 
by solving directly the Schro\"{e}dinger equation of the four-body
problem \cite{Petrov04} and also in terms of Feynmann diagrams
in Refs. \cite{Kagan05,Levinsen05} by using a single-channel model.
In the latter approach, the authors of Refs. \cite{Kagan05,Levinsen05}
have identified the exact equation for the generalized molecule-molecule 
vertex ${\rm{\Gamma}}_{\rm m}$ that contains the dominant fluctuations in the BEC limit
leading to the value 
found in the experiment.
Considering the same class of diagrams in our two-channel formulation
for the vertex ${\rm{\Gamma}}_{\rm m}$ introduced in Eq.~(\ref{eq:molmolTborn})
would also lead to a dressed molecule-molecule ${\rm T}$-matrix
${\Gamma}_{\rm m}\left(\textbf{0},\textbf{0},
\textbf{0},0\right)
\simeq4\pi\hbar^2 ~0.6~a(B)/2m$.
For example, summing over all the ladder diagrams in the particle-particle channel
beyond the Born approximation described by ${\Gamma}^{\rm Born}_{\rm m}$
one would recover the result ${\Gamma}_{\rm m}\left(\textbf{0},\textbf{0},
\textbf{0},0\right)\simeq 4\pi\hbar^2~0.75~a(B)/2m$ found by Pieri and Strinati in \cite{PieriStrinati03}.

The interaction between dressed molecules introduced in Eq.~(\ref{eq:selfencorrzerob})
does not cause any shift on the critical temperature of the Bose-Einstein
condensation of the dressed molecules
because it does not shift the effective mass of the dressed molecules. 
Therefore, we have to consider the 
momentum and frequency dependence in the interaction vertex $\Gamma^{\rm Born}_{\rm m}$
in Eq.~(\ref{eq:molmolTborn}). Following the analysis 
of the BEC limit given by Haussmann \cite{Haussmann94},
we consider the ansatz
\begin{eqnarray}
\label{eq:moleprop2c}
\hbar G^{-1}\left(\mathbf{K},\Omega_n\right)&\simeq\frac{1}{Z~Z_1}\left[
i\hbar\Omega_n-\frac{{\hbar^2\mathbf{K}}^2}{4~m^*}+\mu_{\rm m}
\right]
\end{eqnarray}
for the dressed-molecular propagator. 
The Nozieres-Schmitt Rink approximation corresponds clearly to the case 
$Z_1=1$ and $m^*=m$. To find the first order corrections to $Z_1$ and $m^*$,
we substitute the ansatz of Eq.~(\ref{eq:moleprop2c})
in Eqs.~(\ref{eq:selfencorr}-\ref{eq:moleprop2b}),
we perform the frequency sum and the momentum integral
in Eq.~(\ref{eq:molmolTborn})
and finally we expand the resulting expression up to the first
order in $\mathbf{K}^2$ and $\Omega_n$. 
Ultimately, in the case of a broad resonance, we find \cite{Haussmann94}
\begin{eqnarray}
\label{eq:ultima}
m^*\simeq &m\left[1+2\pi(n_{\rm m}a(B)^3)\right]\\
Z_1\simeq &1+6\pi ~n_{\rm m} a(B)^3\\
\mu_{\rm m}\simeq &2
{\Gamma}^{\rm Born}_{\rm m}\left(\textbf{0},\textbf{0},
\textbf{0},0\right)n_{\rm m}(T). 
\end{eqnarray} 
The critical temperature in the BEC limit is determined by the equation of state.
In a broad resonance, one generally has still $Z\simeq 0$ when 
the gas enters the BEC limit of
the crossover. In that case, the number of particles in the equation of state
of Eq.~(\ref{eq:TcStEqSelfCons}) is determined mainly by the fermionic
Green's function
\begin{align}
\label{eq:TcStEqSelfConsfin}
n\simeq
\frac{1}{\hbar \beta V}\sum_{\sigma}{\rm{Tr}}\left[{G_{\sigma,0}}
\right].
\end{align}
The evaluation of this term given 
in Eq.~(\ref{eq:TcStEqSelfConsT}) is calculated taking 
the fermionic propagator ${G_{\sigma}}$
only the first-order in the selfenergy. 
Expanding the fermionic Green's function 
at the second order in its selfenergy, one can show
\cite{Haussmann94} that the result of Eq.~(\ref{eq:TcStEqSelfConsT})
is modified as 
\begin{align}
\label{eq:TcStEqSelfConsfina}
n\simeq-\frac{1}{Z_1}
2~\frac{1-Z}{Z}\frac{1}{\hbar \beta V}
{\rm{Tr}}\left[{G}\right]\simeq-\frac{1}{Z_1}
2~\frac{1}{Z}\frac{1}{\hbar \beta V}
{\rm{Tr}}\left[{G}
\right].
\end{align} 
Substituting in this equation the propagator of Eq.~(\ref{eq:moleprop2c})
the $Z$ and  $Z_1$ factor drop out. Therefore the equation of state
describes effectively an ideal gas of dressed molecules with renormalized
mass $m^*$.
The critical temperature is calculated from the relation 
\begin{eqnarray}
\label{eq:numbermstarmol}
n\simeq2n_ {\rm m}(T)=&\frac{1}{(2\pi)^3} 
\int d^3\textbf{K} N_B\left(\frac{\hbar^2\mathbf{K}^2}{4m^*}\right)\nonumber\\
=& 2\zeta(3/2)\left[2m^*k_B T_c/2\pi\hbar^2\right].
\end{eqnarray} 
This leads to the negative shift 
of the critical temperature $T_C$ respect to the 
ideal Bose-Einstein critical temperature $T_{BEC}$
\begin{eqnarray}
\label{eq:TcShift}
\Delta T_C/T_{BEC}=m/m^*-1=-\frac{1}{(3\pi)\left(k_F a(B)\right)^3}
\end{eqnarray} 
found by Haussmann.

Although the self consistent calculation is more general than the Nozieres-Schmitt Rink
approximation, we think that the correct theory should
lead to a positive shift and then to 
a maximum in the curve of $T_C$. 
Infact, according to the 
theory of dilute weakly-interacting Bose gases
\cite{Gruter97,Bijlsma96,Baym99},
the repulsive interactions between the molecules in 
asymptotic BEC limit are expected to enhance the value of the critical temperature.
The physical origin
of this maximum, therefore, is different than that described by the Nozieres-Schmitt Rink
approach, as the latter neglects the interactions between the noncondensed molecules.


\section{BEC-BCS crossover below $T_C$.}
\label{crossover}

In this section, we consider the BEC-BCS crossover in the broken symmetry state.
The Gross-Pitaevskii theory at $T=0$ of the dressed-molecule condensate
and the Bogoliubov theory at finite temperature
are derived in the BEC limit by means of analytical methods.
As before, the single channel results are derived as a special case in the limit of very broad resonances.
Moreover, the relation between the composite boson of the single-channel and the dressed-molecule of
the two-channel model is clearly formulated. 

\subsection{Mean-field theory at $T=0$.}
\label{Leggett}
At $T=0$ a set of two self-consistent equations for the BCS energy gap
and the chemical potential
can be derived generalizing the variational approach introduced by Leggett 
\cite{Leggett80}
to our two-channel atom-molecule Hamiltonian given by Eq.~({\ref{eq:Ham}}).
The zero-temperature ground state 
can be written \cite{Levin04} as a product state of both a bare molecular and an atomic contribution, i.e.,
\begin{equation}
\label{eq:groundstate}
|{{\Psi}}_0\rangle = |\Phi_0^F \rangle\otimes|\Phi_0^B\rangle,
\end{equation}
where the normalized fermionic wavefunction is the standard crossover ground state
\cite{Eagles69,Leggett80}
\begin{equation}
|\Phi_0^F\rangle=\prod_{\bf k>\mathbf{0}}(\uk+\vk c_k^{\dagger} c_{-k}^{\dagger})\,|0\rangle,
\label{eq:lggttGS}
\end{equation}
and the normalized bare molecular contribution $|\Phi_0^B\rangle$
is given by 
\begin{equation}
 |\Phi_0^B\rangle=
e^{-{n}_0/2+{\sqrt{n}_0} ~b_0^{\dagger}}|0\rangle \,.
\end{equation}
The real variational parameters are $\uk$, $\vk$ and $\sqrt{{n_0}}$.
Minimizing the grand-canonical energy with respect to the parameter $\sqrt{{n_0}}$,
we find the constraint
\begin{equation}
\langle b_0 \rangle\equiv
\sqrt{{n_0}}
=\frac{g_{{\rm{bare}}}~ \Delta_{\rm{bg}}}
{(\delta_{{\rm{bare}}}-2\mu)~ {\rm{V}}_{\rm{bg}}},
\label{eq:extra}
\end{equation}
where $\Delta _{\rm{bg}}$ is the order parameter related to the attractive
background interaction
\begin{equation}
 \Delta _{{\rm{bg}}}\equiv -\frac{1}{\it{V}}\sum_{\bf k} {\rm{V}}_{\rm{bg}}\uk\vk
 =- \frac{1}{\it{V}}\sum _{\bf k}{\rm{V}}_{\rm{bg}}\langle a_{-{\bf k}\downarrow}a_{{\bf
  k}\sigma1}\rangle \,.
  \label{eq:bcsgap}
  \end{equation}
The variation with respect to $\vk$, under the normalization constraint $\vk^2+\uk^2=1$,
yields
\begin{align}
u_{\bf k}^2 = \frac{1}{2}\left(1+\frac{\epsilon_{\bf k}-\mu }
{\sqrt{\left(\epsilon_{\bf k}-\mu \right)^2+ |\Delta|^2}}\right)\nonumber\\
v_{\bf k}^2 = \frac{1}{2}\left(1-\frac{\epsilon_{\bf k}-\mu }
{\sqrt{\left(\epsilon_{\bf k}-\mu \right)^2+ |\Delta|^2}}\right),
\end{align}
where the total gap of the theory $\Delta$ is defined as 
\begin{align}
\label{eq:totgap}
\Delta=\Delta _{\rm{bg}}-g_{{\rm{bare}}}\sqrt{{n_0}}.
\end{align}

\begin{figure*}
    \begin{tabular}{cc}
      \resizebox{70mm}{!}{\includegraphics{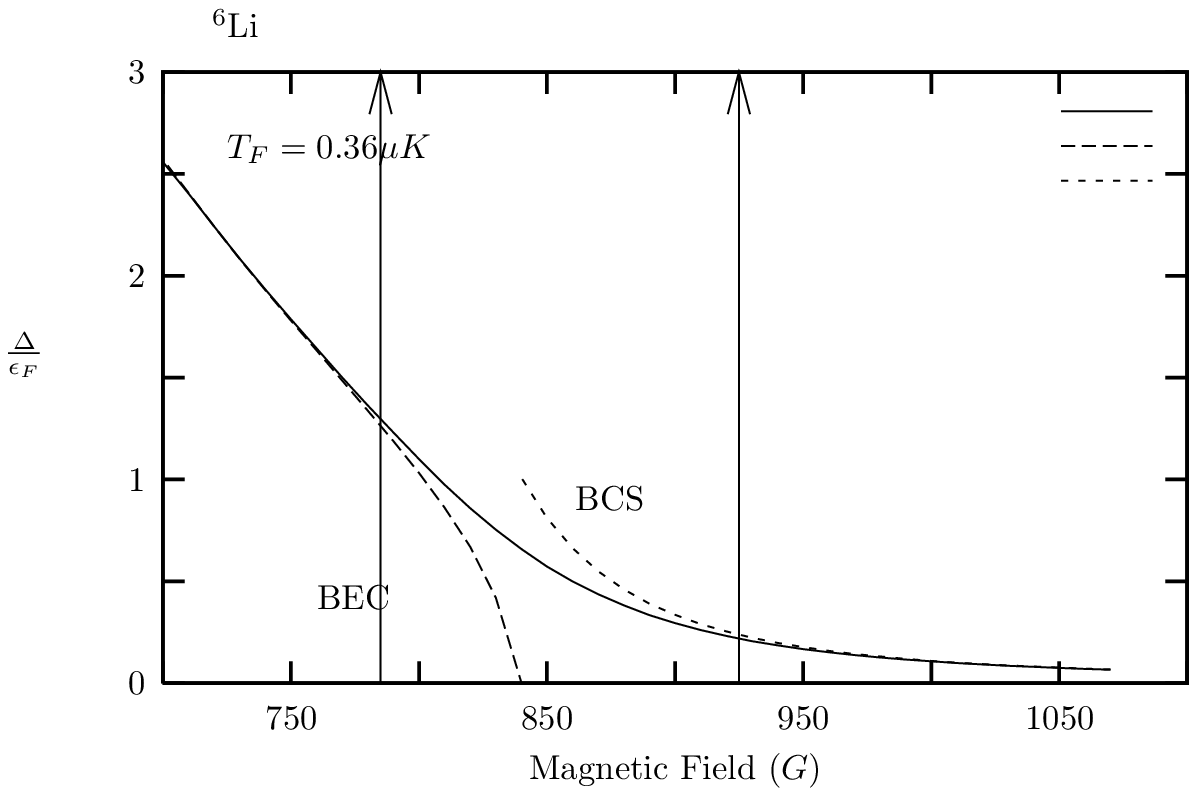}} & \hspace{1 cm}
      \resizebox{70mm}{!}{\includegraphics{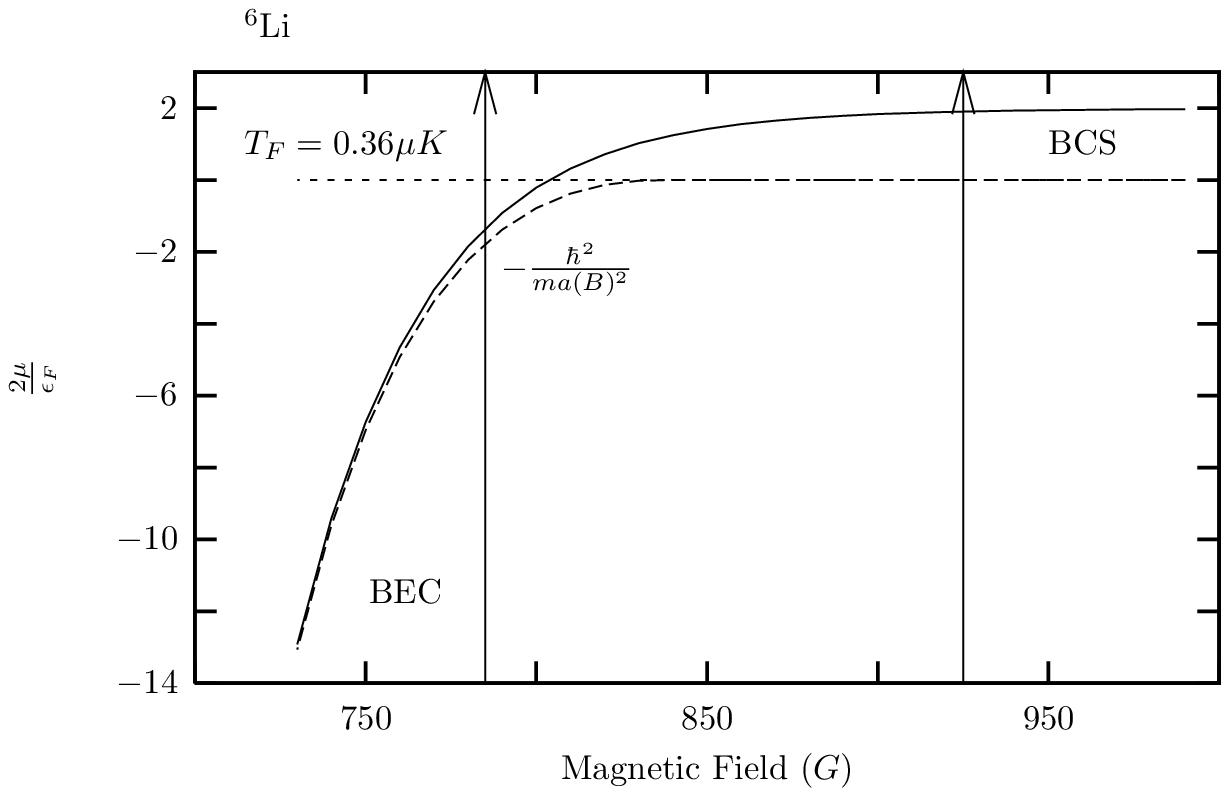}} \\
    \end{tabular}
    \caption{BCS energy gap and chemical potential as a function
  of the magnetic field for the very broad resonance at $834~G$
in a $^6$Li mixture at low density.
    }
    \label{Lig1LiTC2mu1}
\end{figure*}

\begin{figure*}
    \begin{tabular}{cc}
      \resizebox{70mm}{!}{\includegraphics{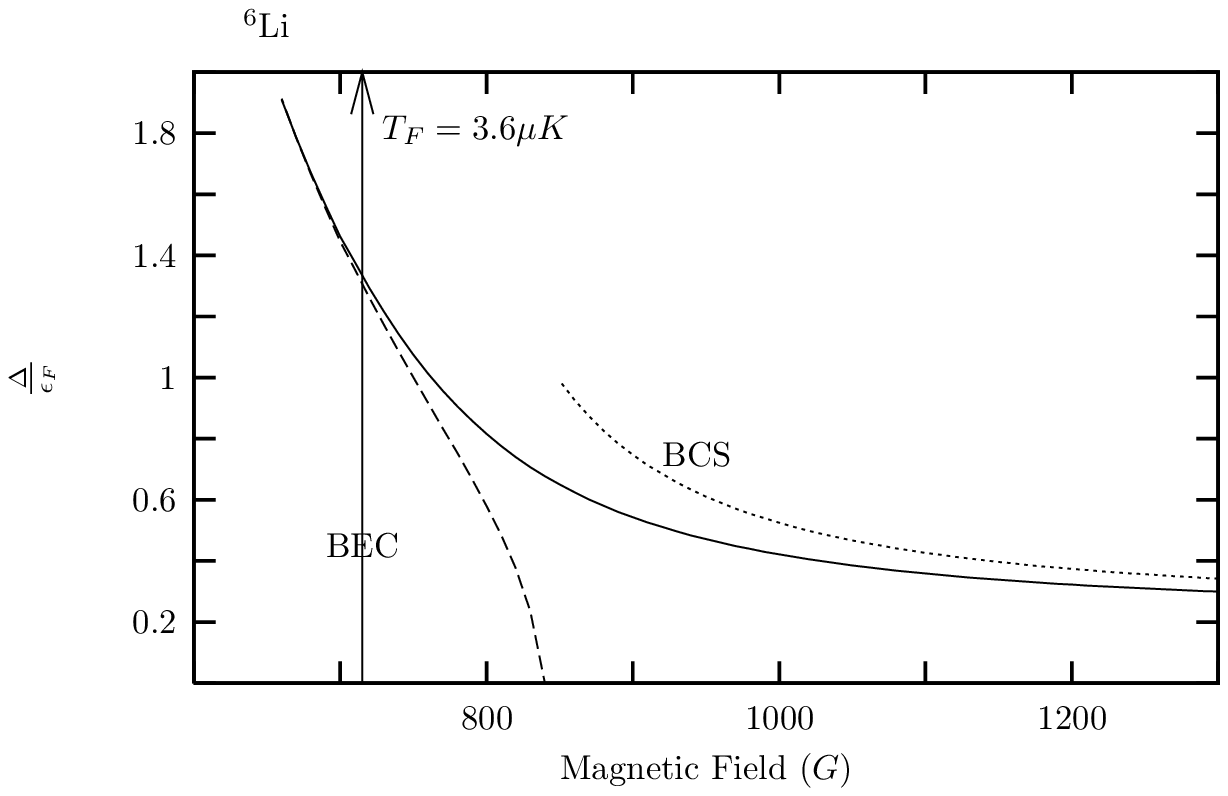}} & \hspace{1 cm}
      \resizebox{70mm}{!}{\includegraphics{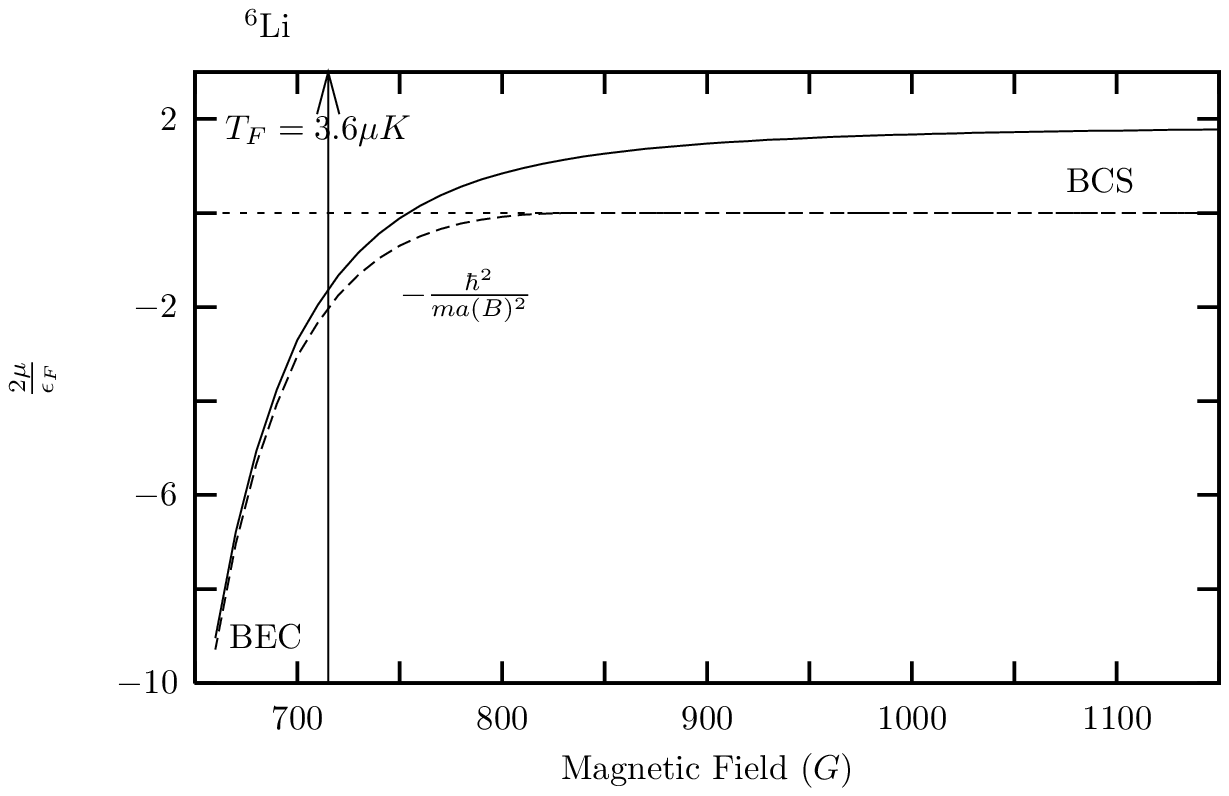}} \\
    \end{tabular}
    \caption{BCS energy gap and chemical potential as a function
  of the magnetic field for the very broad resonance at $834~G$
in a $^6$Li mixture at high density.
    }
    \label{LiTCgap2LiTC2mu2}
\end{figure*}

\begin{figure*}
    \begin{tabular}{cc}
      \resizebox{70mm}{!}{\includegraphics{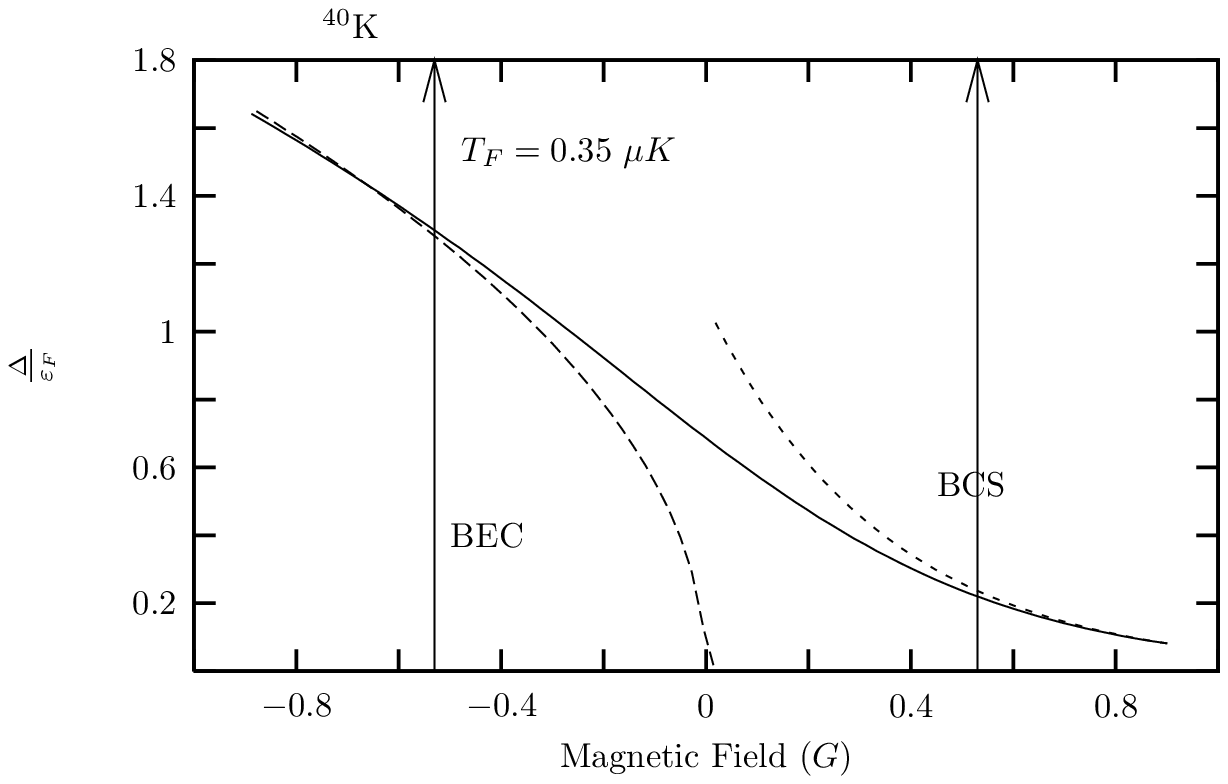}} & \hspace{1 cm}
      \resizebox{70mm}{!}{\includegraphics{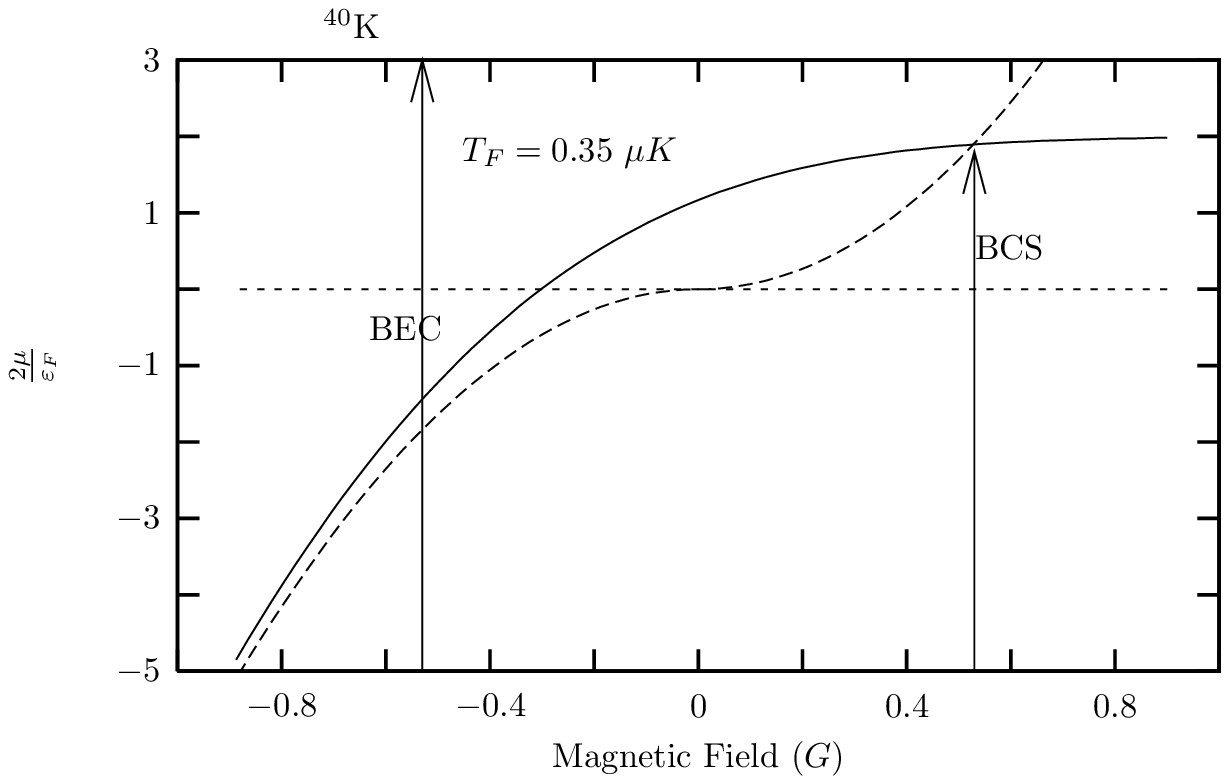}} \\
    \end{tabular}
    \caption{BCS energy gap and chemical potential as a function
  of the magnetic field for the broad resonance at $202.1~G$
  in a $^{40}$K mixture at low density.
    }
    \label{KTCgapKTC2mu}
\end{figure*}

Substituting these results in the definition in Eq.~({\ref{eq:bcsgap}})
and using the constraint in Eq.~({\ref{eq:extra}}),
we arrive at the gap equation in the compact form
\begin{eqnarray}
-\frac{1}{{\rm{V}}_{\rm{bg}}-\frac{g_{\rm{bare}}^2}{\delta_{\rm{bare}}-2\mu}} 
=\frac{1}{\it{V}}\sum_{\bf k} 
\frac{1}{2 \sqrt{\left(\epsilon_{\bf k}-\mu \right)^2+ |\Delta|^2}}\,.
\end{eqnarray}
This equation can be rewritten in terms of the renormalized quantities as
\begin{equation}
\label{eq:varrengap}
-\frac{1}{{\rm{T}}^{2B}_{\rm{bg}}-\frac{g^2}{\delta-2\mu}} 
=\frac{1}{\it{V}}\sum_{\bf k}\left( 
\frac{1}{2 \sqrt{\left(\epsilon_{\bf k}-\mu \right)^2+ |\Delta|^2}}
-\frac{1}{2\epsilon_{\bf k}}\,
\right),
\end{equation}
where now
\begin{eqnarray}
\label{eq:varrengap1}
\Delta \equiv &
\Delta_{{\rm{bg}}}-g\sqrt{{n_0}}
\end{eqnarray}
and
\begin{eqnarray}
\label{eq:varrengap2}
\Delta_{\rm{bg}}=&
\frac{\sqrt{{n_0}}~(\delta-2\mu)~ {\rm{T}}^{2B}_{\rm{bg}}}{g}
\label{eq:extra1},
\end{eqnarray}
as a consequence of the general result 
for the many-body ${\rm T}-$matrix in Eq.~(\ref{eq:theo}),
combined with the fact that the gap
equation in Eq.~(\ref{eq:varrengap}) must be
equivalent to a sum over all ladder diagrams. 
The description of the BEC-BCS crossover at the mean-field level of approximation
is completed by the equation of state
\begin{equation}
\label{eq:stateeq}
n=2 n_0+\frac{1}{(2\pi)^3}\int d^3{\bf{k}} \left(
1-\frac{\epsilon_{\bf{k}}-\mu}{\sqrt
{|\Delta|^2+(\epsilon_{\bf{k}}-\mu})^2}\right).
\end{equation}
The Eqs.~({\ref{eq:varrengap}})-({\ref{eq:varrengap2}}) and~({\ref{eq:stateeq}}) represent
a set of coupled equations in $\Delta$ and $\mu$ at fixed density.
Note that in contrast to the normal state result, the analysis at zero temperature in the
broken symmetry state does not require any modification of the equation of state in order 
to get the proper two-body physics of the dressed molecules in the BEC limit. This is because the 
equation of state of BCS mean-field theory at $T=0$
sums automatically over the ladder diagrams. As a result, we expect also to have smooth crossover curves for
$\Delta$ and $\mu$ already at the mean-field level.  

At positive detuning, 
in the weak-coupling regime 
$k_F|a(B)|\leq 1$, the binding is a cooperative effect in the vicinity of
the Fermi surface. The gas consists of largely overlapping
weakly-bound Cooper pairs. The chemical potential 
is fixed at $\mu=\epsilon_F$
by the equation of state in Eq.~({\ref{eq:stateeq}}).
The solution of Eq.~({\ref{eq:varrengap}}) is then easy calculated
analytically with the result
\begin{equation}
\label{eq:BCSresgap}
\Delta \simeq \frac{8\epsilon_F}{k_{\rm B} } e^{-2}
e^{\frac{2}{\pi}\frac{\sqrt{2\mu}}{\eta}}\ 
             \exp \left\{ -\frac{\pi}{2k_F|a(B)|}  \right\}.~
\end{equation} 
The energy gap is related to the BCS 
transition critical temperature~({\ref{eq:resBCSTcnarw}})  
by the relation $T_C=\left(
e^{\gamma}/\pi\right)\Delta$, as is expected
for weak-coupling superconductivity.
Note that the energy gap of Eq.~(\ref{eq:BCSresgap})
coincides formally with the shift in the ground state energy of Eq.~(\ref{eq:shift}).
This is not surprising because the singularity in molecular selfenergy
discussed in section ~\ref{sec:Kondo}
and the Cooper pairing occur in the same diagram. 

At negative detuning, the system evolves to a Bose-Einstein condensate
of tightly bound dressed molecules.
The roles of the gap equation and the number equation are exchanged.
In the BEC limit, the chemical potential becomes large and negative. 
The crossover equations can be expanded in powers of $|\Delta|/|\mu|$ as
\begin{align}
\label{eq:expvarrengap}
\frac{1}{{\rm{T}}^{2B}_{\rm{bg}}-\frac{g^2}{\delta-2\mu}} 
=&\left(\frac{2 m}{\hbar^2}\right)^{3/2}
\frac{\sqrt{|\mu|}}{8 \pi} 
\left[1+\frac{1}{16} \frac{|\Delta|^2}{\mu^2}\right], ~~{\mbox{and}}
\\
\label{eq:expstateeq}
n=&2 n_0+|\Delta|^2 \frac{  m^2} {4 \pi \sqrt{2m |\mu|} \hbar^3}.
\end{align}
Neglecting the term quadratic in $|\Delta|/|\mu|$, Eq.~({\ref{eq:expvarrengap}})
becomes equivalent to Eq.~({\ref{eq:tottwobodymu}}). 
Thus the chemical potential approaches half the binding energy of Eq.~(\ref{eq:analroot}) 
of a Bose-Einstein condensed dressed molecule, as expected.

The curves for $\mu$ and $\Delta$ obtained by solving
numerically the mean-field crossover 
Eqs.~({\ref{eq:varrengap}})-({\ref{eq:varrengap2}}) and~({\ref{eq:stateeq}}) are 
shown in Figs.~{\ref{Lig1LiTC2mu1}} and~{\ref{LiTCgap2LiTC2mu2}} for a gas of $^6$Li atoms at two
different densities and in  Fig.~{\ref{KTCgapKTC2mu}} for a gas of $^{40}$K atoms. 
In both cases we have $\eta^2\gg \epsilon_F$.
In each figure the vertical lines represent the boundary of the strong 
coupling regime $k_F |a(B)|>1$. In the $^6$Li gas with high density
of Fig.~{\ref{LiTCgap2LiTC2mu2}}, the gas never enters the 
weak-coupling regime $k_F |a(B)|<1$ in the range of the magnetic field
considered in the figure.

The mean-field curves for $\Delta$ and $\mu$ do not differ quantitatively from the solutions obtained
by the single-channel approximation \cite{Leggett80,SadeMelo93}
based on the equations
\begin{align}
\label{eq:singlechangapeq}
-&\frac{1}{{\rm{T}}^{2B}_{\rm{bg}}-\frac{g^2}{\delta}}=
-\frac{1}{\frac{4\pi \hbar^2 a(B)}{m}} 
=\\&=\frac{1}{\it{V}}\sum_{\bf k}\left( 
\frac{1}{2 \sqrt{\left(\epsilon_{\bf k}-\mu \right)^2+ |\Delta_{\rm{sc}}|^2)}}
-\frac{1}{2\epsilon_{\bf k}}\,
\right)\nonumber,
\end{align}
and
\begin{equation}
\label{eq:Singlechstateeq}
n=\frac{1}{(2\pi)^3}\int d^3{\bf{k}} \left(
1-\frac{\epsilon_{\bf k}-\mu}{\sqrt
{|\Delta_{\rm{sc}}|^2+(\epsilon_{\bf k}-\mu})^2}\right),
\end{equation} 
where the single-channel gap is defined as $\Delta_{\rm{sc}}
\equiv({\rm{V}}_{\rm{bg}}+{\rm{V}}_{\rm{res}})\langle
\psi_{\uparrow}(\mathbf{x})\psi_{\downarrow}(\mathbf{x})\rangle$. 
These equations are obtained from  
Eqs.~({\ref{eq:varrengap}}) and~({\ref{eq:stateeq}})
neglecting explicitly the contribution of the bare molecular boson.
They can be derived also by integrating out the molecular field 
in the action of Eq.~(\ref{eq:action})
from the
very beginning.
 
At negative detuning the chemical
potential goes to half the binding energy $\epsilon_{\rm{m}}\simeq-\hbar^2/m a^2(B)$  
described in 
Figs~{\ref{Lig1LiTC2mu1}}(b),~{\ref{LiTCgap2LiTC2mu2}}(b) and~{\ref{KTCgapKTC2mu}}(b)
by dotted lines. 
In Fig.~{\ref{KTCgapKTC2mu}}(b) the curve of the binding energy of the dressed molecules
is continued at positive detuning by the location of the 
maximum in the two-body density of states  given in Eq.~({\ref{eq:posdetmax}}). This illustrates the
main idea of
our modified mean-field picture of  
Sec.~\ref{sec:meanfield}. The magnetic field at which this curve 
crosses $2\mu$ is indicated by a vertical line. At this magnetic field the 
energy of the two-body dressed molecules approaches the chemical potential of the gas \cite{Falco04a}.

In Figs.~{\ref{Lig1LiTC2mu1}},~{\ref{LiTCgap2LiTC2mu2}} and~{\ref{KTCgapKTC2mu}} the dotted lines represent
the two extreme limits
of the crossover. The lines denoted as BCS correspond to the BCS
solution of Eq.~({\ref{eq:varrengap}}). The lines denoted as BEC correspond to the 
curve $\Delta=\sqrt{\left(16/3\pi\right)}\epsilon_F/\sqrt{k_F a(B)}$, which is 
the asymptotic solution for the energy gap according to the single-channel 
gap equation in Eq.~({\ref{eq:singlechangapeq}}).

\subsection{Asymptotic limit of $Z$ in the deep BEC limit.}
\label{sec:prettyneat}

The analysis of the equation of state in the BEC limit requires some 
special consideration. The correct picture of the crossover across the Feshbach resonance
is based on the dressed molecule, which is the true energy eigenstate of the 
diatomic molecule in the presence of the atom-molecule coupling. Unfortunately, however, 
the dressed
molecule density does not appear explicitly in the coupled equations  
of our mean-field treatment and mean-field theory is therefore not able
to extract this quantity. 

Nevertheless, a gas of weakly-interacting closed-channel bare molecules is realized
at sufficiently large negative detunings.
More precisely, the total density amounts to 
\begin{align}
\label{eq:BEClimeqstat0}
n
\simeq2 n_0,
\end{align}
when $k_F a(B)\leq 1$ and $|\delta(B)|\geq \hbar^2/m a_{\rm{bg}}^2$.
For less negative detunings we must
use \cite{Romans05}
\begin{align}
\label{eq:guess}
n_0=Z n_{\rm{mc}},
\end{align}
where 
$n_{\rm{mc}}$ is defined as the condensate density of the dressed molecules.
A careful analysis of the equation of state in Eq.~(\ref{eq:expstateeq}) 
shows that this definition is indeed consistent. The proof proceeds as follows.
First we assume that
in the deep BEC limit the total density can be approximated
by the density of the dressed molecules
\begin{align}
\label{eq:BEClimeqstat}
n \simeq 2 n_{\rm{mc}},
\end{align}
with binding energy $\epsilon_{\rm{m}}\simeq 2\mu $ given by Eq.~(\ref{eq:analroot}).
Then, we eliminate the explicit dependence  on $\delta$ and $\Delta_{\rm{bg}}$
by using Eqs.~(\ref{eq:extra1}) and~({\ref{eq:tottwobodymu}}), to rewrite the equation of state
as 
\begin{equation}
2 n_{\rm{mc}} \simeq 2 Z n_{\rm{mc}}+ \frac{2 Z n_{\rm{mc}} \eta}{2 \sqrt{|\epsilon_{\rm{m}}|}\left(
1+\sqrt{\frac{|\epsilon_{\rm{m}}|}{\epsilon_{bg}}}\right)^2}.
\end{equation} 

\begin{figure}
\vspace{1.4 cm}
\epsfig{figure=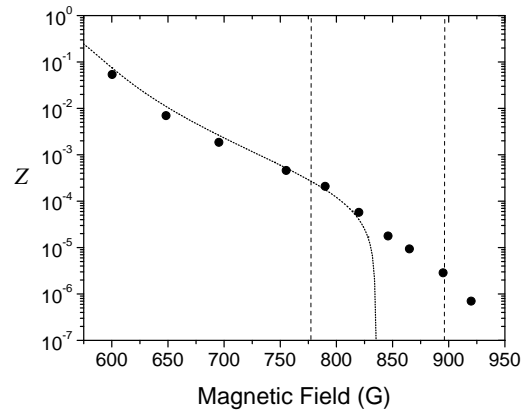,width=8.cm} 
\caption{The value of $Z$ measured in \cite{Partridge05}
is plotted as function of the magnetic field (dots).
The continous line represents the two-body $Z$ given by Eq.~(\ref{eq:zeta})
The vertical dashed lines represent the boundary of the strong coupling regime
$k_F |a(B)|>1$. Below $800$ Gauss, we have $2\mu\simeq\epsilon_{\rm{m}}$
and the two-body $Z$
matches perfectly with the experimental value. The Fermi energy of the gas is 
at $T_F=380nK$. 
} 
\label{Randy}
\end{figure}
Solving for $Z$ we recover the correct two-body limit expression 
of Eq.~(\ref{eq:zeta}), i.e.,
\begin{equation}
Z=\left[1+\frac{\eta}{2 \sqrt{|\epsilon_{\rm{m}}|}\left(
1+\sqrt{\frac{|\epsilon_{\rm{m}}|}{\epsilon_{bg}}}\right)^2}\right]^{-1}.\nonumber
\end{equation}
Putting back this $Z$ in the equation of state we find again Eq.~(\ref{eq:BEClimeqstat}).

It is important to stress the fact that the approximation introduced in
Eq.~(\ref{eq:BEClimeqstat})
is valid only deep in the BEC limit of
strongly-bound dressed molecules. 
This can be understood as follows. According to the BCS theory, at zero temperature
all the atoms are paired far below resonance. In our case we have a gas of dressed molecules which
share an amplitude on the open-channel Cooper pairs wave-function and on the bare closed-channel molecular bound state.
However, even at zero temperature not all the dressed molecules are condensed because interactions deplete
the condensate density as the interactions between the Cooper pairs do in the zero-temperature BCS theory.
Therefore, Eq.~(\ref{eq:BEClimeqstat}) neglects the depletion of the dressed-molecules condensate
and is correct only asymptotically in the deep BEC limit when 
the depletion  descreases progressively because the gas
becomes more and more dilute.    

This means that Eq.~(\ref{eq:BEClimeqstat}) is valid only at magnetic fields 
such that $k_F a(B)\leq 1$, when
the many-body part of the BEC-BCS crossover has already taken place,
and where the chemical potential approaches half the molecular 
two-body binding energy in Eq.~(\ref{eq:analroot}). 
When $k_F |a(B)|\geq 1$,
the fluctuations in the atomic fields becomes important and also at $T=0$ the gas cannot be
approximated only by a condensate of dressed molecules \cite{Romans05}. 
As a result, 
the bare molecular probability $Z$ is not given by $2n_0/n$ \cite{Mackey05,Chen05}
as in the asymptotic BEC limit.
Moreover, $Z$ deviates from its two-body expression of Eq.~(\ref{eq:zeta}) 
and
the two mean-field equations are not sufficient to calculate its value,
because they can not calculate the depletion of condensate of the dressed-molecules.
A more refine approach is needed beyond the saddle point solution \cite{Romans05}.
The quantity $Z$ has been recently measured by Partridge {\it{et al.}} \cite{Partridge05}.
Their data are illustrated in Fig.~\ref{Randy}. The experimental data are in very good agreement with 
our two-body $Z$ of Eq.~(\ref{eq:zeta}) in the BEC limit.  

\subsection{Gross-Pitaevskii theory of the dressed-molecule condensate at $T=0$.}
 
We show in this section that,
in contrast to the normal state analysis,
the Born
interactions between the dressed molecules in the BEC limit of a broad resonance
described in the previous section, 
arises already at the mean-field level in the crossover theory below $T_C$ based on 
Eqs.~(\ref{eq:varrengap}) and~(\ref{eq:stateeq}).
The effective chemical potential for a dilute weakly-interacting gas
of dressed molecules, is defined, at the leading order, as 
\begin{equation}
\mu_{\rm{m}}\equiv\frac{4\pi\hbar^2 a_{\rm{m}}^{\rm{Born}} n_{\rm{mc}}}{m_{\rm{m}}},
\end{equation}
where $a_{\rm{m}}$ is scattering length of two
dressed molecules with mass $m_{\rm{m}}=2 m$.
The quantity $\mu_{\rm{m}}$ is given by
\begin{equation}
\mu_{\rm{m}}=2\mu-\epsilon_{\rm{m}},
\end{equation}
where $2\mu$ is calculated 
retaining the quadratic term $|\Delta|^2/|\mu|^2$
in the gap equation in Eq.~(\ref{eq:expvarrengap}).
Using Eqs.~(\ref{eq:guess}),~(\ref{eq:BEClimeqstat}) and that
$2\mu\simeq\epsilon_{\rm{m}}$
the gap equation in Eq.~(\ref{eq:expvarrengap})
can be rewritten as
\begin{align}
\label{eq:usami1}
&\epsilon_{\rm{m}}=\\&2\mu\!+\!
n_{\rm{mc}}\frac{4\pi\hbar^2 a_{\rm eff}(B)}{m} Z\frac{\eta}
{2 \sqrt{|\epsilon_{\rm{m}}|}\left(
1+\sqrt{\frac{|\epsilon_{\rm{m}}|}
{\epsilon_{\rm{bg}}}}\right)^2}\!+\!O\left(\frac{\Delta}{2\mu}\right)^4,\nonumber
\end{align}
or 
\begin{align}
\label{eq:usami11}
\mu_{\rm{m}}\simeq n_{\rm{mc}}\frac{4\pi\hbar^2 a_{\rm eff}(B)}{m} Z\frac{\eta}
{2 \sqrt{|\epsilon_{\rm{m}}|}\left(
1+\sqrt{\frac{|\epsilon_{\rm{m}}|}
{\epsilon_{\rm{bg}}}}\right)^2},
\end{align}
where the binding energy $\epsilon_{\rm{m}}$
is given by the general solution of Eq.~(\ref{eq:analroot}),
and the quantity $a_{\rm eff}$ has the dimensions of a length and is given by
\begin{align}
\label{eq:aeff}
a_{\rm eff}(B)\equiv\frac{\hbar}{\sqrt{m |\epsilon_{\rm{m}}|}}.
\end{align}
Using in Eq.~(\ref{eq:usami1})
the definition of $Z$ in Eq.~(\ref{eq:zeta}),
we find for the effective chemical potential of the dressed molecules
at the mean-field level 
\begin{equation}
\label{eq:mubare00}
\mu_{\rm{m}}\simeq n_{\rm{mc}}\frac{4\pi\hbar^2}{m_{\rm{m}}}~2a_{\rm eff}(B)(1-Z).
\end{equation}
that implies
\begin{equation}
\label{eq:mubare0}
a_{\rm{m}}^{\rm{Born}}=2a_{\rm eff}(B)(1-Z).
\end{equation} 
However, for very broad resonances such as that at $834 G$ in the $^6$Li gas, 
when the gas enters the weak-coupling regime $k_F a(B)\leq 1$,
the wavefunction in Eq.~(\ref{eq:wavefctmol})
of the dressed molecule still contains only a very small amplitude in the closed bare
molecular channel.
This implies that the dependence of
the binding energy on the magnetic field 
is still in the quadratic regime by Wigner's formula
$\epsilon_{\rm{m}}\simeq-\hbar^2/m a^2(B)$.
Therefore, in this regime Eq.~(\ref{eq:mubare00}) can be approximated as
\begin{equation}
\label{eq:mubare}
\mu_{\rm{m}}\simeq n_{\rm{mc}}\frac{4\pi\hbar^2}{m_{\rm{m}}}~2a(B)(1-Z).
\end{equation}
This result is equivalent to that of a Gross-Pitaevskii theory \cite{Pitaevski61,Gross61}
for a gas of Bose-Einstein condensed dressed molecules interacting with an effective
scattering length 
\begin{equation}
\label{eq:abare0}
a_{\rm{m}}^{\rm{Born}}\simeq2a(B)(1-Z).
\end{equation}
However, because 
in this regime we have $Z\simeq 0$, the scattering length between the dressed molecules
in the Born approximation in the mean-field approximation 
is given essentially by 
\begin{equation}
\label{eq:abare}
a_{\rm{m}}^{\rm{Born}}\simeq 2a(B).
\end{equation}
Such a situation is typical of a very broad resonance.

\subsection{Dressed molecules versus composite bosons.}
\label{sec:piero}

Our derivation in the previous section,
has to be considered as an extension of the
mean-field analysis of the BEC limit for 
a single-channel model crossover discussed by Sa de Melo {\it{et al.}} in Ref.
\cite{SadeMelo93} 
(see also Ref. \cite{RanderiaBook95} )
based on Eqs.~(\ref{eq:singlechangapeq}) and~(\ref{eq:Singlechstateeq}).
In the mean-field single-channel model the Cooper pairs evolve in the BEC limit
towards a dilute gas of 
weakly-interacting  
{\it{composite bosons}} with scattering length $2a(B)$.
In the BEC limit, the equation of state of the mean-field single-channel model
at zero temperature can be approximated by \cite{Pieri04a,Pieri04b,Haussmann94}
\begin{equation}
\label{eq:SinglechstateeqBEC}
n\simeq\frac{m^2 a(B)}{4\pi\hbar^4}|\Delta_{\rm{sc}}|^2.
\end{equation}
The composite boson condensate $\Psi_0^B$ of the single-channel model is 
defined from \cite{Pieri04a,Pieri04b,Haussmann94}
\begin{equation}
\label{eq:compositeboson}
\Delta_{\rm{sc}}\equiv \sqrt{\frac{8\pi\hbar^4}{m^2 a(B)}}\Psi_0^B.
\end{equation}
such that the BEC limit of Eq.~(\ref{eq:SinglechstateeqBEC}) is characterized by a 
gas of composite bosons 
\begin{equation}
\label{eq:aux00}
n\simeq 2|\Psi_0^B|^2.
\end{equation}
The equation of state in Eq.~(\ref{eq:SinglechstateeqBEC}) must be compared 
with its two-channel analogue in the BEC regime 
where
$2\mu\simeq\epsilon_{\rm{m}}$. We have
\begin{align}
\label{eq:aux0}
n&\simeq2 n_0+|\Delta|^2 \frac{  m^2} {4 \pi \sqrt{2m |\mu|} \hbar^3}\\
&= 2 Z n_{\rm{mc}} +\frac{2 Z n_{\rm{mc}} \eta}{2 \sqrt{|\epsilon_{\rm{m}}|}\left(
1+\sqrt{\frac{|\epsilon_{\rm{m}}|}{\epsilon_{bg}}}\right)^2}
= 2n_{\rm{mc}}\nonumber.
\end{align}
Moreover, for a very broad resonance, we have   
$2\mu\simeq\epsilon_{\rm{m}}\simeq-\hbar^2/m a^2(B)$ and 
$Z\simeq 0$. Therefore it is $\epsilon_{\rm{m}}\ll\epsilon_{\rm{bg}}$, and we find 
\begin{equation}
\label{eq:aux1}
g^2 Z\simeq g^2 \frac{2\sqrt{|\epsilon_{\rm{m}}|}}{\eta}=\frac{8\pi\hbar^4}{m^2 a(B)}.
\end{equation}

This relation between the residues of the 
poles associated with the composite boson and with the dressed molecule of the
two-channel model in the case 
of a very broad resonance can now be better understood if we
remember the physical meaning of $Z$ \cite{Duine04} and of $8\pi\hbar^4/m^2 {a(B)}$
\cite{Haussmann94,Pieri04a,Pieri04b}.
The factor $8\pi\hbar^4/m^2 a(B)$ reflects the difference between the bosonic
propagator of the single-channel composite boson and the particle-particle
ladder propagator in Eq.~(\ref{eq:apprTMBren}) in the BEC limit. In that limit
we have $2|\mu|\simeq|\hbar^2/m a(B)^2|\gg k_{\rm B} T $ and 
the latter can be approximated as
\begin{align}
\label{eq:aux2}
\hbar G_{\Delta}^{-1}&\left(\mathbf{K},\Omega_n\right)\!\!
={\rm T}_{\rm{MB}}^{-1}\left(\mathbf{K},\Omega_n\right)
\\&
\simeq 
\frac{m}{4\pi\hbar^2}\left[\frac{1}{a(B)}-\sqrt{\frac{m}{\hbar^2}}
\sqrt{-i\hbar\Omega_n+\frac{\hbar^2\mathbf{K}}{4 m}-2\mu}
\right]\nonumber.
\end{align}
Expanding for small $a(B)$ this can be rewritten in the {\it{polar}} 
form \cite{Pieri04b}
\begin{align}
\label{eq:aux3}
 G_{\Delta}\left(\mathbf{K},\Omega_n\right)/\hbar\!\!
=&{\rm T}^{\rm{MB}}\left(\mathbf{K},\Omega_n\right)
\\
\simeq 
&\frac{8\pi\hbar^4}{m^2 a(B)}~\frac{1}{i\hbar\Omega_n-\frac{\hbar^2\mathbf{K}}{4 m}-\mu_B}\nonumber,
\end{align}
where the chemical potential of the composite boson is defined as $\mu_B=2\mu-
\hbar^2/m a(B)^2$ and can be calculated from the single-channel
gap equation~(\ref{eq:singlechangapeq}) in the BEC limit. We find \cite{Pieri04b,Haussmann94}
\begin{equation}
\label{eq:chempotcompbos}
\mu_B\simeq\frac{4\pi\hbar^2 2a(B)}{2m}\simeq\frac{|\Delta_{\rm{sc}}|^2}{4|\mu|}.
\end{equation}
To be rigorous, we should have considered the particle-particle
ladder propagator in the superfluid state. However, in the
BEC limit, $\hbar\Omega_n$, $\hbar^2\mathbf{K}/4 m$ and $\mu_B$
are of the same order or smaller than $\hbar^2/ma(B)^2$.
Therefore the mean-field
inverse particle-particle propagator $\hbar G_{\Delta}^{-1}$, up to the first order
in $\hbar\Omega_n$ and $\hbar^2\mathbf{K}/4 m$, is the same as in the normal state.

The factor $Z$ reflects the difference between the bare and the dressed molecule
\cite{Duine04}. 
The full molecular propagator in Eq.~(\ref{eq:MBprop})
\begin{align}
\label{eq:remolprop}
\hbar &G^{-1}\left(\mathbf{K},\Omega_n\right)=
i\hbar\Omega_n-\frac{\hbar^2\mathbf{K}}{4 m}+2\mu-\delta(B)+\nonumber\\
-&\frac{1}{V}\sum_{\mathbf{k}}
|g^{\rm{2B}}\left(\mathbf{0},2\epsilon_\mathbf{k}\right)|^2\left[
\frac{1-N_{\mathbf{K}/2+\mathbf{k}}-N_{\mathbf{K}/2-\mathbf{k}}}{i\hbar\Omega_n+2\mu-
2\epsilon_\mathbf{k}-\epsilon_{\mathbf{K}}/2}+\frac{1}{2 \epsilon_\mathbf{k}}
\right],
\end{align}
can be approximated in the BEC regime, where $2\mu\simeq\epsilon_
{\rm{m}}\simeq-\hbar^2/m a^2(B)$ and 
$Z\simeq 0$, as
\begin{align}
\label{eq:remolpropbis}
\hbar &G^{-1}\left(\mathbf{K},\Omega_n\right)=
i\hbar\Omega_n-\frac{\hbar^2\mathbf{K}}{4 m}+2\mu-\delta(B)+\nonumber\\
-&\frac{1}{V}\sum_{\mathbf{k}}
|g^{\rm{2B}}\left(\mathbf{0},2\epsilon_\mathbf{k}\right)|^2\left[
\frac{1}{i\hbar\Omega_n+2\mu-
2\epsilon_\mathbf{k}-\epsilon_{\mathbf{K}}/2}+\frac{1}{2 \epsilon_\mathbf{k}}
\right].
\end{align}
However, 
for broad resonances, in this limit we have also that
$|\delta(B)| \gg 2|\mu|$, $\hbar^2\mathbf{K}/4 m$ and 
$\hbar\Omega_n$, $\hbar^2\mathbf{K}/4 m\ll\hbar
\Sigma\left(\mathbf{K},\Omega_n\right)$. Therefore, we
can approximate the dressed-molecular propagator further as
\begin{align}
\label{eq:remolproptris}
&\hbar G^{-1}\left(\mathbf{K},\Omega_n\right)\simeq
-\delta(B)\\&
-\frac{1}{V}\sum_{\mathbf{k}}
|g^{\rm{2B}}\left(\mathbf{0},2\epsilon_\mathbf{k}\right)|^2\left[
\frac{1}{i\hbar\Omega_n+2\mu-
2\epsilon_\mathbf{k}-\epsilon_{\mathbf{K}/2}}+\frac{1}{2 \epsilon_\mathbf{k}}
\right].
\end{align}
Calculating the integral this can be rewritten as
\begin{align}
\label{eq:remolproptrisbis}
\hbar G^{-1}&\left(\mathbf{K},\Omega_n\right)\simeq
\\&
\frac{1}{g^2}\cdot
\frac{m}{4\pi\hbar^2}\left[\frac{1}{a(B)}-\sqrt{\frac{m}{\hbar^2}}
\sqrt{-i\hbar\Omega_n+\frac{\hbar^2\mathbf{K}}{4 m}-2\mu}
\right]\nonumber
\end{align}
from which, expanding as in Eq.~(\ref{eq:remolproptris}), we obtain 
\begin{align}
\label{eq:remolpropquadris}
\hbar^{-1} G\left(\mathbf{K},\Omega_n\right)&\simeq \frac{1}{g^2}\cdot
\frac{8\pi\hbar^4}{m^2 a(B)}~\frac{1}{i\hbar\Omega_n-\frac
{\hbar^2\mathbf{K}}{4 m}-\mu_{\rm{m}}},
\end{align}
where $\mu_{\rm{m}}=\mu_{B}$.
However, expanding Eq.~(\ref{eq:remolproptris}) around 
the real pole at $\epsilon_{\rm{m}}\simeq\hbar^2/m a^2(B)$
the propagator can also be written as
\begin{align}
\label{eq:remolpropcinqris}
\hbar^{-1} G\left(\mathbf{K},\Omega_n\right)\simeq Z(B)~\frac{1}{i\hbar\Omega_n-\frac
{\hbar^2\mathbf{K}}{4 m}-\mu_{\rm{m}}}.
\end{align}
Comparing Eqs.~(\ref{eq:remolpropquadris}) and~(\ref{eq:remolpropcinqris}) we find again
the relation in Eq.~(\ref{eq:aux1}).
Our above analysis is based on a mean-field approach and due to this reason
the propagators of Eq.~(\ref{eq:remolpropquadris}) 
and~(\ref{eq:aux3}) describe free bosons. 
This is because we are considering only the bosonic propagators
only up to first order in $\hbar\Omega_n$ and $\hbar^2\mathbf{K}/4 m$.
\vspace{1 cm}
\subsection{Gaussian fluctuations around the saddle-point solution.}
\label{sec:Gauss}

In order to get in the BEC limit a superfluid of interacting bosons,
we have to extend our analysis beyond the saddle point approximation.
The nonzero temperature molecular propagator in the broken-symmetry state at the level of Gaussian
fluctuations can be approximated in the BEC limit by the matrix
\begin{widetext}
 \begin{align}
-\hbar {\mathbf{G}}
^{-1}
&\simeq 
\left( \begin{array}{cc} -i\hbar\omega_{n}-2\mu+\epsilon_{\mathbf K}/2 +\delta(B)
 &  0\\
 0 & i\hbar\omega_{n}-2\mu+\epsilon_{\mathbf K}/2
 +\delta(B)
 \end{array} \right)
+\left( \begin{array}{cc} \hbar\Sigma_{11}\left(
 \mathbf K,\omega_{n}\right)
 &  \hbar\Sigma_{12}\left(
 \mathbf K,\omega_{n}\right)\\
 \hbar\Sigma_{21}\left(
 \mathbf K,\omega_{n}\right) & 
 \hbar\Sigma_{22}\left(
 \mathbf K,\omega_{n}\right)
 \end{array} \right),
 \label{eq:broksymprop}
\end{align}
where the molecular selfenergies are given by \cite{Romans05}
\begin{align}
\label{eq:finttempmolselfsuper}
\hbar\Sigma_{11}({\bf q},i\omega_n) &\simeq
     \frac{
     g^2}{1+|a_{\rm bg}|\sqrt{-2\mu }}    
     \frac{1}{V}~\sum_{\mathbf{k}}
     \Bigg\{\frac{u_{\bf k}^2 u_{{\bf k}-{\bf q}}^2}
       {i\hbar\omega_n-\hbar\omega_{\bf k}-\hbar\omega_{{\bf k}-{\bf q}}}-
     \frac{v_{\bf k}^2 v_{{\bf k}-{\bf q}}^2}
       {i\hbar\omega_n+\hbar\omega_{\bf k}+\hbar\omega_{{\bf k}-{\bf q}}}+
     \frac{1}{2 \epsilon_{\bf k}}\Bigg\}~, \\
\hbar\Sigma_{12}({\bf q},i\omega_n) &\!\simeq\! 
    \frac{
     g^2}{1+|a_{\rm bg}|\sqrt{-2\mu}}
     \frac{2}{V}\sum_{\mathbf{k}} \Bigg\{\!
       u_{\bf k}v_{\bf k}u_{{\bf k}-{\bf q}}v_{{\bf k}-{\bf q}}\left(\!
     \frac{1}
       {i\hbar\omega_n-\hbar\omega_{\bf k}-\hbar\omega_{{\bf k}-{\bf q}}}
       \!+\!\frac{1}
       {+i\hbar\omega_n+\hbar\omega_{\bf k}+\hbar\omega_{{\bf k}-{\bf q}}}
           \!\right)\!\Bigg\},~\nonumber 
\end{align}
\end{widetext}
and $\hbar\Sigma_{22}({\bf q},i\omega_n)=\hbar\Sigma_{11}({\bf q},-i\omega_n)$,
$\hbar\Sigma_{12}({\bf q},i\omega_n)=\hbar\Sigma_{21}({\bf q},i\omega_n)$
and $\omega_{\bf k}=\sqrt{\Delta^2+(\epsilon_{\bf k}-\mu)^2}$
is the atomic spectrum in the BCS state. Note that the we have
neglected the thermal Fermi factors $N_{{\bf k}-{\bf q}}$ and $N_{\bf k}$ and that
the corrections due to the 
background interactions 
to the selfenergy (see also footnote \cite{explanation})
are approximated in the two-body normal state limit. This is justified in the BEC limit
where $|\mu|\gg \Delta$, $k_B T$.
To the lowest order in perturbation theory the dressed-molecule propagator
given by Eqs.~(\ref{eq:broksymprop}) and ~(\ref{eq:finttempmolselfsuper})
can be approximated by the matrix 
\begin{widetext}
\begin{align}
\label{eq:G-inverse-strong-coupling}
-\hbar {\mathbf{G}}^{-1}
&\simeq \frac{1}{Z} 
\left( \begin{array}{cc} -i\hbar\omega_{n}+\epsilon_{\mathbf K}/2 +
\frac{|\Delta|^{2}}{4|\mu|} \left(1-Z\right)\left(1+4\sqrt{\frac{|\epsilon_{\rm m}|}{\epsilon_{\rm bg}}}\right)
&  
\frac{|\Delta|^{2}}{4|\mu|} \left(1-Z\right)\left(1+4\sqrt{\frac{|\epsilon_{\rm m}|}{\epsilon_{\rm bg}}}\right)
\\
\frac{{|\Delta|}^{2}}{4|\mu|} \left(1-Z\right)\left(1+4\sqrt{\frac{|\epsilon_{\rm m}|}{\epsilon_{\rm bg}}}\right)
 & i\hbar\omega_{n}+\epsilon_{\mathbf K}/2
+\frac{|\Delta|^{2}}{4|\mu|}\left(1-Z\right)\left(1+4\sqrt{\frac{|\epsilon_{\rm m}|}{\epsilon_{\rm bg}}}\right)
 \end{array} \right)
\nonumber\\
&\simeq \frac{1}{Z} 
\left( \begin{array}{cc} -i\hbar\omega_{n}+\epsilon_{\mathbf K}/2 + {{\rm{T}}_{\rm m}^{\rm Born}}n_{\rm mc}
 &  
 {{\rm{T}}_{\rm m}^{\rm Born}}n_{\rm mc} 
 \\
 {{\rm{T}}_{\rm m}^{\rm Born}}n_{\rm mc} 
& 
i\hbar\omega_{n}+\epsilon_{\mathbf K}/2
+{{\rm{T}}_{\rm m}^{\rm Born}}n_{\rm mc}
 \end{array} \right) 
\end{align}
\end{widetext}
with 
\begin{align}
\label{eq:nonuniamm}
&{{\rm{T}}_{\rm m}^{\rm Born}}\equiv
{{\Gamma}_{\rm m}^{\rm Born}}\left(\mathbf 0,\mathbf 0,\mathbf 0,0\right)\nonumber\\
&=\left(4\pi\hbar^2/m_{\rm{m}}\right)2a_{\rm eff}(B)
\left(1-Z\right)^2\left(1+4\sqrt{\frac{|\epsilon_{\rm m}|}{\epsilon_{\rm bg}}}\right).
\end{align}
The derivation of Eq.~(\ref{eq:G-inverse-strong-coupling})
follows essentially from
an expansion in $|\Delta|/|\mu|$ 
of the normal and anomalous molecular selfenergies
at finite energy and momentum retaining only the linear terms 
in $\hbar\omega_n$ and $\epsilon_{\mathbf K}/2$~\cite{Andrenacci03}.
A sketch of the derivation is given in Appendix A.
Note that in  
Eq.~(\ref{eq:nonuniamm}),
we have made use of 
$|\Delta|^{2}/4|\mu|= 
\left(4\pi\hbar^2/m_{\rm{m}}\right)2a_{\rm eff}(B)(1-Z)n_{\rm mc}
\simeq\left(4\pi\hbar^2/m_{\rm{m}}\right)2a(B)n_{\rm mc}$,
where $a_{\rm eff}(B)$ has been defined in Eq.~(\ref{eq:aeff}).  
Clearly, Eq.~(\ref{eq:nonuniamm}) implies that the molecule molecule scattering
length in the Born approximation is given by 
\begin{align}
\label{eq:nonuniamm1}
a_{\rm m}^{\rm Born }=2a_{\rm eff}(B)
\left(1-Z\right)^2\left(1+4\sqrt{\frac{|\epsilon_{\rm m}|}{\epsilon_{\rm bg}}}\right).
\end{align}
In the BEC regime investigated experimentally until now
for the broad resonance in $^6{\rm Li}$, where $\epsilon_
{\rm{m}}\simeq-\hbar^2/m a^2(B)$ this essentially reduces to
\begin{align}
\label{eq:nonuniamm2}
a_{\rm m}^{\rm Born}=2a(B)
\left(1-Z\right)^2\left(1+4\sqrt{\frac{|\epsilon_{\rm m}|}{\epsilon_{\rm bg}}}\right).
\end{align}
However, we know that in that regime we have $Z\simeq 0$ and ${\epsilon_{\rm bg}}\gg|\epsilon_{\rm m}|$,
such that 
\begin{align}
\label{eq:nonuniamm3}
a_{\rm m}^{\rm Born}\simeq2a(B)
\end{align}
in a very good approximation.
The results of Eqs.~(\ref{eq:nonuniamm1},\ref{eq:nonuniamm2}) and~(\ref{eq:nonuniamm3})
has to be compared with the mean-field results we obtained in Eqs.~(\ref{eq:mubare0},\ref{eq:abare0})
and~(\ref{eq:abare}).
Moreover, it can be shown
that the non universal correction 
induced by the factor $\left(1-Z\right)^2
\left(1+4\sqrt{\frac{|\epsilon_{\rm m}|}{\epsilon_{\rm bg}}}\right)$
is independent of the 
approximation used in the calculation of the molecule-molecule
vertex interaction.
Therefore we can anticipate that
the full dressed-molecule scattering length 
should obey 
\begin{align}
a_{\rm m}(B)\simeq 0.6 ~a(B)~\left(1-Z\right)^2
\left(1+4\sqrt{\frac{|\epsilon_{\rm m}|}{\epsilon_{\rm bg}}}\right), 
\end{align}
in the BEC limit of the crossover region
after
including higher-order corrections \cite{Kagan05,Levinsen05}.

For $\delta\leq -\epsilon_{\rm{bg}} $,
the approximation $\epsilon_
{\rm{m}}\simeq-\hbar^2/m a^2(B)$ is not valid, and
the dressed-molecules scattering length is expected to be
\begin{align}
a_{\rm m}(B)\simeq 0.6 ~a_{\rm eff}(B)~\left(1-Z\right)^2
\left(1+4\sqrt{\frac{|\epsilon_{\rm m}|}{\epsilon_{\rm bg}}}\right), 
\end{align}
which vanishes as $|\epsilon_{\rm m}|^{-3}$ at larger negative detunings
when the wavefunction renormalization factor $Z$
goes fast to $Z\simeq 1$.
Unfortunately, the molecule-molecule scattering length of $^6$Li$_2$ molecules
has not been 
investigated yet at such large negative magnetic fields ($B<650 G$).

From the Bogoliubov propagator of Eq.~(\ref{eq:G-inverse-strong-coupling})
we find an energy spectrum 
for the dressed molecules \cite{Romans05} which is linear at low momenta
\begin{align}
E_{\rm{m}}\left(\mathbf{K}\right)=\sqrt{\left(\hbar^2{\mathbf K}^2/2 m_{\rm{m}}\right)^2
+\left(\hbar^2{\mathbf K}^2/2 m_{\rm{m}}\right)2{{\rm{T}}_{\rm m}^{\rm Born}}n_{\rm m}}.
\end{align}

In the remaining part of this section
we want to show that in this regime the gas is mainly constituted by a gas of interacting dressed
molecules. The mean-field equation of state
in Eq.~(\ref{eq:stateeq}), however, needs to be modified after 
the introduction of the fluctuations that lead to the 
superfluid dressed-molecule
propagator of Eq.~(\ref{eq:G-inverse-strong-coupling}). 
In a self-consistent approach \cite{Haussmann94}, the total number of particles
is given by the formula
\begin{eqnarray}
\label{eq:Serene}
n=2Zn_{\rm{mc}}-2\frac{1}{\hbar \beta V}{\rm{Tr}}\left[{{\mathbf{G}}}\right]+
2\frac{1}{\hbar \beta V}{\rm{Tr}}\left[{\mathbf{G}_{\rm f}}\right],
\end{eqnarray}
where the fermionic single-particle propagator $\mathbf{G}_{\rm f}$ is given by
the matrix
\begin{widetext}
\begin{align}
\left( \begin{array}{cc} G_{\rm f,11}^{-1}(\mathbf{k},i\omega_n) 
& G_{\rm f,12}^{-1}(\mathbf{k},i\omega_n)
  \\ G_{\rm f,21}^{-1}(\mathbf{k},i\omega_n)
& G_{\rm f,22}^{-1}(\mathbf{k},i\omega_n) \end{array} \right)
\!\! = \!\! \left( \begin{array}{cc} {G}_{\uparrow,0}(\mathbf{k},i\omega_n)^{-1} & 0 \\ 0 & -
{G}_{\downarrow,0}(-\mathbf{k},i\omega_n)^{-1} \end{array} \right)  
\!\!- \!\!
\left( \begin{array}{cc} \hbar\Sigma_{11}^{\rm f}(\mathbf{k},i\omega_n) & 
\hbar\Sigma^{\rm f}_{12}(\mathbf{k},i\omega_n)+\Delta \\
\hbar\Sigma^{\rm f}_{21}(\mathbf{k},i\omega_n)+\Delta^{\star} & \hbar\Sigma_{22}^{\rm f}(\mathbf{k},i\omega_n) \end{array}
\right) \label{Dyson-equation},         
\end{align}
\end{widetext}
where the fermionic selfenergies $\hbar\Sigma^{\rm f}_{ij}(\mathbf{k},i\omega_n)$
contain the feedback effects of the dressed-molecules
on the atoms.
The term ${\rm{Tr}}\left[{{\mathbf{G}}}\right]$ is easily evaluated according to the theory
of a weak-interacting Bose gas. We have 
\begin{align}
\label{eq:Bog}
&2 Z n_{\rm{mc}}-2\frac{1}{\hbar \beta V}{\rm{Tr}}\left[{{\mathbf{G}}}\right]=2 Z n_{\rm{mc}}+\\
&+
2 Z \int \! \frac{d {\mathbf q}}{(2 \pi)^{3}} \left[
u_{\rm{m}}^{2}({\mathbf q}) N_B(E_{\rm{m}}({\mathbf q}))
- v_{\rm{m}}^{2}({\mathbf q}) N_B(-E_{\rm{m}}({\mathbf q})) \right]\nonumber\\
&=2 Z\left(n_{\rm{mc}}(T)+n'_{\rm{m}}(T)\right)\nonumber
\end{align}
where
\begin{equation}
v_{\rm{m}}^{2}({\mathbf q}) \, = \, u_{\rm{m}}^{2}({\mathbf q}) - 1 \, = \,
\frac{\frac{{\mathbf q}^{2}}{2m_{\rm{m}}} \, + {{\rm{T}}_{\rm m}^{\rm Born}}n_{\rm{mc}} \, - \,
E_{\rm{m}}({\mathbf q})}{2 E_{\rm{m}}({\mathbf q})}   
\label{u-v-Bogoliubov}
\end{equation}
are the standard bosonic factors of the Bogoliubov transformation \cite{Fetter64}
and $n'_{\rm{m}}(T)$ is the noncondensed density of dressed molecules
at temperature $T$.
The density of Eq.~(\ref{eq:Bog}) corresponds to the
bare-molecular contribution to the total density. It is very small because in the BEC limit
that we are considering
in a broad resonance, we have $Z\simeq0$. therefore, we show that the main contribution
in the equation of state in Eq.~(\ref{eq:Serene}) comes 
from ${\rm{Tr}}\left[{\mathbf{G}_{\rm f}}\right]$. To calculate this trace we have first
to calculated the fermionic selfenergies. 
Upon neglecting higher order contributions, we ultimately find \cite{Pieri04a} 
\begin{widetext}
\begin{align}
\label{eq:Fermipseudo}
\noindent \hbar\left( 
 \begin{array}{cc} G_{\rm f,11}^{-1}(\mathbf{k},i\omega_n) 
& G_{\rm f,12}^{-1}(\mathbf{k},i\omega_n)
  \\ G_{\rm f,21}^{-1}(\mathbf{k},i\omega_n)
& G_{\rm f,22}^{-1}(\mathbf{k},i\omega_n) \end{array} 
 \right)
  =&\left( \begin{array}{cc} i\hbar\omega_n+\mu-
 \epsilon_{\mathbf{k}}-\frac{{\Delta'}^2}{i\hbar\omega_n-\mu+\epsilon_{\mathbf{k}}} 
  &\Delta \\ \Delta^{\star}  & -
i\hbar\omega_n-\mu+\epsilon_{\mathbf{k}}
 -\frac{{\Delta'}^2}{i\hbar\omega_n-\mu+\epsilon_{\mathbf{k}}}\end{array} \right)
\end{align}
\end{widetext}
where 
\begin{equation}
{\Delta'}^2=\frac{g^2 Z n_{\rm{m}}'}{\left(
1+\sqrt{\frac{|2\mu|}{\epsilon_{bg}}}\right)^2}.
\end{equation}
Note that we have neglected the off-diagonal fermionic selfenergies because in
the limit under consideration one can show that 
$\hbar\Sigma^{\rm f}_{12}(\mathbf{0},0)\simeq |\Delta|(|\Delta|^2/2|\mu|^2)$. 
From Eq.~(\ref{eq:Fermipseudo}) we get the expression for $G_{\rm f,11}$
in the BEC limit \cite{Pieri04a}
\begin{equation}
\label{eq:pseudogap1}
G_{\rm f,11}({\mathbf k},\omega_{s}) \, \simeq \, \frac{\hbar}{i\hbar\omega_{s}+\mu- 
\epsilon_{\mathbf k} \, - \,
\frac{|\Delta|^{2} \, + \, {\Delta'}^{2}}{i\omega_{s}-\mu+
\epsilon_{\mathbf k}}}                 
\end{equation}
where we have neglected a term of order ${\Delta'}^{2}/|\mu|$ with respect
to $|\mu|$.
Note that Eq.~(\ref{eq:pseudogap1}) has the same formal structure of the corresponding
BCS expression \cite{Fetter64}, with the replacement
$\Delta^2\rightarrow(\Delta^2+{\Delta}'^2)$.
Accordingly, the trace of the fermionic propagator can be rewritten in 
the BCS-like form
\begin{align}
\label{eq:fictiousBCS}
&\frac{1}{\hbar \beta V}~2~{\rm{Tr}}\left[{\mathbf{G}_{\rm f}}\right]\simeq\\ &\simeq
\frac{1}{(2\pi)^3}\int d^3{\bf{k}} \left(
1-\frac{\epsilon_{\bf k}-\mu}{\sqrt
{(|\Delta|^2+{\Delta}'^2)+(\epsilon_{\bf k}-\mu})^2}\right)\nonumber.
\end{align}
Expanding the right-hand side of this equation as we did for the mean-field
BCS equation of state in Eq.~(\ref{eq:expstateeq}) we have 
\begin{equation}
\label{eq:fictiousBCSexp}
\frac{1}{\hbar \beta V}~2~{\rm{Tr}}\left[{\mathbf{G}_{\rm f}}
\right]\simeq\left(|\Delta|^2+{\Delta}'^2\right) \frac{  m^2} {4 \pi \sqrt{2m |\mu|} \hbar^3}.
\end{equation}
Moreover, using $2\mu\simeq |\epsilon_{\rm{m}}|$ and the definition
of two-body $Z$ we find 
\begin{align}
\label{eq:usa0}
2\frac{1}{\hbar \beta V}{\rm{Tr}}\left[{\mathbf{G}_{\rm f}}
\right]&\simeq\frac{2 Z \left(n_{\rm{mc}}+n_{\rm{m}'}\right) \eta}{2 \sqrt{|\epsilon_{\rm{m}}|}\left(
1+\sqrt{\frac{|\epsilon_{\rm{m}}|}{\epsilon_{bg}}}\right)^2}\\&
\simeq 2\left(1-Z\right)\left(n_{\rm{mc}}+{n_{\rm{m}}'}\right)\nonumber.
\end{align}
Joining the two contributions of Eqs.~(\ref{eq:usa0}) and~(\ref{eq:Bog})
we have
\begin{equation}
\label{eq:finale}
n\simeq 2 \left(n_{\rm{mc}}(T)+n'_{\rm{m}}(T)\right).
\end{equation}
This result holds asymptotically for $k_B T\ll|\epsilon_{\rm{m}}|$ and it
represents the extension
to a two-channel model in the case of a broad resonance,
of the description of the BEC limit in Ref. \cite{Pieri04a,Pieri04b}
based on the composite bosons of the single-channel model.
However, it is important to stress that these results are only valid
in the lowest order of perturbation theory. 
Furthermore, we use a definition of the probability $Z$ that is only true
in the asymptotic BEC limit. 
The calculation of $Z$ in the strong-coupling region
requires 
a more general approach as shown in Ref.~\cite{Romans05}.

Nevertheless,  
our analysis of the BEC limit, shows the way to connect the two models
in the limit of a very broad resonance,  
where the single-channel
model accounts fairly well for the thermodynamics of the gas inside the 
BEC-BCS crossover region.

\section{Conclusions and outlook.}

In this paper we have shown that the dressed-molecule method represents an
effective and consistent
approach in order to describe the physics of ultracold Fermi gases
near a Feshbach resonance.
The dressed molecule is the real physical entity of the BEC-BCS
crossover in atomic Fermi gases near a Feshbach resonance, because the wavefunction
of the Feshbach molecules is always a coherent superposition.
In this formulation
the information about the mixing of the two-channels
is preserved at each stage in the many-body calculation.
Therefore it has the advantage that it can be applied  
also to medium and narrow reconances, when $\eta^2\leq \epsilon_F$,
where the single-channel approximation
is expected to fail.
However, a proper treatment of the narrow resonance case requires
the inclusion of the finite range corrections in the theory \cite{Duine04}.
This will be addressed elsewhere.

\section*{Acknowledgements}
We are most grateful to R. Duine, B. Farid, P. Pieri, P. Coleman, F. Nogueira, M. Haque,
F. Marchetti, B. Marcelis, S. Kokkelmans, T. K{\"o}hler and R. G. Hulet
for stimulating discussions and criticisms.
We would like also to thank Professor H. Kleinert and his group for the hospitality at
the Institute of Theoretical Physics of the Free University of Berlin.

\section*{Appendix A.}

In this appendix we present the detailed 
calculation which leads to the dressed-molecule propagator of Eq.~(\ref{eq:G-inverse-strong-coupling})
from Eq.~(\ref{eq:broksymprop}). In order to make the derivation more transparent we discuss first the case 
when $a_{\rm bg}=0$, because the inclusion of the background interactions 
does not change qualitatively the discussion.
Neglecting the background scattering length corrections,
the molecular selfenergies of Eq.~(\ref{eq:finttempmolselfsuper})  reduce to  
\begin{widetext} 
\begin{align}
\label{eq:finttempmolselfsuperabgzero}
\hbar\Sigma_{11}({\bf q},i\omega_n) &=
     \frac{g^2}{V}~\sum_{\mathbf{k}}
     \Bigg\{\frac{u_{\bf k}^2 u_{{\bf k}-{\bf q}}^2}
       {i\hbar\omega_n-\hbar\omega_{\bf k}-\hbar\omega_{{\bf k}-{\bf q}}}-
     \frac{v_{\bf k}^2 v_{{\bf k}-{\bf q}}^2}
       {i\hbar\omega_n+\hbar\omega_{\bf k}+\hbar\omega_{{\bf k}-{\bf q}}}+
     \frac{1}{2 \epsilon_{\bf k}}\Bigg\}~, \\
\hbar\Sigma_{12}({\bf q},i\omega_n) &\!=\! 
     \frac{2g^2}{V}\sum_{\mathbf{k}} \Bigg\{\!
       u_{\bf k}v_{\bf k}u_{{\bf k}-{\bf q}}v_{{\bf k}-{\bf q}}\left(\!
     \frac{1}
       {-i\hbar\omega_n+\hbar\omega_{\bf k}+\hbar\omega_{{\bf k}-{\bf q}}}
       \!+\!\frac{1}
       {i\hbar\omega_n+\hbar\omega_{\bf k}+\hbar\omega_{{\bf k}-{\bf q}}}
           \!\right)\!\Bigg\},~\nonumber 
\end{align}
\end{widetext}
where, in the definition of the gap of Eq.~(\ref{eq:totgap}) entering the definition of the coherence factors
$v_{{\bf k}}$ and $u_{{\bf k}}$,
the contribution due to the background interaction has to be neglected as well.
In the BEC limit, where $2 \mu\simeq\epsilon_{\rm{m}}\gg k_B T,\Delta$, we can expand the 
selfenergies of Eq.~(\ref{eq:finttempmolselfsuperabgzero}) for small values of $|{\bf q}|$ and $\omega_n$ as
\begin{align}
\label{eq:appExp}
\hbar\Sigma_{11}({\bf q},i\omega_n)=& A_0+ A_1 ~i\hbar\omega_n+A_2 ~\epsilon_{\bf q}/2+O(\omega_n^2,|{\bf q}|^4)\nonumber\\
\hbar\Sigma_{12}({\bf q},i\omega_n)=& B_0+O(\omega_n^2,|{\bf q}|^2).
\end{align}
with the coefficients of the expansion given by
\begin{widetext}
\begin{align}
\label{eq:appcoeff}
&A_0=\frac{g^2}{V}~\sum_{\mathbf{k}}\frac{1}{2\epsilon_{\bf k}}-\frac{1}{2\omega_{\bf k}}
+\frac{g^2}{V}~\sum_{\mathbf{k}}\frac{|\Delta|^2}{4~\omega_{\bf k}^3},
&A_1=-\frac{g^2}{V}~\sum_{\mathbf{k}}\frac{\epsilon_{\bf k}-\mu}{4~\omega_{\bf k}^3},
\nonumber\\
&A_2=\frac{g^2}{V}~\sum_{\mathbf{k}}\frac{1}{8}\Bigg\{
\frac{\left(\epsilon_{\bf k}-\mu\right)\left(2\left(\epsilon_{\bf k}-\mu\right)^2-|\Delta|^2\right)}{4\omega_{\bf k}^5}
+
\frac{2\epsilon_{\bf k}
|\Delta|^2\left(8\left(\epsilon_{\bf k}-\mu\right)^2+3|\Delta|^2\right)}{3~\omega_{\bf k}^7}
\Bigg\},
&B_0=\frac{g^2}{V}~\sum_{\mathbf{k}}\frac{|\Delta|^2}{4~\omega_{\bf k}^3}.
\end{align}
Substituting these results in Eq.~(\ref{eq:broksymprop})
and using the gap equation of Eq.~(\ref{eq:varrengap}) the dressed-molecule propagator
can be rewritten as
 \begin{align}
-\hbar {\mathbf{G}}
^{-1}
&\simeq 
\left( \begin{array}{cc} -i\hbar\omega_{n}+\epsilon_{\mathbf K}/2 
-\frac{g^2}{V}~\sum_{\mathbf{k}}\frac{|\Delta|^2}{4~\omega_{\bf k}^3}
+A_1 i\hbar\omega_n+A_2 ~\epsilon_{\bf K}/2
 &  \frac{g^2}{V}~\sum_{\mathbf{k}}\frac{|\Delta|^2}{4~\omega_{\bf k}^3}\\
 \frac{g^2}{V}~\sum_{\mathbf{k}}\frac{|\Delta|^2}{4~\omega_{\bf k}^3} 
 & i\hbar\omega_{n}+\epsilon_{\mathbf K}/2-\frac{g^2}{V}~\sum_{\mathbf{k}}\frac{|\Delta|^2}{4~\omega_{\bf k}^3}
-A_1 i\hbar\omega_n+A_2 ~\epsilon_{\bf K}/2
 \end{array} \right).
 \label{eq:broksympropqq}
\end{align}
This expression can be simplified further by noting that in the BEC limit one has
\begin{align}
&\frac{g^2}{V}~\sum_{\mathbf{k}}\frac{|\Delta|^2}{4~\omega_{\bf k}^3}\simeq\left(\frac{1-Z}{Z}\right)
\frac{|\Delta|^{2}}{4|
\mu|},
&A_1\simeq A_2\simeq\left(\frac{Z-1}{Z}\right),
\end{align}
in the leading order of $|\Delta|^2/|\mu|$.
Therefore, we get
\begin{align}
\label{eq:G-inverse-strong-couplingApp}
-\hbar {\mathbf{G}}^{-1}\!
\simeq \!\frac{1}{Z} 
\left( \!\!\begin{array}{cc} -i\hbar\omega_{n}+\epsilon_{\mathbf K}/2 +
\frac{|\Delta|^{2}}{4|\mu|} \left(1-Z\right)
&  
\frac{|\Delta|^{2}}{4|\mu|}\left(1-Z\right) 
\\
\frac{{|\Delta|}^{2}}{4|\mu|} \left(1-Z\right) & i\hbar\omega_{n}+\epsilon_{\mathbf K}/2
+\frac{|\Delta|^{2}}{4|\mu|}\left(1-Z\right)
 \end{array} \!\!\right).
\end{align}
\end{widetext}

This derivation can be generalized rather straightforwardly in order to
include the effects of the background scattering length in the molecular selfenergies 
of Eq.~(\ref{eq:finttempmolselfsuper}).
The coefficients of  Eq.~(\ref{eq:appcoeff}) 
differ only by an overal factor
$1/\left(1+|a_{\rm bg}|\sqrt{-2\mu }\right)$,  
and $Z$, $\epsilon_{\rm m}$, $\Delta$ and the gap equation,
must be replaced everywhere with  
the more general expressions  
that include the background scattering length
corrections.     
Note, however, that 
the background scattering corrections in 
the selfenergies of Eq.~(\ref{eq:finttempmolselfsuper})
are valid only in the BEC limit.
Therefore, in order to
semplify the dressed-molecule propagator as we have done in Eq.~(\ref{eq:broksympropqq}),
we cannot use the general gap equation of Eq.~(\ref{eq:varrengap}).
Alternatively, we expand the coefficients of Eq.~(\ref{eq:appcoeff})
in powers of $\Delta/|\mu|$ and then we use the expanded version of the 
gap equation of Eq.~(\ref{eq:expvarrengap}).
\begin{widetext}
Ultimately we find
\begin{align}
\label{eq:G-inverse-strong-couplingApp1}
-\hbar {\mathbf{G}}^{-1}\!
\simeq \!\frac{1}{Z} 
\left( \!\!\begin{array}{cc} -i\hbar\omega_{n}+\epsilon_{\mathbf K}/2 +
\frac{|\Delta|^{2}}{4|\mu|} 
\left(1-Z\right)\left(1+4\sqrt{\frac{|\epsilon_{\rm m}|}{\epsilon_{\rm bg}}}\right)
&  
\frac{|\Delta|^{2}}{4|\mu|}\left(1-Z\right)\left(1+4\sqrt{\frac{|\epsilon_{\rm m}|}{\epsilon_{\rm bg}}}\right) 
\\
\frac{{|\Delta|}^{2}}{4|\mu|} 
\left(1-Z\right)\left(1+4\sqrt{\frac{|\epsilon_{\rm m}|}{\epsilon_{\rm bg}}}\right) 
& i\hbar\omega_{n}+\epsilon_{\mathbf K}/2
+\frac{|\Delta|^{2}}{4|\mu|}\left(1-Z\right)\left(1+4\sqrt{\frac{|\epsilon_{\rm m}|}{\epsilon_{\rm bg}}}\right)
 \end{array} \!\!\right).
\end{align}
\end{widetext}
as given in Eq.~(\ref{eq:G-inverse-strong-coupling}).

\bibliographystyle{apsrev}

\begin{thebibliography}{104}
\bibitem{GriffinComm}
This appears to be contradicted by the results obtained
in Ref. \cite{Ohashi0203}, which uses a Gaussian approximation for the two-channel
model and finds a monotonic curve for the critical temperature.
However, we believe that this is a result of the soft cut-off
introduced to regulate the ultraviolet divercencies in the theory.
\bibitem{explanation}
The molecular selfenergy of Eq.~(\ref{eq:MBSelfenergy}) contains the 
many-body corrections through the energy dependence of the atom-molecule
coupling. In Ref.\cite{Ohashi0203} the many-body molecular selfenergy is 
derived formally in a RPA-like structure. We use also this latter form systematically
in Sec.~\ref{crossover} to avoid introducing an energy-dependent
gap as in Ref. \cite{Romans05} that would complicate unnecessarily the variational
calculation at $T=0$. 
We have seen in Eq.~(\ref{eq:self}) that the two formulations
are essentially equivalent in the two-body limit. It can also be shown 
that in the two extremes of the crossover they are essentially equivalent.
It must be pointed out, however, that 
in Ref.~\cite{Ohashi0203} Ohashi and 
Griffin use a different cut-off procedure to regularize the ultraviolet
divergencies in the momentum integrals.
As a final remark, let us observe that the effect of using an energy-dependent 
atom-molecule coupling as in the definition of the selfenergy of Eq.~(\ref{eq:MBSelfenergy})
consists in shifting the high-momentum cut-off of the theory
from infinity \cite{SadeMelo93} 
to $1/a_{\rm bg}$. When the background scattering length is very small this 
of course does not make any difference. However, when it is rather large
as in $^6$Li mixtures it can lead to important differences in the 
strong coupling region \cite{Romans05}.
\end{thebibliography}

\end{document}